\documentclass[final,3p,times]{elsarticle}
\pdfoutput=1

\usepackage[hidelinks]{hyperref}

\usepackage{amssymb}
\usepackage{amsmath}
\usepackage[font=footnotesize,labelfont=bf]{caption}
\usepackage{subcaption}
\usepackage{placeins}
\usepackage[]{algorithm2e}
\usepackage[utf8]{inputenc}

\usepackage[usenames, dvipsnames]{color}

\newcommand{\bvec}[1]{\mathbf{#1}}
\newcommand{\x}[0]{\bvec{x}}

\newcommand{\q}[0]{\bvec{q}}
\newcommand{\VP}[0]{\bvec{P}}
\newcommand{\T}[0]{\bvec{T}}
\newcommand{\N}[0]{\bvec{N}}
\newcommand{\B}[0]{\bvec{B}}

\newcommand{\bm}[1]{{\bvec{#1}}}
\newcommand{\vertex}[1]{v_{#1}}
\newcommand{\edge}[2]{e_{#1#2}}
\newcommand{\facet}[3]{f_{#1#2#3}}
\newcommand{\simplex}[4]{s_{#1#2#3#4}}
\newcommand{\vertexN}[1]{v}
\newcommand{\edgeN}[2]{e}
\newcommand{\facetN}[3]{f}
\newcommand{\simplexN}[4]{s}

\newcommand{\contribN}[4]{C_{#1}\left(\vertexN{#2},\edgeN{#2}{#3},\facetN{#2}{#3}{#4}\right)}

\newcommand{\sumAdjTwoN}[2]{\sum_{\left(\vertexN{#1},\edgeN{#1}{#2}\right)}}
\newcommand{\sumAdjThreeN}[3]{\sum_{\left(\vertexN{#1},\edgeN{#1}{#2},\facetN{#1}{#2}{#3}\right)}}

\newcounter{myalgo}
\newenvironment{myalgo}[4][]{\refstepcounter{myalgo}\par\medskip\noindent%
  \textbf{Algorithm~\themyalgo.} #2
  \begin{itemize}
  \item[] \textbf{Input:} #3
  \item[] \textbf{Output:} #4
    \rmfamily}{
  \end{itemize}\medskip}%

\begin{document}

\begin{frontmatter}



\title{{\tt ColDICE}: a parallel Vlasov-Poisson solver using moving adaptive simplicial tessellation}
\author[IAP,UT,RESCEU]{Thierry Sousbie\corref{cor}}
\ead{tsousbie@gmail.com}
\author[IAP]{St\'{e}phane Colombi}
\ead{colombi@iap.fr}
\address[IAP]{Institut d'Astrophysique de Paris, CNRS UMR 7095 and UPMC, 98bis, bd Arago, F-75014 Paris, France}
\address[UT]{Department of Physics, The University of Tokyo, Tokyo 113-0033, Japan}
\address[RESCEU]{Research Center for the Early Universe, School of Science, The University of Tokyo, Tokyo 113-0033, Japan}
\cortext[cor]{Corresponding author}

\begin{abstract}
Resolving numerically Vlasov-Poisson equations for initially cold systems can be reduced to following the evolution of a three-dimensional sheet evolving in six-dimensional phase-space. We describe a public parallel numerical algorithm consisting in representing the phase-space sheet with a conforming, self-adaptive simplicial tessellation of which the vertices follow the Lagrangian equations of motion. The algorithm is implemented both in six- and four-dimensional phase-space. Refinement of the tessellation mesh is performed using the bisection method and a local representation of the phase-space sheet at second order relying on additional tracers created when needed at runtime.  In order to preserve in the best way the Hamiltonian nature of the system, refinement is anisotropic and constrained by measurements of local Poincar\'e invariants. Resolution of Poisson equation is performed using the fast Fourier method on a regular rectangular grid, similarly to particle in cells codes. To compute the density projected onto this grid, the intersection of the tessellation and the grid is calculated using the method of Franklin and Kankanhalli \cite{Franklin83,Franklin87,Franklin93} generalised to linear order. As preliminary tests of the code, we study in four dimensional phase-space the evolution of an initially small patch in a chaotic potential and the cosmological collapse of a fluctuation composed of two sinusoidal waves. We also perform a ``warm'' dark matter simulation in six-dimensional phase-space that we use to check the parallel scaling of the code. 
\end{abstract}


\end{frontmatter}
\section{Introduction}
\label{sec:intro}
Stars in galaxies and dark matter in the Universe can be described as a smooth self-gravitating collisionless fluid following Vlasov-Poisson equations,
\begin{eqnarray}
\frac{\partial f}{\partial t} 
+ \bm{u}.\nabla_{\bm{r}}f - \nabla_{\bm{r}}\phi.\nabla_{\bm{u}}f=0, \label{eq:vlaeq} \\
\Delta_{\bm{r}} \phi=4\pi G\rho=4\pi G 
\int f(\bm{r},\bm{u},t)\ {\rm d} \bm{u}, \label{eq:poieq}
\end{eqnarray}
where $f(\bm{r},\bm{u},t)$ represents the phase-space density at position $\bm{r}$, velocity $\bm{u}$ and time $t$, $\phi$ is the gravitational potential and $G$ is the gravitational constant.

In this article, we focus on the cold case, relevant to the dynamics of cold dark matter. In the concordant model of large scale structure formation \cite{Planck2014,Planck2015}, the matter content in Universe is indeed dynamically dominated by a cold and collisionless component, designated by ``dark'' matter as it does not emit detectable light or radiation.  The cold nature of this component implies that the phase-space distribution function is initially concentrated on a phase-space sheet: at the macroscopic level, the thickness of the this sheet is virtually null:
\begin{equation}
f(\bm{r},\bm{u},t=t_{\rm i})=\rho_{\rm i}(\bm{r})\ \delta_{\rm
  D}[\bm{u}-\bm{u}_{\rm i}(\bm{r}) ],
\label{eq:inicondfroid}
\end{equation}
where $\rho_{\rm i}(\bm{r})$ is the initial density distribution, $\bm{u}_{\rm i}$ the initial velocity field and $\delta_{\rm D}$ the Dirac distribution function. In this case, the matter is initially concentrated on a $D=3$ hypersurface in 2$D=6$-dimensional phase-space. 

Liouville theorem states that the phase-space density is conserved along trajectories,
\begin{equation}
f[\bm{r}(t),\bm{u}(t),t]={\rm constant},
\label{eq:liouville}
\end{equation}
for any point $[\bm{r}(t),\bm{u}(t)]$ following the equations of motion. This means that topological properties of the phase-space distribution function are conserved during motion, in particular that the phase-space sheet, i.e. the region where $f$ is non null remains a non self-intersecting three-dimensional hypersurface at all times. 

In our understanding of large scale structure formation, the initial density field $\rho_{\rm i}(\bm{r})$ is close to constant and the initial velocity field, when subtracted from the expansion of the Universe, is very small: the large scales structures observed today, such as clusters of galaxies, filaments and underdense regions nearly empty of galaxies \citep[see, e.g.,][]{Gott2005}, grew from small initial fluctuations in the density field through gravitational instability \cite{Peebles1980}. 

Vlasov-Poisson equations are traditionally resolved numerically with a $N$-body approach \cite[see, e.g.,][for reviews]{Bertschinger1998,Colombi2001,Dolag2008,Dehnen2011}. There exist many kinds of $N$-body codes, among those one can list direct summation codes (PP for particle-particle interactions) \cite{Aarseth1979,Aarseth1999}, particle-mesh (PM) codes \cite{Melott1982,Melott1983,Bouchet1985} coming initially from plasma physics \cite{Hockney1981}, treecodes \cite{Appel1985,Barnes1986,Hernquist1987,Bouchet1988,Springel2001} as well as hybrid codes such as P$^3$M (PM combined with PP for local interactions) \cite{Hockney1981,Efstathiou1985}, treePM (PM combined with treecode for local interactions) \cite{Xu1995,Bagla2002,Springel2005}, adaptive mesh refinement codes (AMR) \cite{Villumsen1989,Suisalu1995,Splinter1996,Gelato1997,Kravtsov1997,Knebe2001,Teyssier2002,Bryan2014} and AP$^3$M (P$^3$M with AMR) \cite{Couchman1991}. In all these methods, which mainly differ from each other by the way Poisson equation is solved, the phase-space distribution function is represented by an ensemble of particles, that is a set of Dirac functions in phase-space interacting with each other through gravitational forces. To avoid numerical instabilities due to close collisions, the gravitational force is smoothed at scales smaller than a softening parameter $\varepsilon$ which corresponds to the local grid resolution in mesh based methods such as PM and AMR.

The representation of a smooth distribution with a discrete set of macro-particles can have non trivial consequences on the numerical behaviour of the system.\footnote{We do not discuss here the self-consistent
  field method \cite{Clutton1972,Vanalbada1977,Hernquist1992,Hozumi1997}, 
because it is seldomly used. In this approach, the
  projected density and the gravitational potential are represented on a finite set of carefully chosen
  smooth functions of which the respective weights are computed from a set of
particles following the equations of motion. There is no softening
needed in this method and the noise introduced by the
tracers is different from what is expected in standard $N$-body simulations but is still unquestionably present
\cite{Hernquist1990}.}  Close $N$-body encounters and collective effects due to the shot noise of the particles may drive the system away from the expected solution in the mean field limit \cite{Aarseth1988,Hernquist1990,Goodman1993,Splinter1998,Binney2002,Boily2002,Binney2004,Diemand2004,Joyce2009}. For instance, shot noise of the particles can introduce significant distorsions of the phase-space sheet as well as nonlinear resonant instabilities \cite{Alard2005,Colombi2014,Colombi2015},  which can have dramatic consequences  on the numerical behaviour of the system, particularly in the cold case \cite{Melott1997,Melott2007,Angulo2013}. Hence, the fine structure of dark matter halos is still the object of debates despite multiple convergence studies with $N$-body simulations
\cite{Navarro1996,Navarro1997,Moore1998,Jing2000,Jing2002,Power2003,Springel2008,Stadel2009,Boylan2009,Klypin2015}.

For all these reasons and because computational power now allows it, it is justified to explore alternative numerical routes and to try solving Vlasov-Poisson directly in phase-space without resorting to particles. This is important to confirm many results obtained with the traditional $N$-body approach and that are used to test the cold dark matter scenario paradigm against observations. 

In the warm case, i.e. where the initial velocity dispersion is non negligible, there exists a very rich literature about Vlasov solvers, mainly coming from plasma physics. Many of these solvers exploit directly Liouville theorem (\ref{eq:liouville}). Among them, one can cite the famous splitting scheme of Cheng and Knorr \cite{Cheng1976} first applied to astrophysical systems by \cite{Fujiwara1981,Watanabe1981,Nishida1981,Fujiwara1983}. In this algorithm, the phase-space distribution function is sampled on a mesh. It is updated between two time steps by following backwards the motion of test particles and by using an interpolation scheme to compute $f$ from the particles positions at previous time step. The procedure is performed  
in a split fashion, by decomposing the Hamiltonian motion into a ``drift'' (e.g., position update using velocity) and a ``kick'' (e.g., velocity update using acceleration). This algorithm is {\em semi-Lagrangian}, in the sense that it relies on the calculation of characteristics. 

Many other grid based Vlasov solvers have been proposed since the seminal contribution of Cheng and Knorr, most of them being of semi-Lagrangian nature \citep[see, e.g.,][but this list is far from being exhaustive]{Shoucri1978,Shlosman1979,Zaki1988,Utsumi1998,Nakamura1999,Sonnendrucker1999,Filbet2001,Aber2002,Besse2003,Alard2005,Umeda2008,Crouseilles2009,Crouseilles2010,Qiu2010,CamposPinto2011,Rossmanith2011,Heath2012,Guclu2014}. For instance one can mention the recent Vlasov-Poisson simulations of Yoshikawa, Yoshida \& Umemura \cite{Yoshikawa2013} in six-dimensional phase-space using the positive flux conservation scheme \cite{Filbet2001}. 

While it can be of interest for describing warm astrophysical systems such as galaxies, or cosmological fluids with significant velocity dispersion such as the neutrino distribution in the Universe, sampling the phase-space distribution function on an Eulerian grid seems unrealistic in the cold case, although not impossible. One way to deal with the cold nature of the system could consist in following a coarse grained version of the phase-space distribution function as proposed by Klimas \cite{Klimas1987,Klimas1994}, originally to fix problems of filamentation in the warm case. Another way would be to approximate the phase-space density by a slightly warm distribution and to perform adaptive refinement in phase-space \cite{Gutnic2004,Alard2005,Mehrenberger2006,Besse2008,CamposPinto2008} to concentrate on the locus of the phase-space sheet. 

An even more direct way to exploit Liouville theorem is to use a pure Lagrangian approach and to exactly transport the phase-space distribution from initial conditions. For instance, one could trace back test particles trajectories to initial conditions to reconstruct, at a given time, the phase-space density at any point \cite{Rasio1989}. An important simplification occurs if one assumes that $f$ is constant inside patches: in this case, it can be seen from equation (\ref{eq:liouville}) that only the boundary of these patches needs to be followed according to the equations of motion, which reduces the effective
number of dimensions of the problem by one. This is the essence of the waterbag method \cite{Depackh1962,Roberts1967,Bertrand1968,Bertrand1970,Janin1971,Cuperman1971,Colombi2008,Colombi2014}.  For instance, in 2-dimensional phase-space, the boundary of the patches, the ``waterbags'', can be drawn with polygons, while in 4-dimensional phase-space, the boundary becomes three-dimensional and needs to be sampled with a tessellation, e.g. of tetrahedra.  Obviously, it seems difficult to apply the waterbag method as just stated in 6-dimensional phase-space, because the cost of sampling a five-dimensional phase-space hypersurface with simplices seems prohibitive. 

Another issue with the waterbag method is that it is needed to add sampling elements on the boundary of the patches during runtime. Indeed,  because of mixing, the waterbags boundaries get elongated in phase-space and form rich structures, for instance spirals or more complex patterns. The cost of the waterbag method thus increases with time and becomes obviously exorbitant in the presence of chaos. This is not the case of grid based methods, which are, at variance with the waterbag method, not entropy conserving: details of the phase-space distribution function are erased below the grid resolution --be it
possibly adaptive-- which allows one to control the cost of the method. In the waterbag approach, limiting the cost of the scheme could theoretically be achieved by employing techniques analogous to contour surgery \cite{Dritschel1989}, but this method was never tested, to our knowledge, in gravitational dynamics and is probably incompatible with the fact that according to Liouville theorem, topology has to be preserved in phase-space. 

Because of its algorithmic complexity and the computational cost issues mentioned above, the waterbag method was never really exploited beyond 2D in phase-space.\footnote{One can however mention the gyrokinetic waterbag model in four-dimensional phase-space \cite[see, e.g.,][]{Besse2009,Coulette2013}, but in the current implementations, the borders of the patches are followed in an Eulerian manner, hence requiring some approximations to deal with shell-crossings in configuration space.} 
 
However, the cold case we consider in this article represents in fact an interesting limiting case of the waterbag method. Indeed, the phase-space sheet is equivalent to an infinitely thin waterbag, reducing furthermore the dimensionality of the problem to following a 3D hypersurface in 6-dimensional phase-space instead of a 5D one in the warm case: this makes the approach feasible despite the computational cost issues mentioned above. This is the algorithm we aim to implement, in the spirit of the waterbag method: we propose to sample the phase-space sheet with an adaptive tessellation of which the vertices follow the equation of motion. Our approach is therefore analogous to the recent scheme proposed by Hahn, Abel and Kaehler \cite{Hahn2013} and Hahn and Angulo \cite{Hahn15}, although completely different in the actual implementation. It also follows closely ideas discussed in \cite{Shandarin12,Abel12}, where the concept of sampling the phase-space density with a Lagrangian tessellation was introduced in the cosmological context. 

More specifically, we propose to sample the phase-space sheet with a conforming simplicial tessellation, that is with an ensemble of joint tetrahedra that coincide exactly on facets when intersecting with each other. To start with, the phase-space sheet follows equation (\ref{eq:inicondfroid}) with $\rho_{\rm i}=$ constant and $\bm{u}_{\rm i}=0$ and is sampled with a homogeneous simplicial mesh. Its vertices are then slightly perturbed using Lagrangian perturbation theory \citep[see, e.g.][]{Zeldovich1970,Bouchet1992,Buchert1992,Buchert1993,Bouchet1995,Bernardeau2002} to create the initial state. Once the initial conditions are set-up, the vertices of the tessellation evolve according to the Lagrangian equations of motion.  

We also consider 4-dimensional phase-space corresponding to $D=2$ dimensions in space and 2 dimensions in velocity. In this case, the phase-space sheet is two-dimensional and is tessellated with triangles; physically, the system under consideration corresponds to a continuum of infinite lines interacting gravitationally with each other, with a force proportional to the inverse of the distance between the lines, or in other words to a logarithmic potential instead of the usual $1/r$ one in the $D=3$ case. 

Because of mixing, it is necessary to add sampling elements when needed, i.e. to introduce local refinement on the tessellation. To do so, we use techniques analogous to finite element methods \citep[see, e.g.][]{Ho1988,Lo2002,Zienkiewicz2005}: the phase-space sheet is refined using bisection \citep[see, e.g.,][and references therein]{Rivara84,Bansch90,Kossaczky94,Liu94,Arnold00,Zhang09}, that is by cutting when required some simplices into two smaller simplices while preserving the conforming nature of the tessellation at all times. New vertices created during this procedure are placed in such a way that the local representation of the phase-space sheet remains accurate at second order. Indeed, because a significant amount of curvature is generated during the course of dynamics, following locally the shape of the phase-space sheet at second order greatly improves the quality of the representation of the system and is nearly a must \citep[see also][]{Hahn15}. 

In order to preserve as well as possible the Hamiltonian nature of the system, the criterion we use to decide when simplices have to be refined as well as the way they are refined relies on the measurement of local Poincar\'e invariants, that is contour integrals of the form
\begin{equation}
I =\oint \bm{u}.{\rm d}\bm{r}(s)
\label{eq:poinc}
\end{equation}
over closed curves in phase-space composed of points following the equations of motion. In Hamiltonian systems, such contour integrals should be conserved during motion \citep[see, e.g.,][]{Meiss92,Binney2008}: similarly to the waterbag
code presented in \cite{Colombi2014}, our refinement criterion tries to set limits to violations of this property. 

To best follow the dynamics, our refinement is {\em anisotropic}, which is made possible with the bisection technique. This is a major difference between our implementation and that of Hahn \& Angulo \cite{Hahn15} who employ regular refinement in Lagrangian space, by cutting each tetrahedron into 8 smaller ones. Using anisotropic refinement can be much more efficient than regular one if there are preferred directions in the dynamics.

To solve Poisson equation, we project the tessellation onto a regular rectangular grid, by adapting the volume calculation method of Franklin \cite{Franklin83,Franklin87,Franklin93} using raytracing and generalising it to first order, that is with a hypersurface density represented at linear order inside each simplex.  This is another major difference between our algorithm and that of Hahn of Angulo, who sample each simplex with a set of regularly distributed particles describing the phase-space sheet shape at second order prior to projection onto the grid. To speed up the process at a small cost in accuracy, we use an AMR technique in such a way that the size of the mesh elements is locally of the same order of that of the simplices under consideration. In this sense, our calculation of the intersections is exact at linear order but only at scales comparable to the size of simplices, while the calculation of Hahn and Angulo is valid up to second order but is not free of noise due to discreteness effects. Gathering the information on the final fixed resolution grid is performed by a simple donor cell procedure. Then force calculation and vertex position and velocity updates are performed just as in standard PM codes. 

Because following the evolution of a 3D adaptive tessellation in 6D phase-space remains a very costly exercise, our code
is fully parallel: at the local level, it exploits shared memory parallelism with OpenMP library, while distributed memory parallelism is achieved at the coarser level with domain decomposition techniques using MPI library.

This article is organised as follows. Section \ref{sec_mesh} details our parallel implementation of the adaptive simplicial tessellation, also designated by simplicial mesh. After introducing some terminology (\S~\ref{sec_meshTerm}), we describe the parallel structure of the tessellation (\S~\ref{sec_meshImpl}) and how refinement and coarsening are performed (\S~\ref{sec_meshRefine}). 

Section \ref{sec_exactProj} deals with projection, i.e. with the calculation of the intersection of the tessellation with a rectangular, possible locally adaptive mesh. After introducing the formalism allowing one to compute integrals of functions inside polyhedral volumes up to linear order (\S~\ref{sec_franklinForm}), our version of the projection algorithm of Franklin is described in \S~\ref{sec_exactProjImpl}. Subtle but nonetheless critical issues related to degenerate cases in the algorithm are discussed and resolved in \S~\ref{sec_robustness} to enforce its full robustness, while possible accuracy problems in the actual calculation of the intersecting masses are fixed in \S~\ref{sec:accu}. Finally, parallelisation is addressed in \S~\ref{sec:paralfrank}.

Section \ref{sec:vlapoi} deals with the Vlasov-Poisson solver itself. We first explain how initial conditions are implemented (\S~\ref{sec:ic}) and describe our locally quadratic representation of the phase-space sheet inside each element of the tessellation (\S~\ref{sec_quadraticElements}). Details on the time step are given in \S~\ref{sec:timest}, followed by a description of the calculation of the acceleration (\S~\ref{sec:projdenscal}), using the projection algorithm of \S~\ref{sec_exactProj} to compute the projected density on a fixed resolution grid over which Poisson equation is solved in Fourier space with the {\tt FFTW} library. We finally discuss our anisotropic refinement procedure based on the measurements of local Poincar\'e invariants (\S~\ref{sec:aniso}). 

Section \ref{sec:exemple} illustrates the performances of our code with various examples, namely the evolution of an initially small patch in the presence of a fixed but chaotic potential (\S~\ref{sec:2dchaos}), the collapse of two intersecting sine waves in two dimensions (\S~\ref{sec:doublesine}) and finally a cosmological simulation in the framework of a fictive warm dark matter (WDM) model (\S~\ref{sec_WDM}). Simple analyses are performed, such as estimates of radial density profiles, tests of total energy conservation, measurements of total number of simplices and phase-space sheet surface/volume as functions of time.  

Section \ref{sec:parallelscaling} tests OpenMP (\S~\ref{sec_ompScaling}) and MPI (\S~\ref{sec_mpiScaling}) parallelism on the WDM simulation. 

Finally, section \ref{sec:conclusion} briefly discusses potential improvements of the code. 

To lighten the presentation, some technical details are deferred when needed to a set of appendices.

The C++ code developed in the framework of this article is publicly available and can be retrieved, together with a few movies and illustrations, from the following website: \href{http://www.vlasix.org/index.php?n=Main.ColDICE}{www.vlasix.org}\footnote{see also \url{https://github.com/thierry-sousbie/dice}}. All major features of the code, including the adaptive tessellation, AMR grid and exact projection algorithms are implemented as a standalone C++ template library, {\tt DICE}. The Vlasov-Poisson solver itself, {\tt ColDICE}, uses these libraries as an application.

\section{The distributed adaptive simplicial mesh}
\label{sec_mesh}
In this section, we present our implementation of the distributed adaptive simplicial mesh, which is provided as the first main component of the standalone library {\tt DICE}. The different choices of design we made along the course of development were mainly driven by the requirements of the Vlasov-Poisson solver but the library can be used for other purposes. We indeed tried to opt for an approach as generic and flexible as possible so that future developments are not too limited by design choices and ended-up with the following guidelines:
\begin{itemize}
\item The mesh can be distributed and load balanced over large computer clusters.
\item The implementation should work in 2D and 3D, with support for an embedding space of arbitrary dimension.
\item A simplicial mesh suffices. We therefore implemented triangular and tetrahedral meshes, without requiring support for more exotic elements (such as pyramids, prisms, ...).
\item Mesh elements can be iterated easily and in parallel (i.e. support shared memory multi-threading such as OpenMP).
\item At least periodic and free boundary conditions should be supported, but other boundary types could be added easily.
\item Elements neighbourhood can be quickly recovered, including along the boundaries of distributed regions, so that demanding algorithms such as the one presented in section \ref{sec_exactProj} can be implemented easily.
\item Anisotropic refinement and (optionally) coarsening are supported, and implemented in a flexible way.
\item The different implementation choices are made to optimize flexibility, speed and memory consumption, in this order.
\end{itemize}

\subsection{Terminology}
\label{sec_meshTerm}
Let us start by introducing the terminology we will use in the rest of this article concerning the unstructured mesh. Let a $k$-simplex be the convex hull of a set of $k+1$ points that we call its vertices: a $0$-simplex is a point, a $1$-simplex a line segment, a $2$-simplex a triangle, a $3$-simplex a tetrahedron, ... A $p$-face of a $k$-simplex ($p<k$) is the $p$-simplex formed from a $p+1$ distinct elements subset of its vertices: the $2$-faces of a tetrahedron are its $4$ facets, the $1$-faces its $6$ edges and the $0$-faces its $4$ vertices. In general, we will designate as the {\em faces} of a $k$-simplex its $(k-1)$-faces (i.e. the faces of a triangle are its edges, while the faces of a tetrahedron are its facets). The concept of $k$-face can be used to define a notion of neighbourhood over simplices. More specifically, a $k$-simplex $K$ and a $q$-simplex $Q$ with $k<q$ are said to be incident if $K$ is a $k$-face of $Q$: the tetrahedra incident to a given edge are all the tetrahedra that contain it entirely. Similarly, $K$ and $Q$ are said to be adjacent if they have at least a vertex in common: the tetrahedra adjacent to a given edge are all the tetrahedra that contain any vertex of that edge, which includes those incident to it.

A $d$-dimensional simplicial mesh designates a set of $d$-simplices (i.e. triangles for $d=2$, tetrahedra for $d=3$) that we will simply call simplices when their dimension is unspecified and matches that of the mesh. Refining the notion of adjacency, we will in general call neighbours two simplices of a mesh that are adjacent (i.e. they share exactly one face): two tetrahedra in a 3D mesh are neighbours if they share one facet. In this article, we will only consider meshes for which any face of a simplex has at most two simplices incident to it\footnote{a necessary but not sufficient condition for the mesh to be a manifold.} so a tetrahedron has at most $4$ neighbours and a $k$-simplex in general can have at most $k+1$ neighbours. Such a mesh is said to be conforming if all the simplices intersect only through shared $k$-faces: a 3D conforming simplicial mesh is a set of tetrahedra that only intersect at their vertices, edges or facets. If a mesh is non-conforming, the non-conforming intersections are called hanging nodes. Note that a $d$-dimensional mesh is only a topological structure defined over a set of vertices, and that it can be embedded in a $D$-dimensional space simply by mapping its vertices to points of that embedding space.

\subsection{Implementation}
\label{sec_meshImpl}
Supercomputers featuring large clusters of shared memory nodes are becoming the norm, with a continuing trend of increasing number of cores per node. Taking advantage of such processing power is challenging, especially for problems such as gravitational dynamics that are by essence non-local and for which significant inter-process communication cannot be avoided. Scaling up to (tens of) thousands of cores using pure MPI communication, the standard message passing interface for distributed memory computers, often results in numerous messages being sent all over the network, triggering traffic contentions that almost invariably end up being the limiting performance factors. One way to alleviate this problem consists in using a hybrid approach, combining coarse grained MPI parallelism with local shared-memory multiprocessing via for instance OpenMP. Indeed, MPI parallelism is oftentimes achieved through domain decomposition, each MPI distributed sub-domain communicating preferentially with its direct neighbors, but also potentially with all other sub-domains via all-to-all type communications. The usage of local OpenMP style parallelism allows for larger and less numerous sub-domains. In this case, neighbor-to-neighbor communications, that are often achieved via buffer regions called ``ghost layers''  locally keeping track of neighboring domains boundaries updates, are therefore reduced, since these regions typically scale like the surface of the sub-domains. Similarly, all-to-all communications are also improved because they are made via larger but less numerous messages, which reduces network contentions. Another potential advantage of the hybrid approach is the possibility of exploiting different types of parallelism in a very efficient way, and we therefore choose this approach here.

In our implementation, a global mesh $M$ is decomposed into $P$ non intersecting sub-meshes distributed among $P$ MPI processes. All computations local to a MPI process are parallelised via OpenMP. In order to reduce inter-process communication while maximizing flexibility, we allow for a locally stored ghost layer of simplices that are only updated on need. Storing such a ghost layer increases per-process memory requirements in a reasonable way as long as the surface to volume ratio of the local sub-meshes can be kept low (i.e. as long as they are compact and as large as possible). On the other hand, this procedure gives us a lot of flexibility in implementing complex algorithms that require a knowledge of the neighborhood of each element of the mesh and can potentially improve performances by reducing inter-process communications. The global distribution of sub-meshes is simplex-based, and given a mesh composed of a large number of simplices, each process is assigned a different sub-set of these simplices and their vertices. Note that only simplices and vertices have to be stored explicitly since intermediate elements can be defined implicitly: in 3D, for instance, tetrahedra facets can be uniquely identified as pairs of a tetrahedron $t$ and an opposite vertex in $t$ while segments are identified as pairs of vertices.

\begin{figure}
\centering
\includegraphics[width=0.95\linewidth]{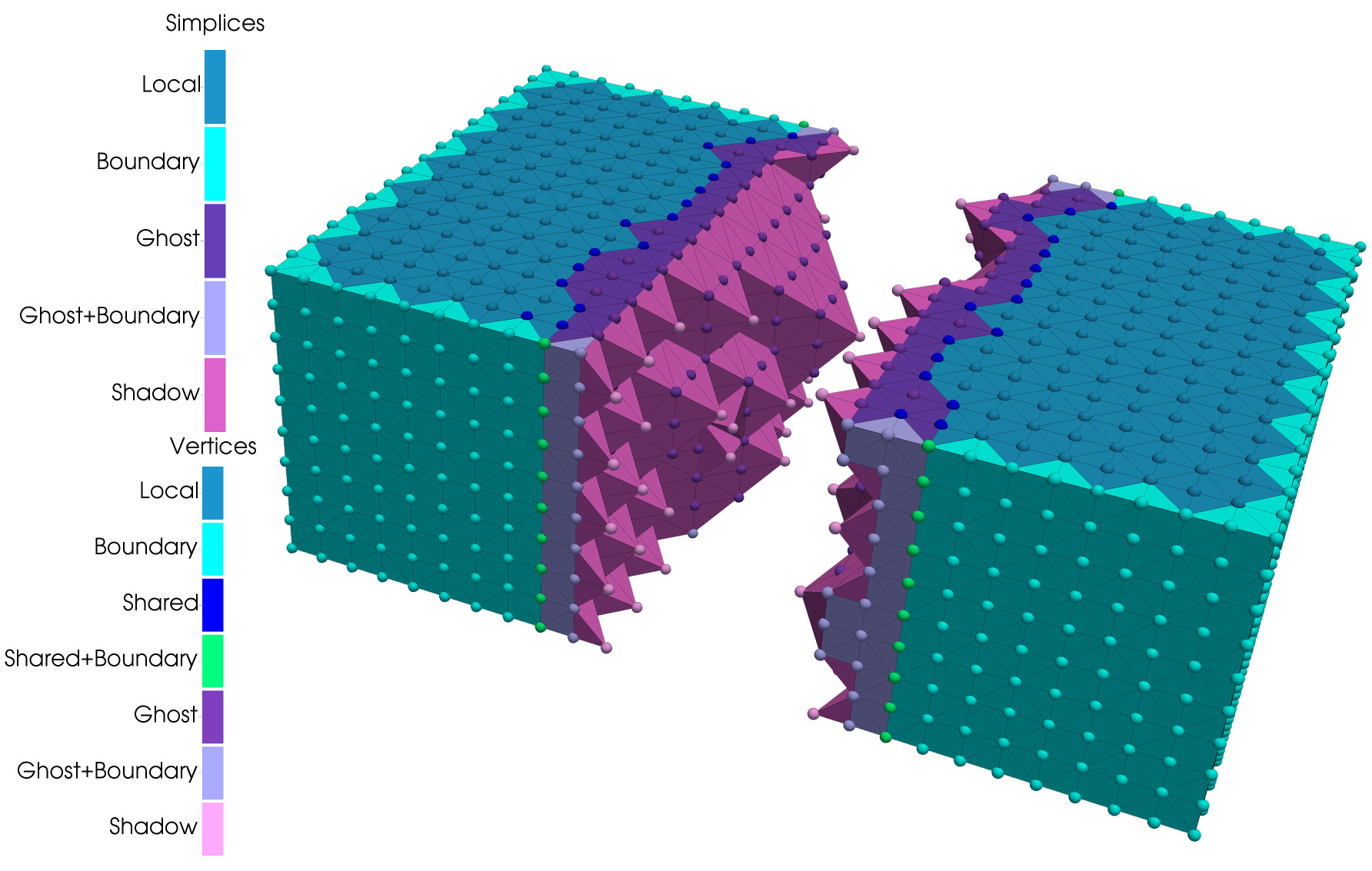}
\caption{Denomination of the different types of simplices and vertices in the distributed mesh. A cubic domain decomposed into a tetrahedral mesh has been divided into two sub-domains and distributed to two MPI processes. The upper parts of each sub-domain has been clipped so that the inner region can be seen. The medium and light blue tetrahedra (labeled local and boundary) are assigned to their respective domain, while other tetrahedra (labeled "ghost" and "shadow") are only images of tetrahedra that belong to a different domain. The vertices have similar classifications except for the dark blue ones labeled as "shared" that belong to the frontier between two domains.\label{fig_3Dmesh}}
\end{figure}

Given these requirements, we divide simplices and vertices stored on a given MPI process into possibly intersecting groups that we label as follows:
\begin{itemize}
\item \textbf{Simplices labels:}
  \begin{description}
  \item[Local:] a simplex that belongs to the local process.
  \item[Boundary:] a $d$-simplex with less than $d+1$ neighbours. A face of a boundary simplex is a boundary face if it has only one simplex incident to it.
  \item[Ghost:] a non-local simplex adjacent to (sharing at least one vertex with) a local simplex.
  \item[Shadow:] a non-local and non-ghost simplex which is the neighbour of a ghost simplex.
  \end{description}
\item \textbf{Vertices labels:}
  \begin{description}
  \item[Local:] a vertex adjacent only to local simplices.
  \item[Boundary:] a vertex that belongs to at least one boundary face (see ``boundary simplex'' item).
  \item[Shared:] a vertex adjacent to at least one local simplex and one ghost simplex.
  \item[Ghost:] a vertex adjacent to at least one ghost simplex and exactly $0$ local simplex.
  \item[Shadow:] a vertex only adjacent to one or more shadow simplices.
  \end{description}
\end{itemize}
According to these generic definitions and as illustrated on figure~\ref{fig_3Dmesh}, the ghost layer therefore contains any non local simplex incident to a vertex that belongs to a local simplex. This allows for a quick and easy local retrieval of the data associated to elements adjacent to any local element. We moreover add a so-called shadow layer, that contains all the missing ghost simplices neighbours so that direct neighbours of any ghost simplex can also be retrieved locally.

From a technical point of view, the mesh has to be generated before it is distributed. However, because it is potentially very large, it may be too expensive to be stored on a single process. We therefore implement the initial tessellation implicitly: every initial simplex and vertex is identified by a unique index and the indices of adjacent elements are retrieved on the fly using arithmetic so that the memory footprint is very small. Such a result can obviously only be achieved for simple initial domain geometries and so far only the tessellated box and torus\footnote{or equivalently, periodic boundary box.} have been implemented, with the possibility however to create multi-component domains (such as e.g. two disconnected boxes). Using such implicit tessellation, it is easy to create an initial explicit tessellation which is arbitrarily distributed over each process without requiring any communication. The quality of this initial distribution can then be improved by re-partitioning it with the help of the open source libraries {\tt METIS} and {\tt ParMetis} \cite{METIS}, that we also use to re-partition the domains to load balance the charge of each MPI process during the calculations. The technique used in ParMetis is called parallel multilevel k-way graph-partitioning~\cite{Karypis99,Karypis00}. It is based only on the connectivity graph of the mesh (although the geometry can also optionally be used to help the process), and allows for the fast generation of high quality connected partitions that minimize the number of cuts or equivalently the surface to volume ratio of each partition. A standard Peano-Hilbert curve partitioning is also optionally available.

Another issue related to the distribution of the mesh across MPI processes is the consistent indexing of vertices and simplices. Let us from now on consider a mesh $M$ with $N^S$ simplices and $N^V$ vertices distributed over $P$ MPI processes. Each process has $N^S_p$, $N^{GS}_p$, $N^{SS}_p$ and $N^V_p$, $N^{GV}_p$, $N^{SV}_p$ local, ghost and shadow simplices and vertices respectively. Then, in our implementation, each of them is given three types of index:
\begin{itemize}
  \item A {\em local index} $i\in [0,N_p[$ identifies each simplex and vertex on the local process. Local, ghost and shadow simplices indices are consecutive integers ranging from $0$ to $N^{S}_p-1$,  $N^{GS}_p-1$, and $N^{SS}_p-1$ respectively and vertices indices are defined in a similar way. 
  \item A {\em global identity} that uniquely identifies each simplex and vertex over the whole cluster. The global identity of each simplex is a 64-bit integer whose first $16$ bits store its local rank $R$, i.e. the rank of the process where the simplex is local, and the remaining $48$ bits store the local index on this process. The global identity of vertices is built in a similar way, except for the shared vertices which are local to several processes. The first $16$ bits of a shared vertex global identity is set to the lowest rank $R_{\rm min}$ among processes that share it. The local index used in the global identity, in this case, corresponds to the local index of the vertex in process $R_{\rm min}$ and might thus differ from the local index in process under consideration, $R$. 
  \item A {\em global index}, only defined for simplices, that can be built on the fly from the global identity. Global indices are consecutive integers ranging from $0$ to $N^S-1$ and are computed as $\sum_{p=0}^{r-1}N^S_p + i$ where $r$ is the local rank of the simplex and $i$ is its local index on the corresponding  process (as stored in the second $48$ bit part of the global identity).
\end{itemize}
Note that, in our implementation, ghost, shadow simplices and all types of vertices need to store their local index and global identity. The global identity of local simplices, on the other hand, can be computed on the fly from their local index and the rank of the process they belong to. 

Given the elements described above and the adaptive nature of the mesh, we adopt a very simple data structure based on the simplices to store the mesh. Simplices are defined by an array of pointers to their vertices, each of them being stored in a consistent order over the whole mesh so that the orientation of the simplices is well defined. Additionally to these pointers, an array of pointers to the neighbours of each simplex is also stored in an order such that the $n^{\rm th}$ neighbour is the one adjacent to the simplex via the facet opposite to its $n^{\rm th}$ vertex, which eases navigation and construction in general. Moreover, each simplex also contains an 8-bit ``flag'' used to record the state of the simplex, a 32-bit integer to store the local index and a 64-bit ``cache'' variable that can be used for convenience by different algorithms to store temporary data. Ghost and shadow simplices also record an additional 64-bit global identity. Vertices, on the other hand, store their respective coordinates as $D$ double or simple precision floating point numbers with $D$ the dimension of the embedding space, as well as a 32-bit local index, a 64-bit global identity and an 8-bit flag variable. Accounting for data structure padding required for memory management, this rounds-up to a total of $80$ bytes per simplex and $64$ bytes per vertex in the case of a 3D mesh embedded in a $D=6$ dimensional space stored on a 64-bit architecture using double precision coordinates, which is only an absolute minimum as the users are also given the possibility of very easily adding their own variables directly inside the simplices and vertices data structure in order to gain direct access to them, depending on their needs.

We finish this section with what is probably the most critical implementation issue: the organisation and management of the different elements in memory. In order to be practical and efficient, the data structure that handles them must indeed fulfil several constraints:
\begin{itemize}
\item It must be possible to insert new elements in a constant time.
\item It must be possible to remove elements in a constant time (for coarsening and load balancing).
\item The address in memory of each element should not change during its lifetime. Indeed, references to simplices and vertices are stored as pointers and such changes would necessitate a complete update of the local sub-mesh, which is unacceptable in terms of performances.
\item It must be possible to iterate easily over the different types of elements. Moreover, it should be possible to do so in parallel, different threads scanning different subsets of elements. In both cases, iterating must take a constant time per element.
\item The user should be given the possibility of randomly accessing elements in constant time through simple indexing as some specific algorithms require it.
\item Memory usage should always be kept as low as possible.
\item Locality in memory should be preserved as much as possible, in order to improve performances by allowing more frequent processor cache hits.
\end{itemize}
We therefore use a specifically designed memory pool structure whose implementation is detailed in \ref{app_mesh}. This particular design allows us to fulfil all the previously mentioned requirements except for the random access to elements. This however can be achieved relatively easily and at a minimal memory cost of one pointer per element by simply storing pointers to simplices and/or vertices in an array, each pointer being stored at index $n$ where $n$ is the local index of the simplex/vertex it points to. Finally, we also point out that the performances of most algorithms can be greatly improved by increasing the locality of data access. In particular, it is often very profitable to store spatially close elements as contiguously as possible in memory so that the usage of the processor cache is maximised. This can for instance be achieved through an occasional sort along a Peano-Hilbert curve as explained in \ref{app_mesh}
.

\subsection{Refinement and coarsening}
\begin{figure}
\begin{centering}
\hfill%
\begin{subfigure}[b]{0.3\textwidth}
\centering
\includegraphics[width=.99\linewidth]{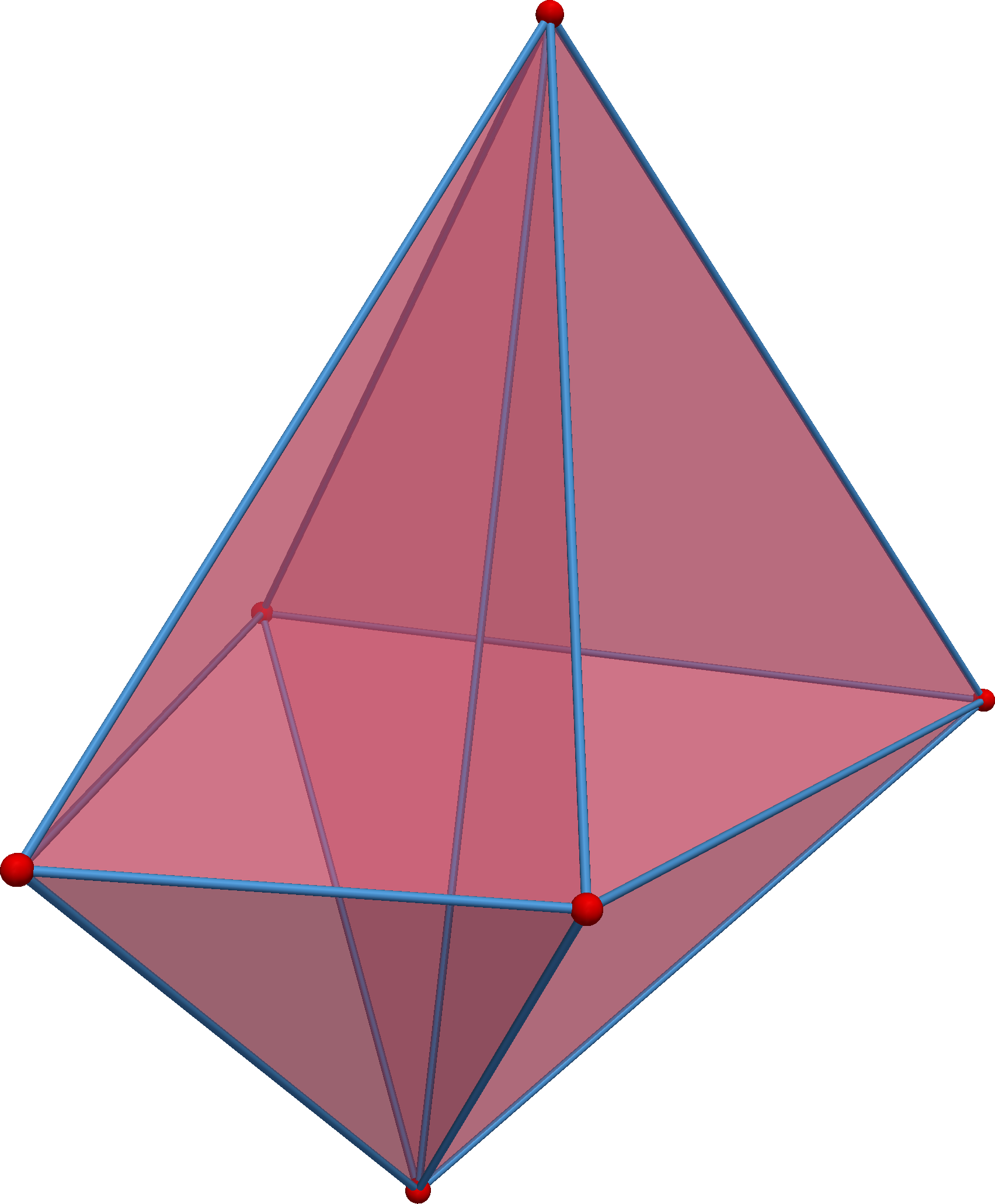}
\caption{Before edge bisection\label{fig_tetRefineBefore}}
\end{subfigure}
\hfill%
\begin{subfigure}[b]{0.3\textwidth}
\centering
\includegraphics[width=.99\linewidth]{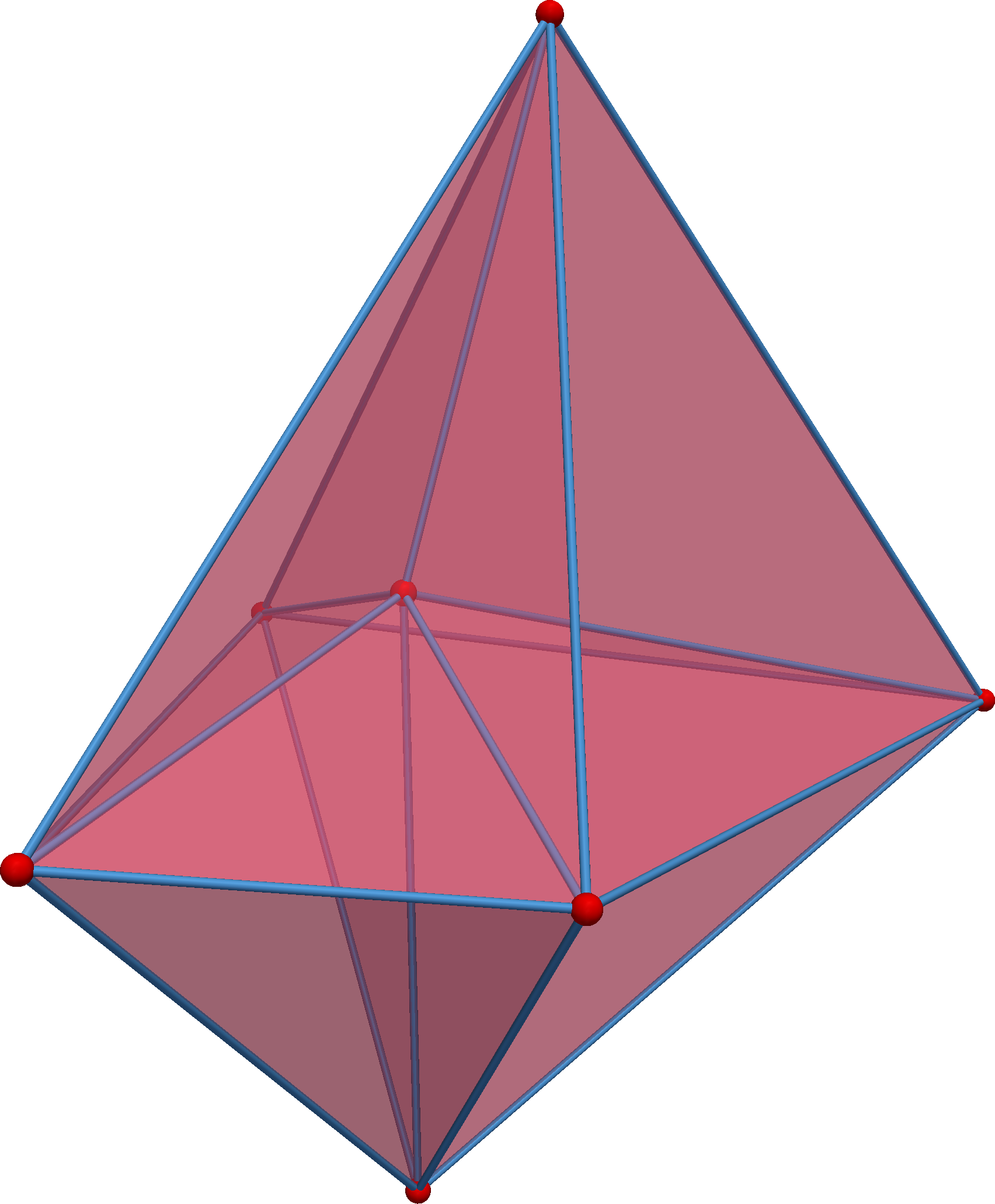}
\caption{After edge bisection\label{fig_tetRefineAfter}}
\end{subfigure}
\hfill%
\end{centering}
\caption{Illustration of the refinement of four tetrahedra incident to a bisected edge. A single new vertex is inserted at a user defined location which uniquely defines the geometry of the eight resulting refined simplices.\label{fig_tetRefine}}
\end{figure}

We opt for a very generic refinement procedure through bisection of individual edges, where every bisected edge implies the bisection of any incident simplex in order to maintain a conforming mesh at all times. The basic underlying idea is to let the user decide whether a given simplex needs refinement based on the computation of a user defined arbitrary quantity computed over each simplex and its local neighbourhood. Whenever the user defined criteria are not met, simplex refinement is triggered and an appropriate edge has to be selected for bisection. The freedom to choose that edge is also given to the user on a simplex basis, each simplex to be refined deciding on their refinement edge. These criteria  may however result in several edges of a given simplex to be selected for bisection at the same time when several incident simplices require refinement. Such refinement conflicts have to be resolved by either allowing simplices to be bisected more than once in a single pass or by using a multi-pass procedure where conflicts are resolved before actually bisecting the simplices. We use the second option, which presents the advantage of flexibility and generality at the expense of execution speed. Practically speaking, whenever a simplex needs refinement, a user defined function is called to decide of the edge to bisect and its priority, possibly based on the actual values of the user defined quantities that triggered refinement. Any two bisection edges are in conflict whenever they refine the same simplex, in which case the bisection of the edge with lower priority is canceled. In practice, a single pass of the refinement algorithm therefore consists in checking user defined functions associated to each simplex, identifying the corresponding bisection edges if needed, eliminating conflicts and actually bisecting remaining edges. The mesh is considered refined whenever enough passes have been performed so that no simplex needs refinement anymore. Finally, we note that all simplices incident to a bisected edge are refined by introducing a new vertex in the mesh as illustrated on figure \ref{fig_tetRefine}, and we also give the user the freedom of arbitrarily defining its coordinates and updating the user defined data associated to it. Similarly, user defined data associated to simplices are updated upon splitting through a user defined function specific to each data type. A more specific description of the algorithm we use for distributed mesh refinement is given in \ref{app_meshRefine}.
\label{sec_meshRefine}

We conclude this section with a brief description of the coarsening procedure implemented in the code but note that we reserve applications of mesh coarsening for future works. The idea is to allow undoing refinement with the aim of greatly lowering the number of mesh elements required to reach a given resolution at all times. Implementing such a feature requires keeping track of the successive simplex bisections, which we do by maintaining a binary tree structure where references to bisected simplices originating from a given simplex are stored as its daughters. The implementation of this tree requires adding two pointers per actual simplex, which act as a leaf of the tree (one to the parent node in the tree and one to the ``partner'' bisected simplex). In addition, a structure of three pointers is used to represent non-leaf nodes of the tree (one pointing to the parent node, and two to the daughter nodes). This leads to a non-negligible but acceptable increase in memory requirements. Practically, the user is given the opportunity for every vertex such that all the incident simplices result from a single split to decide whether the vertex should be removed in order to cancel that refinement. This can for instance be achieved by measuring the same criterion used for triggering refinement over the incident simplices, and triggering coarsening whenever it gets lower than a certain coarsening threshold equal to a fraction of the refinement threshold. Upon coarsening, incident simplices are merged with their sibling daughters in the binary tree and the mesh is therefore locally coarsened. The detailed description of our implementation of coarsening are relatively involved and we leave it for a future article.\footnote{the interested reader is nevertheless encouraged to read it directly from the freely available source code.} Finally, it may be useful to mention that coarsening also imposes to store locally (i.e. on the same MPI process) all simplices resulting from the successive splits of a given initial simplex. A consequence is that load balancing has to be performed over the root nodes, by weighting them proportionally to the number of simplices they have produced. This is however not a problem as long as the initial number of elements in the mesh is sufficiently large to maintain good balance, which was largely the case in all our test problems.

\section{Exact projection}
\label{sec_exactProj}
In \cite{Franklin83} and \cite{Franklin87}, Franklin proposes a formalism to derive properties of polyhedra, including their volume, surface and perimeter, based solely on the knowledge of the location and direct neighbourhood of their vertices. In practice, he shows that explicit information on the global topology of the polyhedra (i.e. the way vertices are connected) is not required, instead only a set of so called ``augmented rays'' defined by an ordered set of unit vectors departing from each vertex. A practical implementation of this formalism is proposed in \cite{Franklin93} and applied to the calculation of the intersecting volumes of overlaying 3-dimensional tetrahedral meshes. We propose here to adapt this formalism and to design an efficient algorithm to solve the problem of exactly projecting a piece-wise linear function defined over an unstructured simplicial mesh onto an adaptively refined regular grid in 2D and 3D. This algorithm represents the second main component of the standalone library {\tt DICE}. 

\subsection{Formalism}
\label{sec_franklinForm}
\begin{figure}
\centering
\includegraphics[width=0.9\linewidth]{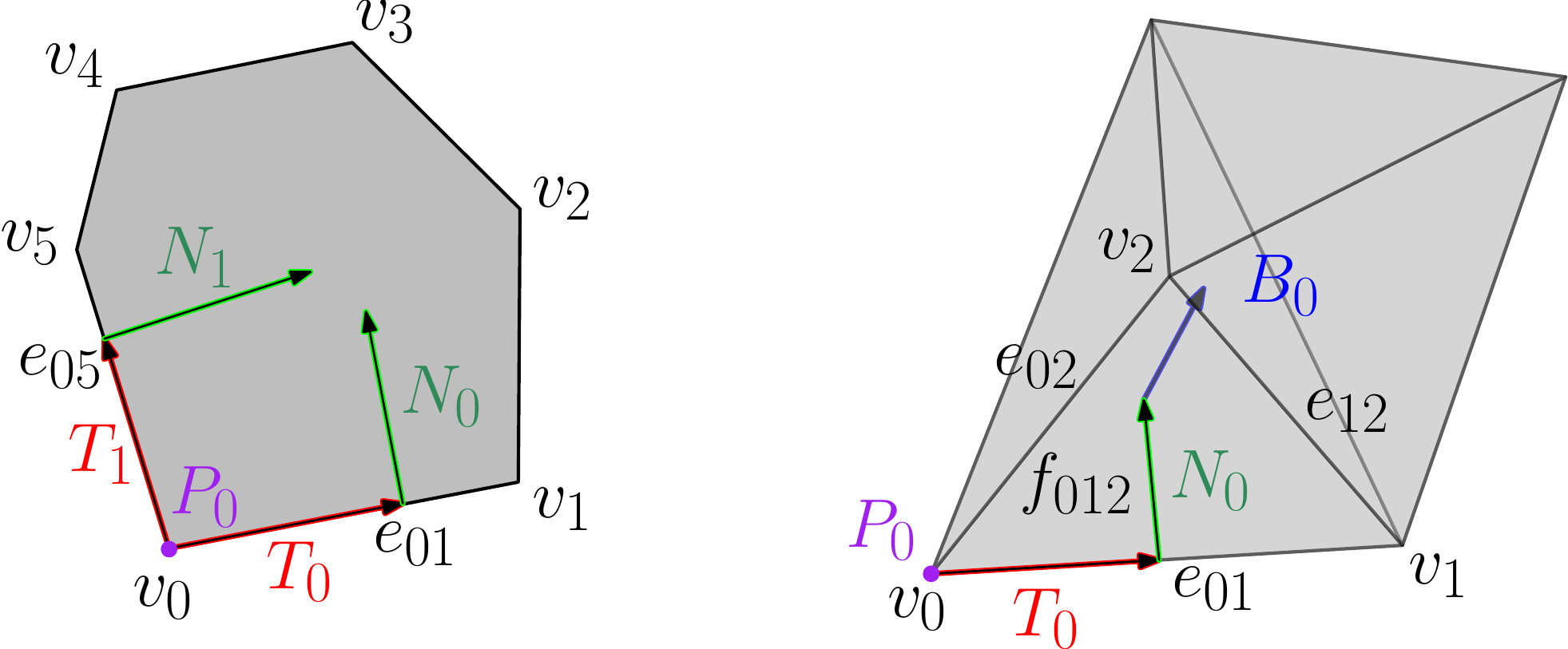}
\caption{Illustration of the definition of the vector triplets $(\VP,\T,\N)$ (2D, left figure) and quadruplets $(\VP,\T,\N,\B)$ (3D, right figure) associated to each polyhedron as defined by \cite{Franklin93}. Such multiplet exists for each possible combination of incident vertex $\vertex{i}$, edge $\edge{i}{j}$ and facet $\facet{i}{j}{k}$ in the polyhedron, and there therefore exist $12$ of them for the 2D polyhedron on the left figure and $36$ of them for the 3D polyhedron of the right figure.\label{fig_ptnb}}
\end{figure}

Following the notations of \cite{Franklin93}, given a polyhedron with vertices $\vertex{i}$, edges $\edge{i}{j}=\lbrace \vertex{i},\vertex{j} \rbrace$ and facets $\facet{i}{j}{k}=\lbrace \vertex{i},\vertex{j},\vertex{k} \rbrace$, a unique set of four vectors $(\VP,\T,\N,\B)$ (quadruplet hereafter) can be conveniently associated to each set of incident vertex, edge and facet $(\vertex{i},\edge{i}{j},\facet{i}{j}{k})$. The elements of the quadruplets are such that:
\begin{description}
\item[$\VP$] is the coordinate vector of vertex $\vertex{i}$,
\item[$\T$] is a unit tangent vector along an edge $\edge{i}{j}$ departing from $\vertex{i}$,
\item[$\N$] is a unit normal vector orthogonal to $\T$ in the plane defined by $\facet{i}{j}{k}$ and pointing inward $\facet{i}{j}{k}$,
\item[$\B$] is the unit bi-normal vector orthogonal to $\T$ and $\N$ (i.e. orthogonal to $\facet{i}{j}{k}$) and pointing inward the polyhedron,
\end{description}
as illustrated on figure \ref{fig_ptnb} in the case of a 2D and 3D polyhedron. Using these quantities, one can simply express properties of polyhedra as sums over all possible combinations of incident vertex, edge and facets $\left(\vertexN{i},\edgeN{i}{j},\facetN{i}{j}{k}\right)$ of functions depending only on the corresponding $\left(\VP,\T,\N,\B\right)$ quadruplets, and the volume for instance is given by:

\begin{align}
  {\rm Vol}^{2D} &=  \frac{1}{2} \sumAdjTwoN{i}{j}\,\mathbf{P}.{\mathbf{T}}\,\mathbf{P}.{\mathbf{N}},\\
  {\rm Vol}^{3D} &= -\frac{1}{6} \sumAdjThreeN{i}{j}{k}\,\mathbf{P}.{\mathbf{T}}\,\mathbf{P}.{\mathbf{N}}\,\mathbf{P}.{\mathbf{B}}.\label{eq_vol}
\end{align}

As noted by \cite{Franklin93}, these formula can be readily used to compute the integral $M^0( {V_j})$ of the piece-wise constant approximation $\rho^0(U)$ of a function $\rho(U)$ defined over an unstructured mesh $U$ whose elements are noted $U_i$ onto another unstructured mesh $V$ with elements $V_j$, and we have for any $V_j \in V$:
\begin{equation}
\label{eq_zerothOrder}
  M^0\left( {V_j}\right) = {\sum_{U_i}}\, \rho^0\left(U_i\right) {\rm Vol}\left({U_i} \cap {V_j} \right),
\end{equation}
where $\rho^0\left(U_i\right)$ is the averaged value of $\rho$ over $U_i$ and ${\rm Vol}({U_i} \cap {V_j})$ is computed using equation (\ref{eq_vol}). As shown in \ref{app_order1}, this result can be extended to a piecewise linear approximation $\rho^1(U)$ of $\rho(U)$ which is linear over each element $U_i$ of $U$ and we obtain :
\begin{align}
\label{eq_firstOrder}
  M^1\left( {V_j}\right) &= {\sum_{U_i}}\, \left[ {\rm Vol}\left({U_i} \cap {V_j}\right)  {\tilde \rho}^0\left(U_i\right) + \mathbf{E}.\mathbf{\nabla}\rho^1\left(U_i\right) \right],\\
 &={\sum_{U_i}}\,\sumAdjThreeN{i}{j}{k}\, \contribN{U_i\cap V_j}{i}{j}{k},
\end{align}
where ${\tilde \rho}_0(U)$ now corresponds to the value of $\rho^1(U)$ extrapolated to the origin of coordinates, $\mathbf{E}=E_0\mathbf{E_1}$, and
$\left(\vertexN{i},\edgeN{i}{j},\facetN{i}{j}{k}\right)$ stands for any combination of incident vertex, edge and facet in ${U_i} \cap {V_j}$, with $\contribN{U_i\cap V_j}{i}{j}{k}$ the contribution associated to it:
\begin{equation}
  \label{eq_contrib}
  \contribN{U_i\cap V_j}{i}{j}{k}=E_0 \left[{\tilde \rho}^0\left(U_i\right) + {\mathbf E_1}.{\mathbf \nabla \rho^1\left(U_i\right)}\right].
\end{equation}
Each contribution depends on the quadruplet associated to $\left(\vertexN{i},\edgeN{i}{j},\facetN{i}{j}{k}\right)$ through the scalar $E_0$ given by
\begin{align}
E_0^{\rm 2D} &= \frac{1}{2} \mathbf{P}.{\mathbf{T}}\,\mathbf{P}.{\mathbf{N}},\label{eq:E02D}\\
E_0^{\rm 3D} &= -\frac{1}{6} \mathbf{P}.{\mathbf{T}}\,\mathbf{P}.{\mathbf{N}}\,\mathbf{P}.{\mathbf{B}},
\label{eq:E03D}\end{align}
and the vector $\mathbf{E}_1$ given by
\begin{align}
\mathbf{E}_1^{\rm 2D} &= \frac{1}{3} \left(\mathbf{P}.{\mathbf{T}}\,\,{\mathbf{T}} + 2 \mathbf{P}.{\mathbf{N}}\,\,{\mathbf{N}}\right),\label{eq:E12D}\\
\mathbf{E}_1^{\rm 3D} &= \frac{1}{4} \left(\mathbf{P}.{\mathbf{T}}\,\,{\mathbf{T}} + 2 \mathbf{P}.{\mathbf{N}}\,\,{\mathbf{N}}+ 3 \mathbf{P}.{\mathbf{B}}\,\,{\mathbf{B}}\right).\label{eq:E13D}
\end{align}
We finally note that computing both functions ${\tilde \rho}^0(U_i)$ and $\nabla \rho^1(U_i)$ is simple if the density $\rho_j$ is known for each vertex $j$ belonging to simplex $U_i$. Given the positions $\bm{x}_j$ of each of these vertices which we identify by their arbitrary index $j=0,\cdots,D$, with $D$ the dimension of configuration space, one can define the three vectors $\delta \bm{x}_j=\bm{x}_j-\bm{x}_0$ as well as a matrix $\bm{M}$ composed by the transpose of these three vectors and the vector $\delta \bm{\rho}$ of which each coordinate is given by $\delta \rho_j \equiv \rho_j-\rho_0$. Then the gradient inside the simplex is easily computed as $\nabla \rho^{1}(U_i)=\bm{M}^{-1} \delta \bm{\rho}$ and ${\tilde \rho}^0$ as e.g. ${\tilde \rho}^0(U_i)=\rho_0-\nabla \rho^1 \bm{x}_0$. 

\subsection{Implementation}
\label{sec_exactProjImpl}
\begin{figure}
\begin{centering}
\begin{subfigure}[b]{0.5\textwidth}
\centering
\includegraphics[width=.99\linewidth]{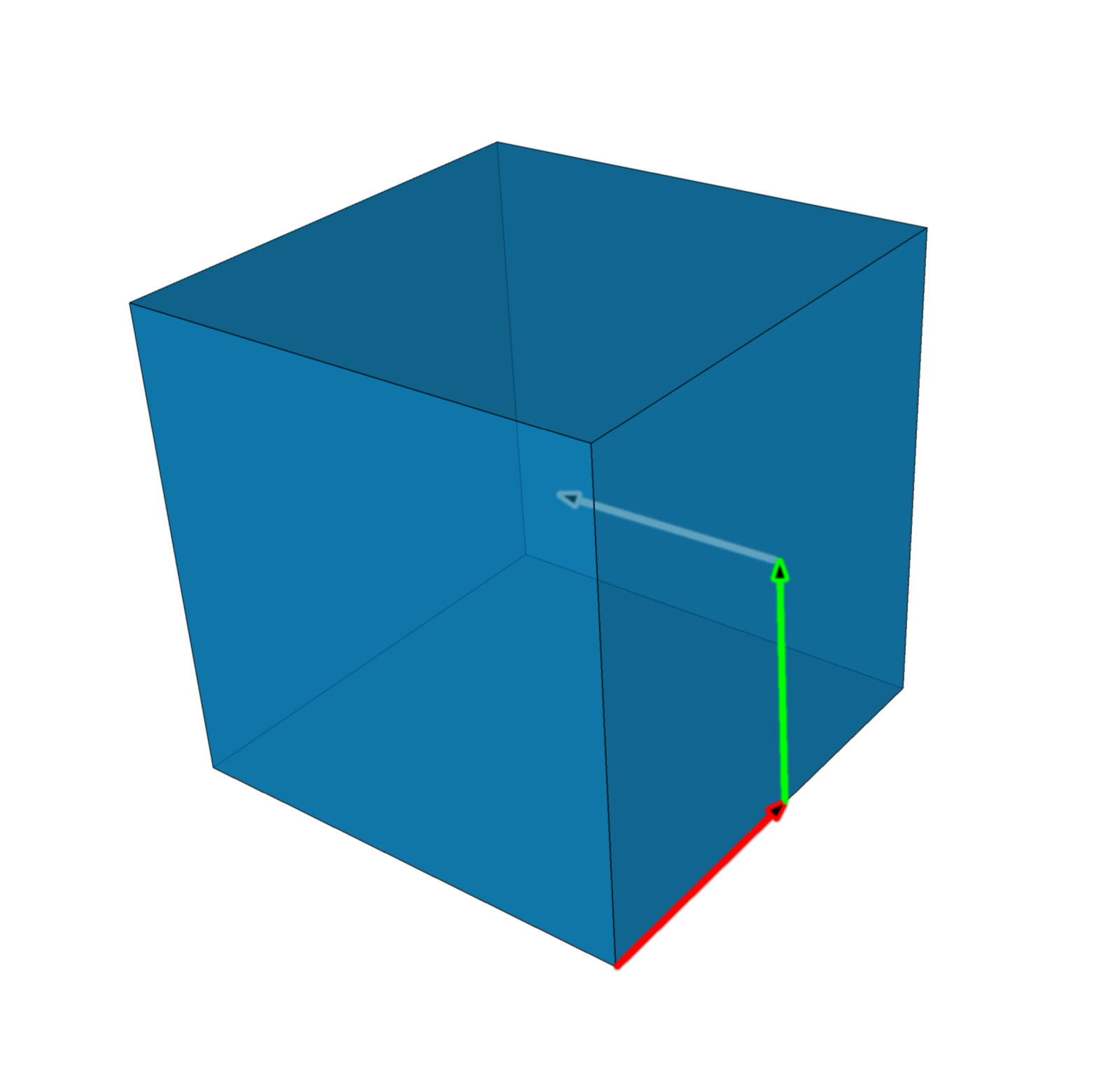}
\caption{Corner of a cube\label{fig_cubeTetInter_a}}
\end{subfigure}
\hfill%
\begin{subfigure}[b]{0.5\textwidth}
\centering
\includegraphics[width=.99\linewidth]{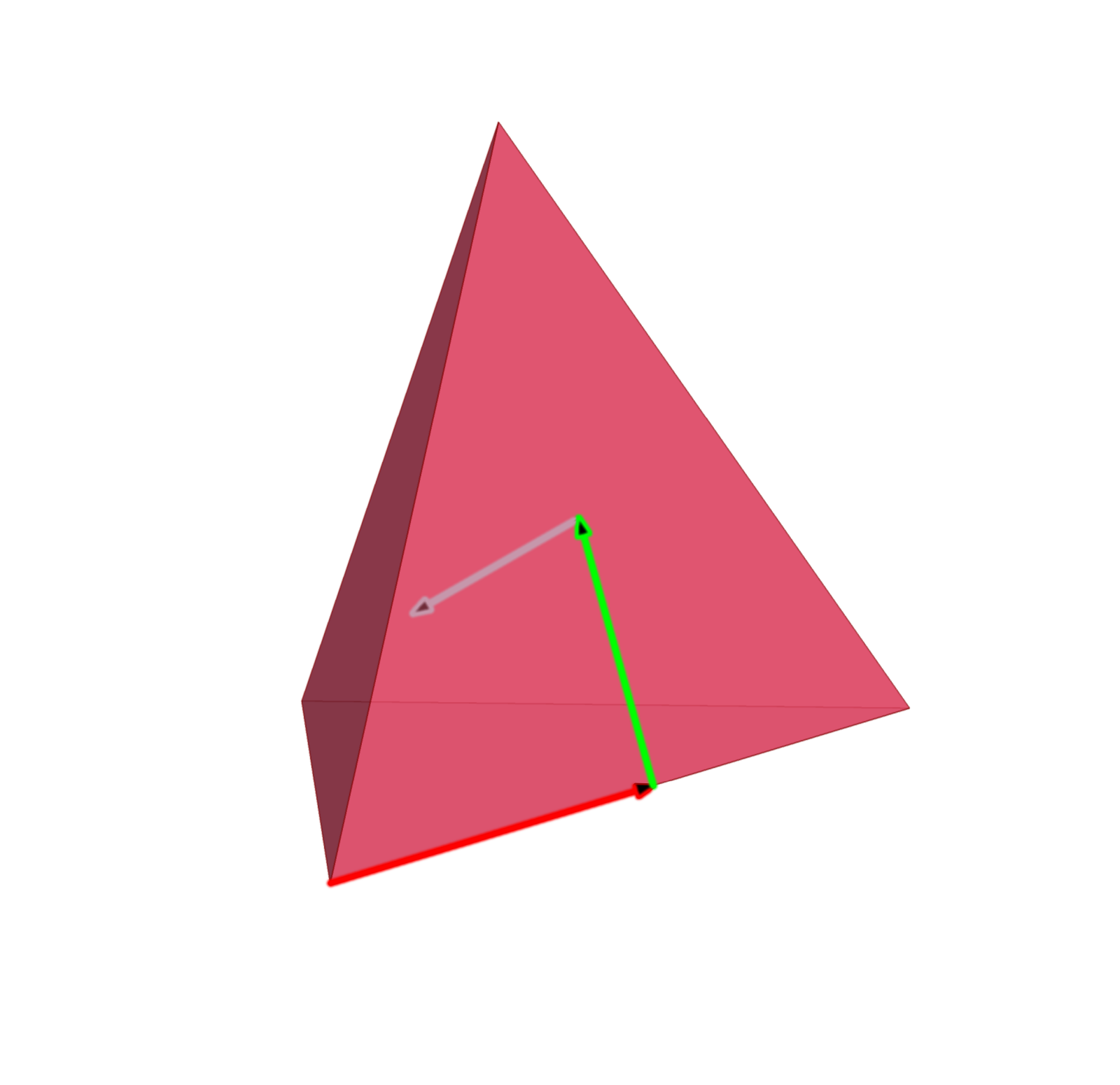}
\caption{Corner of a tetrahedron\label{fig_cubeTetInter_b}}
\end{subfigure}
\end{centering}
\begin{centering}
\begin{subfigure}[b]{0.5\textwidth}
\centering
\includegraphics[width=.99\linewidth]{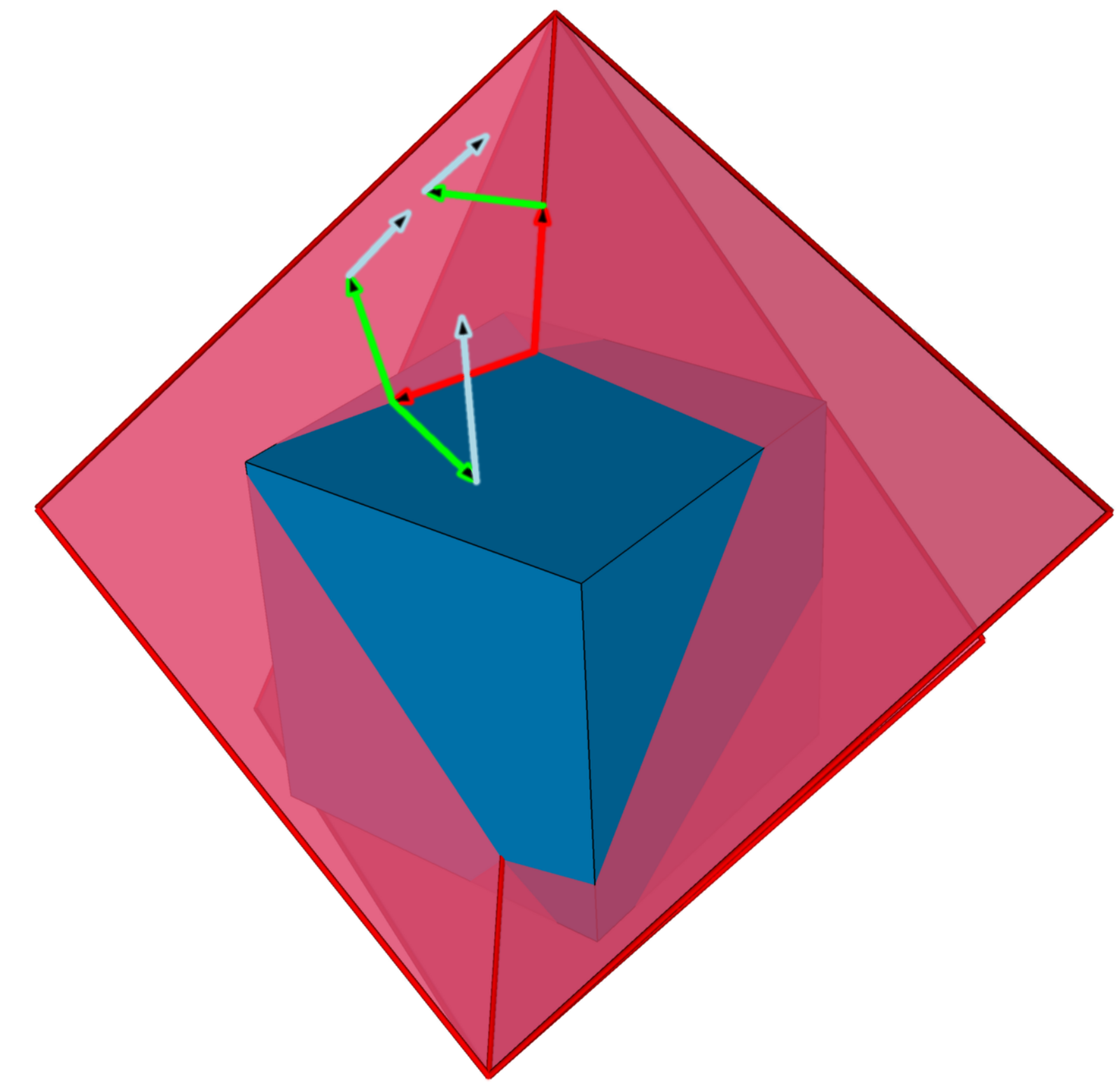}
\caption{Tetrahedron edge / cube facet\label{fig_cubeTetInter_c}}
\end{subfigure}
\begin{subfigure}[b]{0.5\textwidth}
\centering
\includegraphics[width=.99\linewidth]{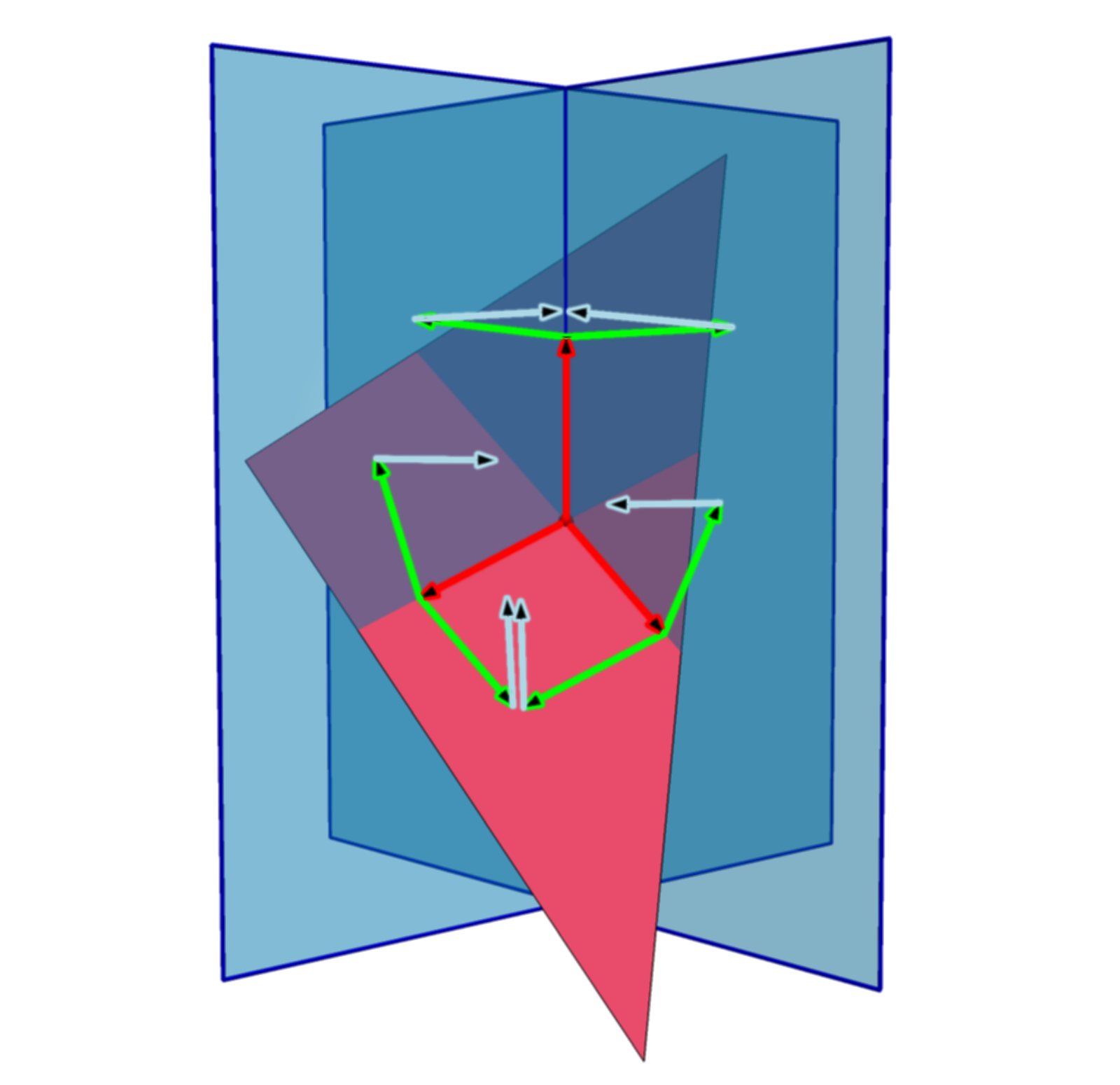}
\caption{Tetrahedron facet / cube edge\label{fig_cubeTetInter_d}}
\end{subfigure}
\end{centering}
\caption{Illustration in the 3D case of the four different types of polyhedron corners created by intersecting a regular (adaptive) grid of voxels (blue cubes) and an unstructured simplicial mesh (pink tetrahedra). The red, green and light blue arrows correspond to the $\bm{T}$, $\bm{N}$ and $\bm{B}$ vectors described in section \ref{sec_franklinForm}. Note that most of these vectors will be involved in the computation of the contributions to distinct polyhedra sharing cubes and/or tetrahedra facets, edges or vertices. For clarity, on (c), the voxel considered for the intersection is the one above the blue cube drawn on the figure. \label{fig_cubeTetInter}}
\end{figure}

We now demonstrate how the formalism introduced in the previous section is used to design a fast algorithm for the projection of a piece-wise linear function defined over an unstructured mesh $M$ onto a regular (and possibly adaptive) grid $G$. For the sake of simplicity, we will consider from now on that $M$ is a simplicial mesh (i.e. it is made of triangles in 2D or tetrahedra in 3D). Performing such a projection amounts to integrating a first order function over all the disjoint polyhedra formed by the intersection of each simplex $M_i$ (triangles in 2D, tetrahedra in 3D) in $M$ and individual voxels\footnote{a voxel is a {\bf vo}lumetric-pi{\bf xel}.} $G_j$ (squares in 2D, cubes in 3D) in $G$. The key observation is that the corners of such polyhedra may only have $4$ distinct origins ($3$ in 2D), each of them producing a specific set of associated $(\VP,\T,\N,\B)$ quadruplets [triplets $(\VP,\T,\N)$ in 2D] used to perform the integration according to equation (\ref{eq_firstOrder}). More specifically, as illustrated on figure \ref{fig_cubeTetInter}, a polyhedron corner may be:

\begin{description}
\item[A.~The corner of a voxel $G_i$ (fig.~\ref{fig_cubeTetInter_a})]\hfill\\ 
  Such a corner produces 6 quadruplets (2 triplets in 2D), each with $\T$, $\N$ and $\B$ vectors aligned with the Cartesian axes and whose orientation depends only on the position of the corner's vertex $v$ with respect to the voxel. The contribution of such a corner is particularly easy to compute since equation (\ref{eq_contrib}) reduces to a very simple expression in this case. It is also most efficient to compute the contributions for all $8$ incident corners ($4$ in 2D) generated by voxels incident to $v$ on the fly as they only differ by the sign due to the anti-symmetry of equation (\ref{eq_contrib}) in $\T$, $\N$ and $\B$.
\item[B.~The corner of a simplex $M_i$ (fig.~\ref{fig_cubeTetInter_b})]\hfill\\ 
  Such a corner produces $6$ quadruplets (2 triplets in 2D) defined by the $3$ segments and $3$ facets incident to the corner vertex $v$. Note that each quadruplet $(\VP,\T,\N,\B)$ associated to a facet of this configuration has a symmetric quadruplet $\left(\VP,\T,\N,-\B\right)$ generated by the neighbouring simplex corner sharing this facet.   
\item[C.~A simplex edge $e_M$ and voxel facet $f_G$ intersection (fig.~\ref{fig_cubeTetInter_c})]\hfill\\ 
  Such a corner produces a total of 3 quadruplets (2 triplets in 2D) for every facet of a simplex incident to the edge traversing the voxel facet. The computation of such contributions can be significantly improved by noting that $e_M$ may intersect different voxels facets of $G$ an undefined number of times and that many terms will be similar for each such intersection. Indeed, for each facet incident to $e_M$, only vector $\VP$ will change the quadruplet for which $\T$ is along $e_M$ and only three configurations exist for the remaining quadruplets (i.e. one for each voxel facet orientation). It is therefore advantageous to pre-compute as much information as possible for the quadruplets associated to each simplex facet $f_M$ incident to $e_M$ and then ray-trace $e_M$ through $G$ while having all this information at hand. Moreover, similarly to the previous case, we note that all quadruplets have symmetric counterparts through the facet of the simplex contributing to the same voxel. 
\item[D.~A simplex facet $f_M$ and voxel edge $e_G$ intersection (fig.~\ref{fig_cubeTetInter_d})]\hfill\\ 
In the 2D case, this type of corner is identical to the previous one and can be advantageously skipped. In $3D$, it produces a total of $3$ quadruplets for every facet of a voxel incident to the edge traversing the simplex facet. The computation of such contributions can be accelerated by taking into account the alignement of one quadruplet with one Cartesian axe as well as symmetries of the quadruplets for corners sharing a simplex or voxel facet. Unlike the previous case, it is more time consuming to raytrace edges into an unstructured mesh than to compute directly the intersections of voxels edges with simplices facets since one can take advantage of the fact that the edges of $G$ are axis aligned. 
\end{description}

To optimize our algorithm, we exploit the symmetries above-mentioned by creating sets of vertices sharing many common properties. We then proceed by iterating over these sets to compute the contributions associated to each vertex. A rough sketch of the algorithm in 3D goes as follow:

\begin{myalgo}{Exact projection (simplified version)}{A simplicial mesh $M$, a function $\rho_i=\rho\left(\vertex{i}\right)$ defined for each vertex $v_i\in M$ and a (adaptively refined) grid $G$.}{The exact projection of the linear interpolation of $\rho_i$ onto the voxels of $G$.}
\label{algo_projectShort}
\item For each vertex $\vertex{i}$ in $M$:
\begin{enumerate}
\item Associate sets $S_{\left(\vertex{i},e,f\right)}\left(\vertex{i}\right)$ of incident vertex, edge and facets triplets $\left(\vertex{i},e,f\right)$ to vertex $\vertex{i}$. 
\item Add contributions associated to $S_{\left(\vertex{i},e,f\right)}\left(\vertex{i}\right)$ to the voxel that contains $\vertex{i}$ and the voxels that contain the other extremity $\vertex{j}$ of any edge $\edge{i}{j}$ in a triplet of the set.
\item Ray-trace all edges $\edge{i}{j}$ of any triplet in $S_{\left(\vertex{i},e,f\right)}\left(\vertex{i}\right)$ through the voxels of $G$, adding corresponding contributions as the ray encounters voxel facets.
\item Associate sets $S_{M}\left(\vertex{i}\right)$ of incident simplices to vertex $\vertex{i}$. 
\item For each simplex $M_i$ in $S_{M}\left(\vertex{i}\right)$: 
  \begin{enumerate}
  \item Find the subset of voxels $S_{G}\left(M_i\right)$ that overlap with the bounding box of $M_i$.
  \item Test the intersection of the facets of $M_i$ with edges of $G_j\in S_{G}\left(M_i\right)$ and add contributions as required.
  \item For each vertex of any voxel $G_j\in S_{G}\left(M_i\right)$, test if it falls inside $M_i$ and if so, add contributions as appropriate.  
  \end{enumerate}
\end{enumerate}
\end{myalgo}

The actual implementation of the algorithm obviously requires a notion of neighbourhood over $M$ and $G$. While vertices incidences on $G$ are relatively straightforward to recover, such information is not usually directly available in the data structure used to store unstructured meshes. Indeed, unstructured meshes are typically represented as an array of vertices locations and an array storing the indices of the vertices forming each element of the mesh. Our algorithm therefore requires as a first step to pre-compute an array containing the set of simplices incident to each vertex. This can be quickly achieved by iterating twice over the simplices, a first time to compute how many simplices are incident to each vertex, and a second time, after proper memory allocation, to store the simplices incident to each vertex. Having pre-computed the incidences, a detailed description of our actual implementation is given in \ref{app_exactProjImpl}.

\subsection{Robustness}
\label{sec_robustness}
\begin{figure}
  \begin{centering}
    \begin{subfigure}[b]{0.32\textwidth}
      \centering
      \includegraphics[width=\linewidth]{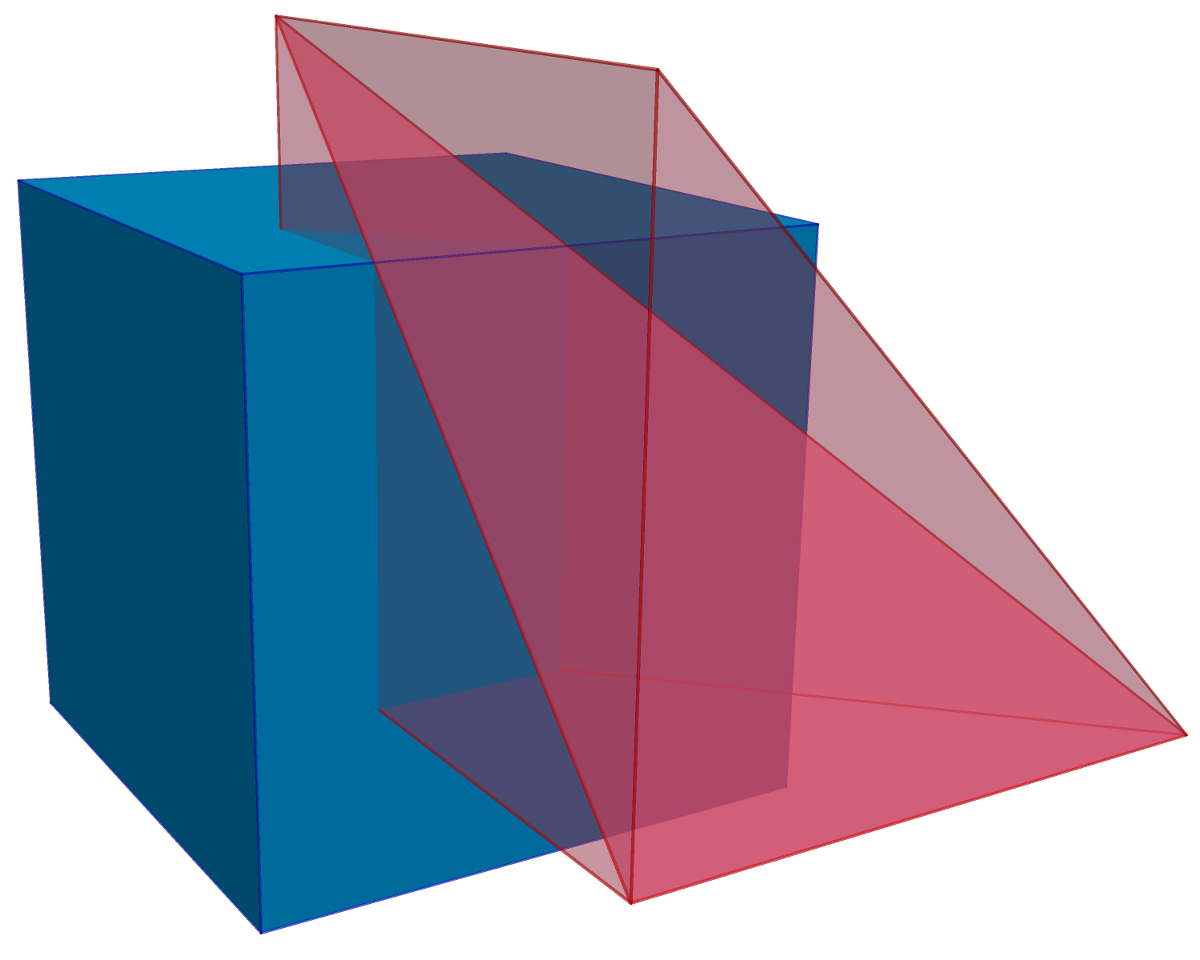}
      \caption{\label{fig_degenerate_dontknow}}
    \end{subfigure}
    \hfill%
    \begin{subfigure}[b]{0.32\textwidth}
      \centering
      \includegraphics[width=\linewidth]{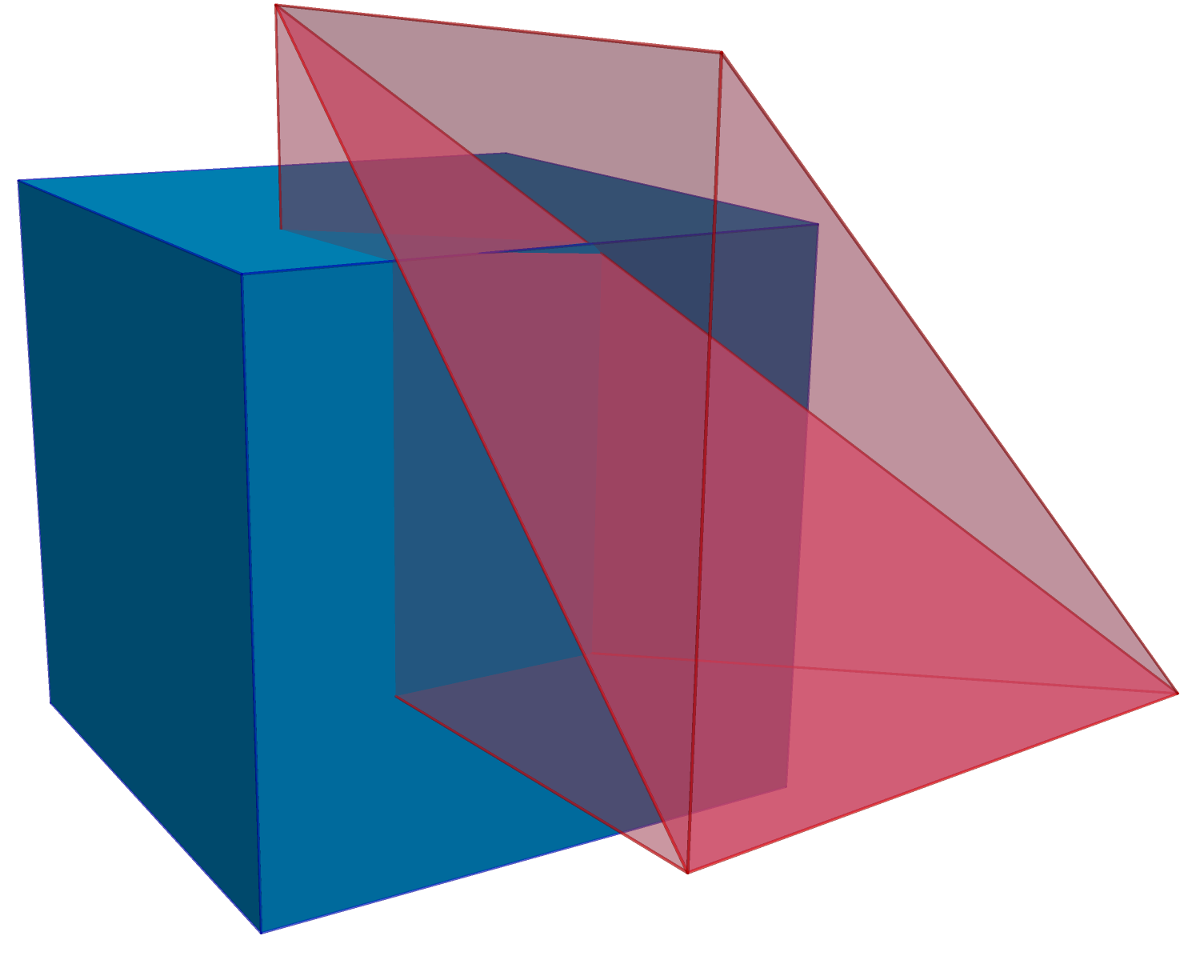}
      \caption{\label{fig_degenerate_crossed}}
    \end{subfigure}
    \hfill%
    \begin{subfigure}[b]{0.32\textwidth}
      \centering
      \includegraphics[width=\linewidth]{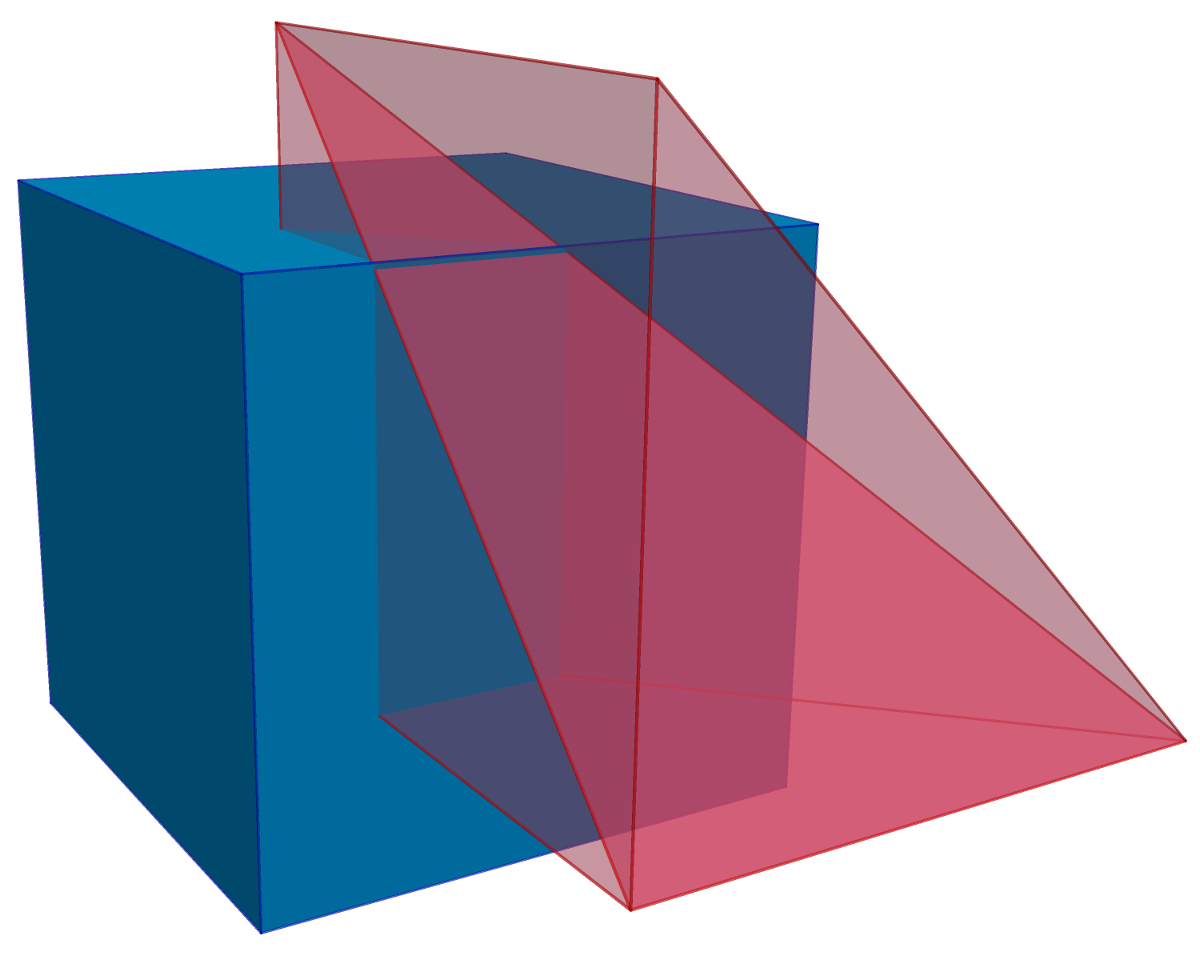}
      \caption{\label{fig_degenerate_below}}
    \end{subfigure}
  \end{centering}
  \caption{Illustration of an ambiguous intersection that may occur when the edge $e$ of a voxel (blue) and the facet $f$ at the interface of two simplices (pink) are almost co-planar. These three cases may very well be indistinguishable within the numerical machine accuracy and the result of geometric predicates for different intersections may therefore become inconsistent : if $e$ is found to intersect $f$ as in case \ref{fig_degenerate_crossed}, then exactly one edge of $f$ should also be found to intersect two facets of the voxel. However, these two types of intersection predicates are computed independently and if numerical precision is lacking, their predictions cannot be trusted. One may therefore very well find that $e$ and $f$ intersect as in case \ref{fig_degenerate_crossed} while two of the edges of $f$ also intersect the voxel as in case \ref{fig_degenerate_below}, which results in too many contributions to the voxel and a wrong result. Note also that even if geometric predicates are made consistent, degenerate cases will still occur when $e$ and $f$ are exactly co-planar and this problem still has to be dealt with.\label{fig_degenerate}}
\end{figure}

Enforcing robustness of geometric predicates is a critical issue in the practical implementation of the algorithm. These geometric predicates consist in testing point inclusion inside a simplex, calculating the intersection of a facet and a segment and raytracing a segment across the voxels of $G$. Any inconsistency in the results of these predicates implies a catastrophic failure of the algorithm, as illustrated on figure \ref{fig_degenerate}. We therefore need a method to enforce consistency of the geometric predicates, as well as a way to deal with the unavoidable degeneracies that will occur whenever geometric predicates are undecidable (e.g. a point lying on a plane is neither above nor below it).

Consider the input coordinates of the voxels and simplices vertices to be exact, even though given at finite precision. One obvious way of enforcing the consistency of any geometric predicate consists in making it exact. This can be achieved using multi-precision arithmetic libraries such as {\tt GMP} \cite{GMP} or {\tt MPFR} \cite{MPFR} provided one can ensure that the result of any computation involved remains exactly representable using a finite number of bits, at any step of the algorithm (this excludes for instance the division operation\footnote{if such operation is needed, {\tt GMP} also provides an arbitrary rational number type implemented using arbitrary sized integers to store the  numerator and denominator.}). But exactness comes with a much greater computational cost, by orders of magnitude, which could be problematic for a computationally intensive algorithm such as exact projection. A good compromise consists in following \cite{Burnikel98,Bronnimann01} and filtering the geometric predicates: the result of finite precision predicates are correct in the general case, so one could try to filter those rare occurrences when the predicates may fail and only then, use exact arithmetic. To be efficient, such an approach requires the filter to provide an accurate estimate of the error margin on the output of the predicate algorithm at a small additional computational cost. Such a method is used for instance in the state-of-the-art geometric library {\tt CGAL} \cite{CGAL_LG} and we also adopt it to design the predicates required by our exact projection algorithm. Measurements show that the impact on performances of using filtered exact predicates is negligible in our implementation, ranging from up to $20\%$ in the very degenerate case of the overlap of two identical or infinitesimally perturbed grids, down to an unnoticeable overhead in the general case.

Having exact predicates does not completely solve our problems. Indeed, even though consistency is enforced, degenerate cases will still occur,  e.g. when a voxel corner happens to lie exactly on a simplex edge. In such cases, the quadruplets associated to polyhedra corners become undefined. A generic way of dealing with this issue is ``Simulation of Simplicity''\cite{Edelsbrunner90} (SoS): instead of trying to identify and solve each and every degeneracy, circumvent them by {\em formally} perturbing the geometry of the problem so that no degeneracy can ever happen. From a practical point of view, SoS consists in replacing the coordinates of vertices by polynomials in $\epsilon$ such that the perturbed set of vertices tend to the original one as $\epsilon$ tends to $0$. In our algorithm, degeneracies originate from the fact that vertices, edges and facets of the voxels of the regular grid $G$ may be in non-generic configurations with respect to the vertices, edges and facets of the overlapping unstructured mesh $M$. Any such non-generic configuration can be suppressed by as simple perturbation of the voxels vertices $v_G$ such as
\begin{equation}
\label{eq_sos}
v_G=(x,y,z)\mapsto v_G(\epsilon)=(x-\epsilon^2,y-\epsilon^3,z-\epsilon^4),
\end{equation}
which is the simulation of simplicity convention we will choose from now on as it suffices to eliminate any uncertainty in the predicates we use, as detailed in \ref{app_predicates}. Note that in equation (\ref{eq_sos}), each coordinate is perturbed by a different power of $\epsilon$ so that $v_G\left(\epsilon\right)$ describes a non straight curve as a function of $\epsilon$: if a configuration is degenerate for a particular value of $\epsilon$, this particular degeneracy is necessarily lifted as $\epsilon \to 0$.

\subsection{Accuracy}
\label{sec:accu}
Using the techniques described in previous section, it is possible to enforce the robustness of the algorithm even in degenerate cases and to compute the correct set of contributions (\ref{eq_contrib}) involved in the projection. The level of accuracy of the result, which is expressed as a sum of individual contributions (equations \ref{eq_zerothOrder} and ~\ref{eq_firstOrder}), is however not guaranteed. It depends on the details of the algorithm, on the precision of the floating point type involved in the calculations and on the actual configuration of the mesh and grid overlap. In particular, the computational errors on individual contributions may become large in some degenerate or almost degenerate configurations. A special attention should therefore be given to the computation of the $\left(\VP,\T,\N\right)$ triplets in order to maximize the number of correct digits and to try to maintain this number bounded even in close to degenerate configurations. In our implementation, this is mainly achieved by assessing the numerical precision of critical computations at run-time\footnote{in particular for the calculation of cross and dot products.} so that we can switch to higher precision floating point number type if required, or even to a different algorithm presenting different sets of degenerate configurations.

Let $P$ be the result of the projection of a function defined over an unstructured mesh onto a given voxel $V$ of a grid. Then $P$ is computed according to equations (\ref{eq_zerothOrder}) and (\ref{eq_firstOrder}) as a sum of individual contributions $C_i$ given by equation (\ref{eq_contrib}):
\begin{equation}
\label{eq_sumContrib}
P=\sum_i C_i.
\end{equation}
Each contribution $C_i$ is internally computed as a base $2$ floating point number, which can always be expressed as
\begin{equation}
C_i={s_i}.2^{e_i},
\end{equation}
where $s_i$ is the value of its $S$-bits significand\footnote{also called mantissa in the literature.} interpreted as a fixed point number such that $0\leq s_i < 1$, and $e_i$ is the value of its $E$-bits base $2$ exponent. 

Let us assume for now that only a fixed number of bits $S-W$ of the significand of $C_i$ are correct and that the trailing $W$ bits are wrong. Since $P$ is a sum over all contributions $C_i$, its value is expected to be accurate down to:
\begin{equation}
\label{eq_accuracy}
\epsilon_P=2^{e_{\rm max}-S+W},
\end{equation} 
where $e_{\rm max}={\rm max}_i (e_i)$.

Equation (\ref{eq_accuracy}) can be used to control the accuracy of the algorithm provided that an upper bound on the number $W$ of wrong trailing bits in the significand $s_i$ of any contribution can be determined. A theoretical prediction being out of the reach of this paper, we resort to an empirical method and evaluate the value of $W$ statistically. From equation (\ref{eq_accuracy}), the configurations that are most likely to suffer from lack of accuracy are those for which there exist very large individual contributions (i.e. with large $e_i$) summing up to much lower values of $P$: regions where the gradient is very steep, or regions where a very small fraction of a very dense simplex intersects the voxel. We therefore proceed to the evaluation of $W$ by testing a large number of such configurations and comparing the results obtained when using double and quadruple precision arithmetic. We find that a conservative value of $W\approx 10$ bits seems to be reliable even in the most extreme cases. Practically speaking, for the value of the projection $P$ to a given voxel to be accurate to at least $A$ bits, its binary exponent $e_p$ should be such that:
\begin{equation}
\label{eq_accuracy2}
e_p \geq e_{\rm max}-S+W+A.
\end{equation}
That is, to reach a guaranteed accuracy of at least $0.1\%$($\sim 2^{-10}$) on the value of the projection using double ($S=52$), extended-double ($s=64$), double-double\footnote{i.e. using a combination of two double precision numbers.} ($S=104$), quadruple ($S=112$), and quadruple-double\footnote{i.e. using a combination of four double precision numbers.} ($S=209$) precision floating point numbers, the maximum contribution $C_{\rm max}$ involved in equation (\ref{eq_sumContrib}) should be lower than $2^{32}$ ($4 \times 10^9$), $2^{44}$ ($2\times 10^{13}$), $2^{84}$ ($2\times10^{25}$), $2^{92}$ ($5 \times 10^{27}$) and $2^{189}$ ($8 \times 10^{56}$) times the resulting value of $P$, respectively. For information, table \ref{tab_floatPrecision} also indicates a worst case estimate of the overhead associated to using each of these multi-precision floating point number types.

\begin{table}[!htbp]
\centering
\begin{tabular}{|c|c|c|}
\hline
 Floating point number type &  Significand precision $S$ (bits) &  Overhead factor \\
\hline
 double & $52$ & $1$\\
\hline
 extended-double & $64$ & $\sim 10$\\
\hline
 double-double ({\tt libqd}) & $104$ & $\sim 20$\\
\hline
 quadruple ({\tt libquadmath}) & $112$ &$\sim 80$\\
\hline
 quadruple-double ({\tt libqd}) & $209$ &$\sim 140$\\
\hline
\end{tabular}
\caption{Precision of different types of multi-precision floating point numbers and corresponding estimated overhead factor.\label{tab_floatPrecision}}
\end{table}

While double precision seems sufficient in most cases, a cold Vlasov-Poisson solver is a particularly demanding application since the projected density field is expected to locally feature huge contrasts around caustics. The accuracy of higher precision floating point arithmetic can therefore be expected to become a necessity around these regions when the required physical resolution is high. Switching to quadruple precision arithmetic has a significant impact on computational performances as shown on table \ref{tab_floatPrecision}, mainly because this data type is not implemented in hardware on most modern architectures. A better choice is to use double-double arithmetic such as in the {\tt libqd} library, in which case the impact on performances is reduced by a factor of $\sim 4$ for a comparable precision. It is also possible to significantly dampen the impact on performances via a per voxel accuracy control mechanism based on equation (\ref{eq_accuracy2}), as implemented in {\tt DICE}: for each voxel $V_j$ in the grid, record the maximum value $e^j_{\rm max}$ of individual contributions to this particular voxel. This value is compared a-posteriori on a per-voxel basis to the result $P_j$ of the projection onto voxel $V_j$ to identify the voxels for which accuracy is lacking. It is then only a matter of tagging the simplices overlapping those voxels and re-projecting them using a higher precision floating point type. Such a technique allows for potentially large savings in computational power. 

\subsection{Parallelisation}
\label{sec:paralfrank}
Shared memory parallelisation of our projection algorithm is relatively straightforward. Indeed, the algorithm is designed in such a way that all the corner contributions to the voxels of $G$ are distributed among the vertices of $M$ and their respective computation is completely independent. However, adding contributions to a given voxel requires caution because of potentially conflicting vertices contributing to the same voxel. We solve this minor issue by creating lists of pairs of voxels and contributions $\left[G_i,C_{G_i}\right]$ stored in pre-allocated buffers associated to individual threads. Whenever one of these buffers is full, we lock the structure containing values associated to voxels for writing and add all the contributions on the fly from the buffer. With such a method, parallel scaling is in practice not affected by conflicts, as shown in section \ref{sec_ompScaling}.

Similarly, distributed memory parallelisation via MPI is not particularly problematic. Considering an already distributed mesh, one can indeed simply set the value of non local simplices to be uniformly $0$ on every MPI process. Each local mesh subset can then be projected onto a local grid, and the resulting grids summed globally, provided, naturally, that the resolution of their voxels is identical wherever the mesh subsets assigned to different MPI processes overlap. This last point may be problematic if the local resolution of the AMR grid itself depends, as for our Vlasov-Poisson solver, on the geometrical properties of the simplicial mesh (\S~\ref{sec:projdenscal} below) if this latter is self-intersecting in projection space. Indeed, it is not trivial in this case to ensure construction of local AMR grids with identical resolution on the edges of the MPI processes without requiring potentially heavy MPI communications. In practice, we solve this problem by enforcing AMR grids to have maximum resolution along the boundaries of local subsets of the global mesh, resulting in a slight augmentation of computational cost when increasing the number of MPI processes, as illustrated in section \ref{sec_mpiScaling}. 

A short account of the performances and parallel scaling of the projection algorithm is given in section \ref{sec:parallelscaling}, where it is used in the context of our Vlasov-Poisson solver.

\section{The Vlasov-Poisson solver: {\tt ColDICE}}
\label{sec:vlapoi}

We now provide details about our Vlasov-Poisson solver, {\tt ColDICE}, designed to follow numerically the phase-space density distribution function $f(\x,\bm{u},t)$ as an hypersurface density defined on the $D$ dimensional phase-space sheet. This phase-space sheet is sampled with an adaptive simplicial tessellation as described in section \ref{sec_mesh}. Its vertices follow the Lagrangian equations of motion that we integrate using a simple second-order predictor corrector scheme, which reduces to standard leapfrog if time step $\Delta t$ is constant. To compute the acceleration, we solve Poisson equation with the fast Fourier transform method on a fixed resolution grid, where the density is computed by projecting the simplicial mesh at linear order using the algorithm described in section \ref{sec_exactProj}. To follow in the best way the details of the phase-space sheet, anisotropic refinement of the simplicial mesh is performed using the bisection method described in section \ref{sec_meshRefine}. 

Refinement is however critical and needs special attention. Firstly, it requires an accurate calculation of phase-space coordinates of newly created vertices. To achieve this, we approximate the phase-space sheet by a quadratic hypersurface at the simplex level through the introduction of special tracers. This second order description also allows one to compute quantities integrated along the phase-space sheet accurately, such as for instance total kinetic energy or total surface/volume of the phase-space sheet. Secondly, it is important to use the best criteria of refinement and the best way of refining to preserve the Hamiltonian nature of the system while minimising computational cost. We use measurements of local Poincar\'e invariants to do so.

The main steps of our solver can thus be described as follows, which sets the structure of this section:
\begin{enumerate}
\item[(o)] Generation of initial conditions (\S~\ref{sec:ic})  for a quadratic simplicial mesh supplemented with additional tracers (\S~\ref{sec_quadraticElements});
\item[(i)] calculation of the value of the next time step $\Delta t$ (\S~\ref{sec:timest});
\item[(ii)] ``drift'' of vertices and tracers positions: 
\begin{equation}
\bm{x}(t+\Delta t/2)=\bm{x}(t)+\bm{u}(t) \Delta t/2;
\label{eq:drift1}
\end{equation}
\item[(iii)] calculation of the projected density from the phase-space sheet on a grid, resolution of Poisson equation using {\tt FFTW} library followed by calculation of acceleration $\bm{a}$ at vertices and tracers positions (\S~\ref{sec:projdenscal});
\item[(iv)] ``kick'' of vertices and tracers velocities followed by a second drift,
\begin{eqnarray} 
\bm{u}(t+\Delta t) &=&\bm{u}(t)+\bm{a}(t) \Delta t, \label{eq:kick} \\
\bm{x}(t+\Delta t)&=&\bm{x}(t+\Delta t/2)+\bm{u}(t+\Delta t) \Delta t/2; \label{eq:drift2}
\end{eqnarray}
\item[(v)] anisotropic refinement of the phase-space sheet where needed, using measurement of local Poincar\'e invariants (\S~\ref{sec:aniso});
\item[(vi)] measurement of the work load balance between MPI processes, re-partitioning of the phase-space sheet if imbalance exceeds a user-defined threshold;
\item[(vii)] start again from step (i) until the simulation is finished. 
\end{enumerate}
To simplify the presentation, all the equations of this section are given in the standard physical framework. Changes brought about by cosmology, in particular by the expansion of the Universe, are discussed in \ref{app:cosmology}. 
Discussions about load balancing (item vi) are deferred to section \ref{sec:parallelscaling}, where actual tests of the performances in terms of parallelism of the code will be performed. 

\subsection{Initial conditions}
\label{sec:ic}
\begin{figure}
\begin{centering}
\hfill%
\begin{subfigure}[b]{0.45\textwidth}
\centering
\includegraphics[width=.99\linewidth]{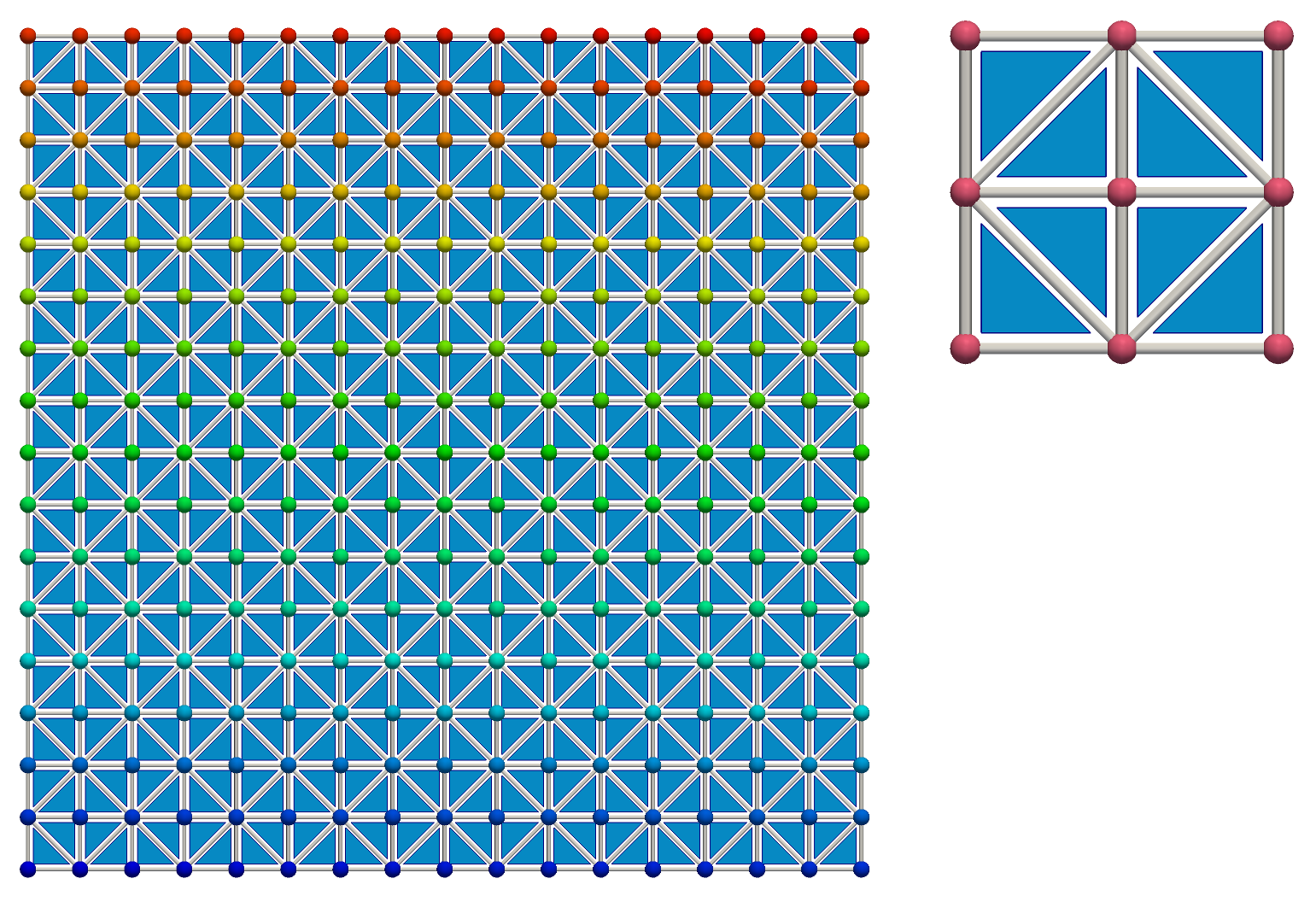}
\caption{2D\label{fig_tesselation2D}}
\end{subfigure}
\hfill%
\begin{subfigure}[b]{0.45\textwidth}
\centering
\includegraphics[width=.99\linewidth]{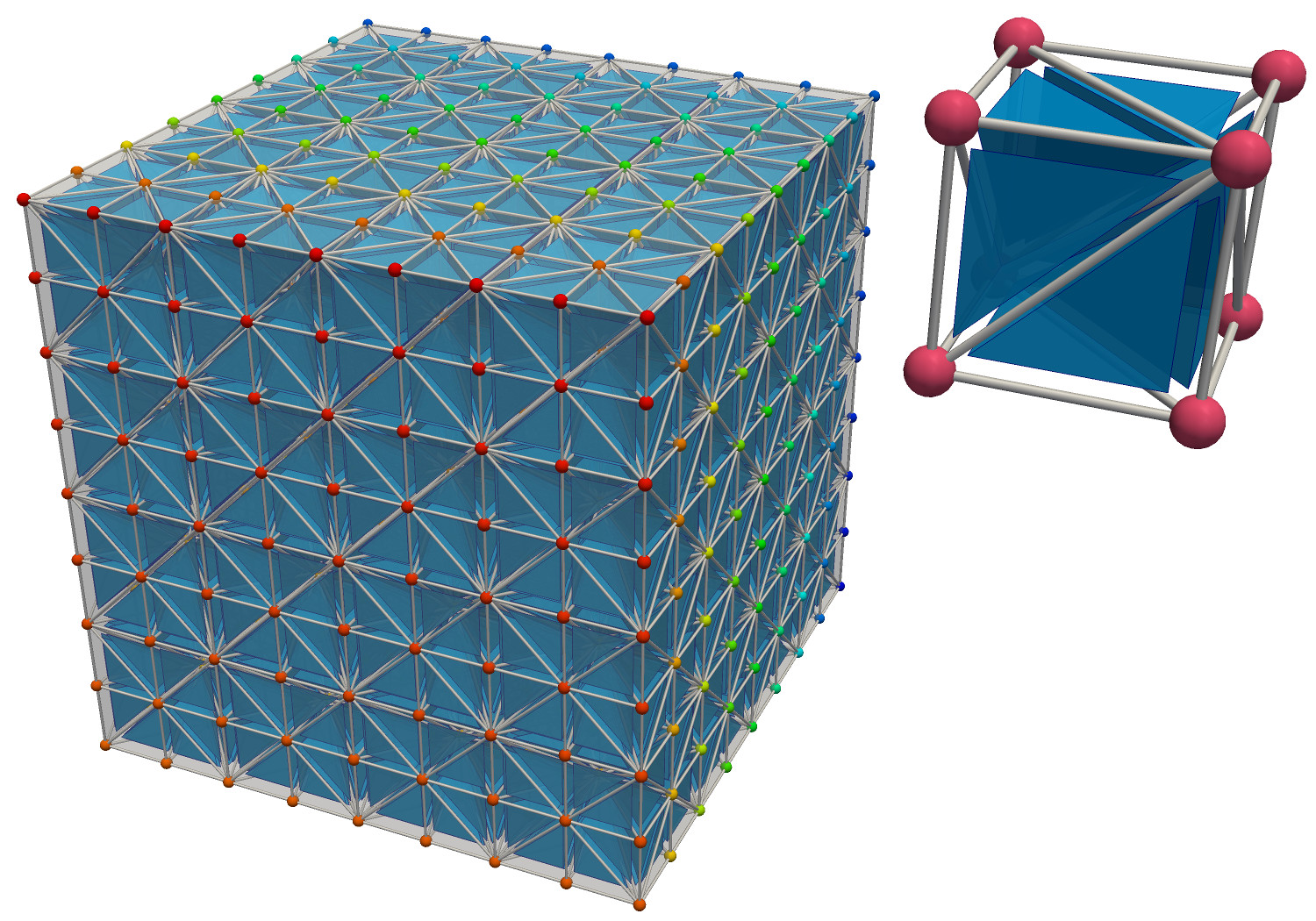}
\caption{3D\label{fig_tesselation3D}}
\end{subfigure}
\hfill%
\end{centering}
\caption{Example of a possible Lagrangian tessellation of the phase-space sheet into simplices in the 2D case (left) and in the 3D case (right).\label{fig_tesselation}}
\end{figure}

As illustrated on figure~\ref{fig_tesselation}, the phase-space sheet is initially tessellated with a regular 3D uniform simplicial grid spanning the $\bm{u}=0$ sub-manifold of the $(\x,\bm{u})$ 6D phase-space on which we define the three-dimensional Lagrangian coordinate $\bm{q}\equiv \bm{x}$. The elements of this tessellation are partitioned into connected patches of roughly equal sizes distributed among MPI processes. Additional tracers, required to obtain a second order description of the local phase-space sheet, are created as mid-points of each segment in the tessellation (see discussion in the next section). From this geometrical setup, initial conditions are then generated by assigning a smoothly varying (or even constant) mass to each simplex and moving the vertices/tracers in 6D space according to a smooth displacement field, which does not need to be small in general. Smoothness is required for a well behaved refinement. 

\begin{figure}
\begin{centering}
\hfill
\includegraphics[width=0.5\linewidth]{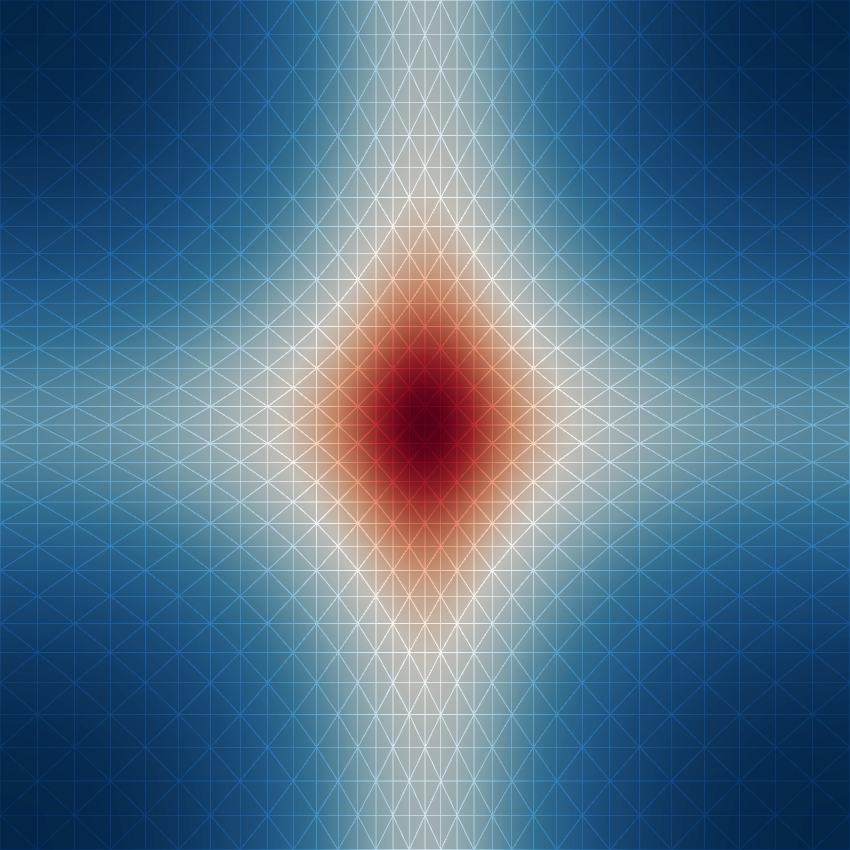}
\hfill
\end{centering}
\caption{Example of initial conditions for a two sinusoidal waves collapse problem in the two-dimensional case. The lines shows the projection in configuration space of a $32^2$ elements phase-space sheet tessellation after applying the following displacement to each coordinate of its vertices: $P_x=(0.4 L/2\pi) \sin( 2\pi q_x/L)$, $P_y=(0.3 L/2\pi ) \sin(2\pi q_y/L)$ where $L$ is the simulation box size. The color scales with the corresponding projected density at order $1$.\label{fig_2sineInit}}
\end{figure}

In the cosmological case, the simulation volume always has periodic boundaries. The initial positions and velocities of the vertices/tracers are usually set according to the fastest growing mode given by Lagrangian perturbation theory of first or second order, as detailed in \ref{app:inic}, for completeness. In this case, all the simplices have the same mass and the initial displacement of the vertices in phase-space is small. Figure~\ref{fig_2sineInit} illustrates the result obtained when generating initial conditions corresponding to two orthogonal sine waves with slightly different amplitudes, as studied below in section \ref{sec:doublesine}. 

\subsection{Quadratic mesh}
\label{sec_quadraticElements}
\begin{figure}
\begin{centering}
\hfill%
\begin{subfigure}[b]{0.49\textwidth}
\centering
\includegraphics[width=.99\linewidth]{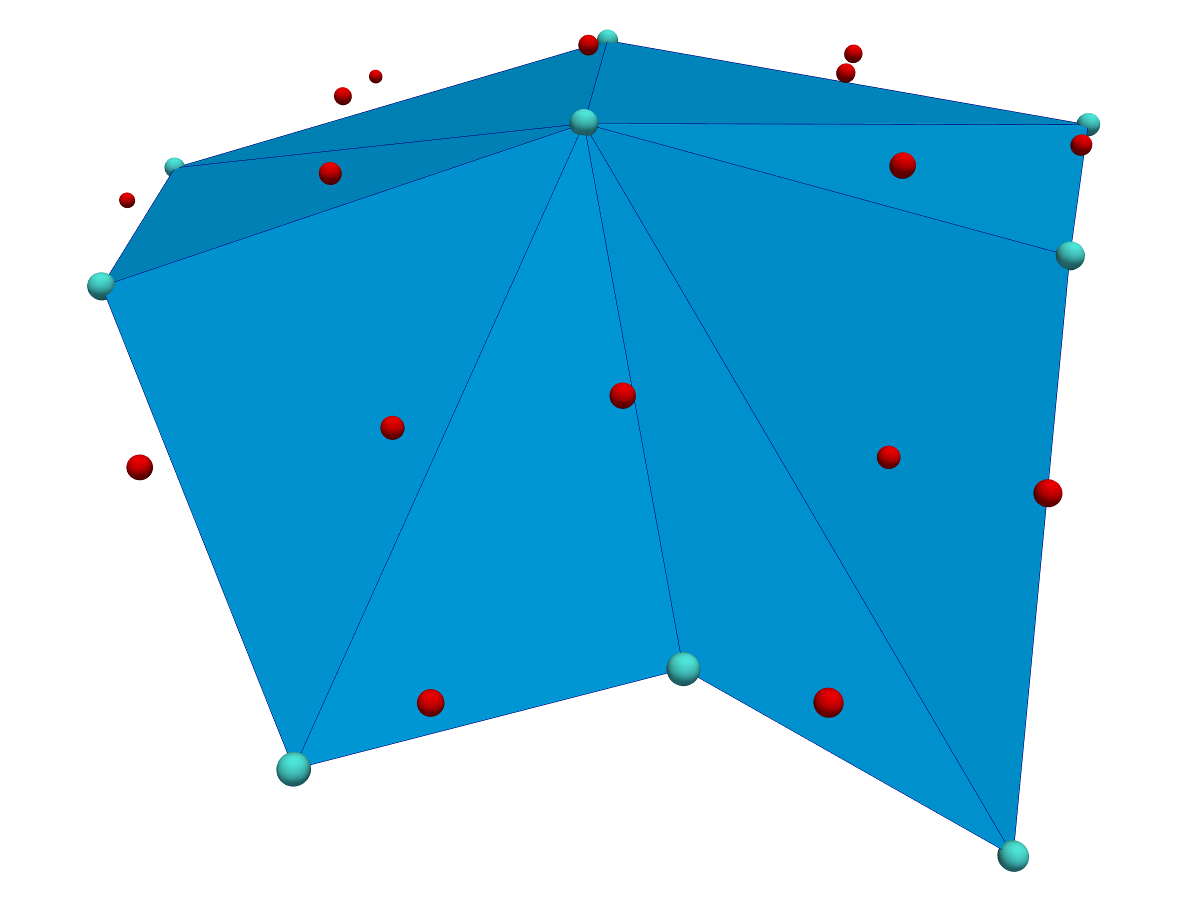}
\caption{First order elements with edge tracers (red)\label{fig_tracersNoFit}}
\end{subfigure}
\hfill%
\begin{subfigure}[b]{0.49\textwidth}
\centering
\includegraphics[width=.99\linewidth]{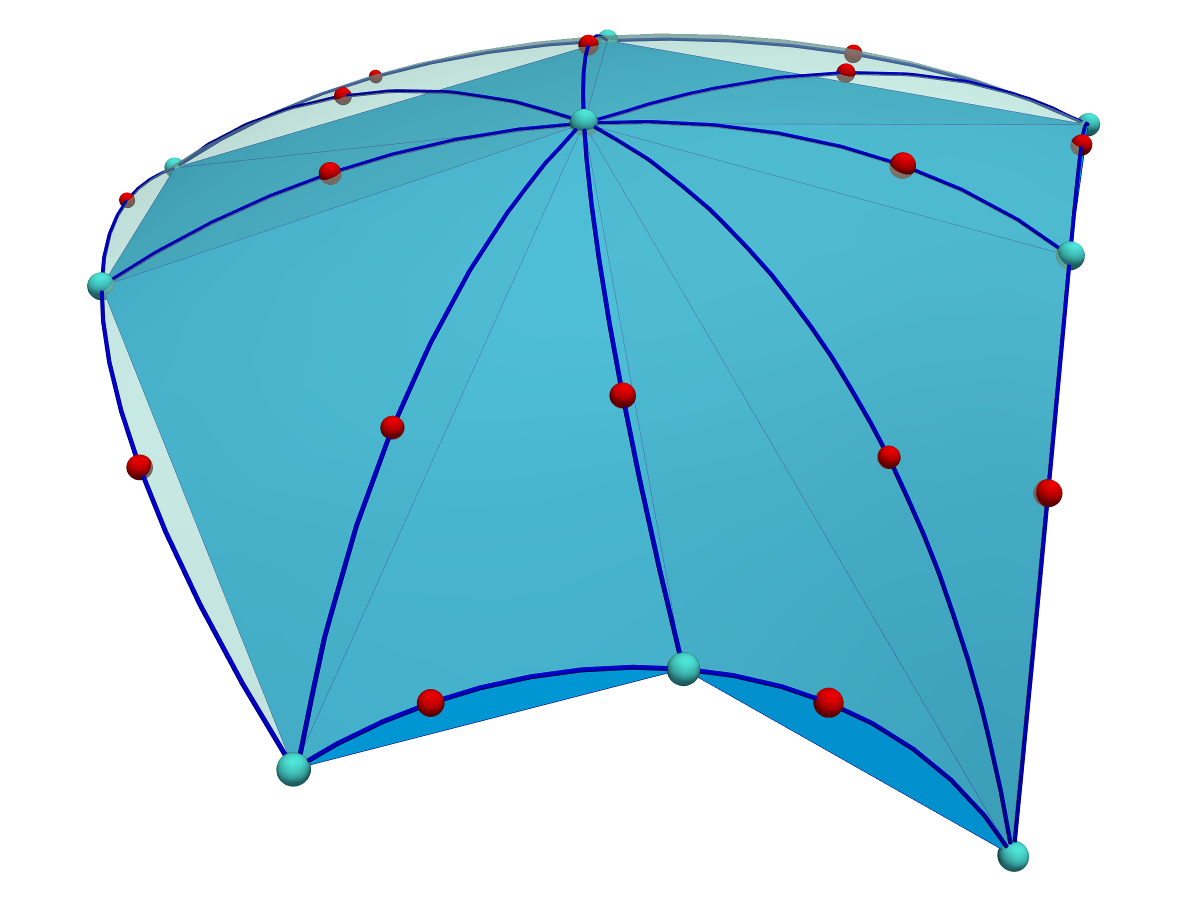}
\caption{Second order elements\label{fig_tracersFit}}
\end{subfigure}
\hfill%
\end{centering}
\caption{Illustration of the usage of edge tracer particles to define quadratic mesh elements. A total of $6$ nodes per element ($3$ tracers and $3$ vertices) is required to uniquely define the quadratic triangular elements of a 2D quadratic mesh. The 3D case is similar, with a total of $10$ nodes ($6$ tracers and $4$ vertices) required to define quadratic tetrahedra.\label{fig_tracers}}
\end{figure}

One simple way of defining higher order elements that is widely used in finite element methods \cite[see, e.g.][]{Zienkiewicz2005,FiniteElements} consists in adding an adequate number of supplementary nodes to each simplex. The additional nodes, that we call tracers, supplement the vertices of the simplex so that a smooth surface of given order passing through all the nodes is defined uniquely. Using this approach and restricting ourselves to second order, a total of $3$ and $6$ tracers ($1$ per segment) are necessary in 2D and 3D respectively to define quadratic triangular and tetrahedral elements, or equivalently exactly one additional node per edge.  A convenient way of placing the tracers consists in choosing their Lagrangian (unperturbed) position to  correspond to the middle of each edge of the tessellation. This way, each tracer is shared between all simplices incident to the edge it is paired with, so that continuity of the mesh is naturally preserved at second order (see figure \ref{fig_tracers} for an illustration in the 2D case). It then suffices to advect the tracers with the flow together with the regular vertices in order to track the phase-space sheet surface at second order. 

More specifically, following e.g.~\cite{Zienkiewicz2005,FiniteElements}, let $\xi_i$ be the $D+1$ barycentric coordinates associated with the $D+1$ vertices $v_i$ of a simplex. Then we have $\sum_i\,\xi_i=1$ and any function $Q$ with value $Q_i$ at vertex $v_i$ can be linearly interpolated at a point $\bm{P}$ with barycentric coordinates $\xi_i$ inside the simplex as:
\begin{equation}
Q\left(\bm{P}\right) = \sum_{i=1}^{D+1}\,\xi_i Q_i.
\label{eq:barycentricinterp}
\end{equation}
Let $\bm{e}_i$ be the barycentric position of vertex $v_i$ and $\bm{V}_i$ its actual position. In three dimensions, $\bm{e}_1=(1,0,0,0)$, $\bm{e}_2=(0,1,0,0)$, etc, and likewise in two dimensions.  Let $v_{ij}$ the additional tracer with position $\bm{V}_{ij}$ associated to the mid-point of edge $[\nu_i,\nu_j]$  of which the barycentric position is $\bm{e}_{ij}=(\bm{e}_i+\bm{e}_j)/2$. One can locally define the hypersurface ${\bm S}(\bm{\xi})$ associated to the quadratic simplex as the quadratic function of $\bm{\xi}$ which passes through the vertices $v_i$ and the tracers $v_{ij}$ 
\begin{eqnarray}
S(\bm{e}_{ij}) & \equiv &\bm{V}_{ij}, \\
S(\bm{e}_i) & \equiv & \bm{V}_{i}.
\end{eqnarray}
One can then define the following conventional shape functions $N_i$ as
 \begin{equation}
N_i \equiv
\begin{cases}
  \xi_i\left(2\xi_i-1\right),& i\leq D+1,\\
  4\,\xi_{j\left(i\right)}\,\xi_{k\left(i\right)},              & D+1 < i \leq N=(D+1)(D+2)/2,
\end{cases}
\end{equation}
with appropriately chosen functions $j(i)$ and $k(i)$ to cover all the combinations $j(i) < k(i)$, $1 \leq j,k \leq D+1$.
Using these shape functions, each point $\bm{P}$ with barycentric position $\bvec\xi=(\xi_i)$ in the linear simplex maps to a unique point $\bm{P}^\prime$ in its quadratic counterpart
\begin{equation}
\bm{P}^\prime=S(\xi)= \sum_{i=1}^{D+1}\,N_i\left(\xi\right) \bm{V}_i+\sum_{i=D+2,N}\,N_i\left(\xi\right) \bm{V}_{j(i),k(i)}.
\end{equation}
To simplify the notation, let $n_i$ be vertex $v_i$ if $i\leq D+1$ and vertex $[v_{j(i)},v_{k(i)}]$ if $i>D+1$. Then, any function with value $Q_i$ at node $n_i$ is interpolated to second order at point $\bm{P}^\prime$ as:
\begin{equation}
Q\left(\bm{P}^\prime\right) = \sum_{i=1}^{N}\,N_i\left(\xi\right) Q_i.
\label{eq:quadintbar}
\end{equation}
This is true in particular for coordinate functions, which yields a direct mapping from Lagrangian coordinates $\q$ to phase-space coordinates $(\x,\bm{u})$ as 
\begin{equation}
\label{eq_lagMapping}
\q = \sum_{i=1}^{D+1}\,\xi_i\, \q_i \mapsto \left(\x,\bm{u}\right)=\left(\sum_{i=1}^{N}\,N_i\left(\bvec\xi\right) \x_i,\sum_{i=1}^{N}\,N_i\left(\bvec\xi\right) \bm{u}_i\right), 
\end{equation}
where $\q_i$ and $(\x_i,\bm{u}_i)$ are the Lagrangian coordinates vector and phase-space coordinates vector at node $n_i$ respectively. 

\subsection{Time step}
\label{sec:timest} 
We use two criteria to compute constraints on the value of $\Delta t$. Firstly, because we employ a grid of fixed resolution $\Delta x$ to solve Poisson equation, we impose the traditional Courant-Friedrichs-Lewy (CFL) condition
\begin{equation}
\Delta t \leq C_{\rm CFL} \frac{\Delta x}{u_{\rm max}}, \quad C_{\rm CFL} < 1
\label{eq:CFL}
\end{equation}
where $u_{\rm max}$ is the maximum of the magnitude of each velocity coordinate and usually $C_{\rm CFL}=0.25$. Secondly, we impose a classic dynamical condition on the value of $\Delta t$
\begin{equation}
\Delta t \leq \frac{C_{\rm dyn}}{\sqrt{ 4\pi G \rho_{\rm max}}}, \quad C_{\rm dyn} \ll 1, 
\label{eq:DYN}
\end{equation}
where $\rho_{\rm max}$ is the maximum projected density and $C_{\rm dyn}$ typically of the order of $10^{-2}$. Equation (\ref{eq:DYN}) states that the time step should be small compared to the dynamical time of the system \citep[see, e.g.][]{Alard2005,Colombi2014}, which can be estimated e.g. by assuming spherical symmetry and a locally harmonic potential. 

In the cosmological case, equation (\ref{eq:DYN}) is slightly different and an additional condition on the relative variation of the expansion factor of the universe between two time steps is set to be able to follow accurately growing modes at early times (see \ref{sec:timestepappcosm} for details). 
\subsection{Calculation of the acceleration}
\label{sec:projdenscal}
An independent AMR grid is constructed over each MPI node, with a local resolution roughly corresponding to the resolution of the component of the phase-space sheet locally stored on the MPI node. In practice, the AMR grid is locally refined as long as the AMR cells are larger than the smallest perpendicular height of any projected simplex intersecting them or until maximum refinement level is reached (corresponding to the resolution of the grid used to solve Poisson equation).  

The projected density contributed from each local phase-space sheet patch is calculated on the corresponding AMR grid using the exact projection algorithm of section \ref{sec_exactProj}.  To do so, we need to estimate the value $\rho_j$ of the projected density on each vertex $j$ while we have chosen to have only access to the mass of each simplex. To compute $\rho_j$, we use the following formula,
\begin{equation}
\rho_j=\frac{\sum_{i\ {\rm incident}} M_{i}}{\sum_{i} V_{i}},
\end{equation}
where the sum is performed over the simplices incident to vertex $j$,  $M_i$ and $V_i$ are respectively the mass and the projected volume/surface of simplex $i$. 

The subsets of projected densities computed for each local AMR grid are then globally gathered and summed to a uniform grid using a standard donor cell procedure and partitioned into slabs to accommodate the distributed fast Fourier transform algorithm implemented in the {\tt FFTW} library. 

Due to finite resolution of AMR cells, our projection is in fact not exactly accurate up to linear order. Some small residual errors are therefore expectable on the projected density sampled on the final grid. While it would go beyond the scope of this paper to perform a full demonstration, we conjecture that these residual errors should remain negligible by construction if the initial surface density of the phase-space sheet presents sufficiently small variations across the initial simplices scale. 

Poisson equation is solved in parallel using {\tt FFTW} to obtain the gravitational potential. Free boundaries, if required, are implemented with the convolution method of Hockney \cite{Hockney1970}. This method is simple to implement but costly in memory as it requires doubling the size of the mesh. Improvements of free boundaries conditions using e.g. the correction charge method of James \cite{James1977} are left for future work. Note that in the cosmological case, the source term of Poisson equation is slightly different from equation (\ref{eq:poieq}) (see \ref{app:supcom}).

The acceleration is then computed on the fly at the location of each vertex/tracer by combining a standard 4 point central difference of the potential field with a second order dual TSC interpolation of the resulting force field \cite{Hockney1981} into a single operation at the vertex/tracer position. 

It is important to mention here that the values of the potential needed to update the velocity of each local vertex/tracer are not necessarily locally available and therefore require to be communicated via a global MPI scatter operation first.

\begin{figure}
\begin{centering}
\hfill
\includegraphics[width=0.90\linewidth]{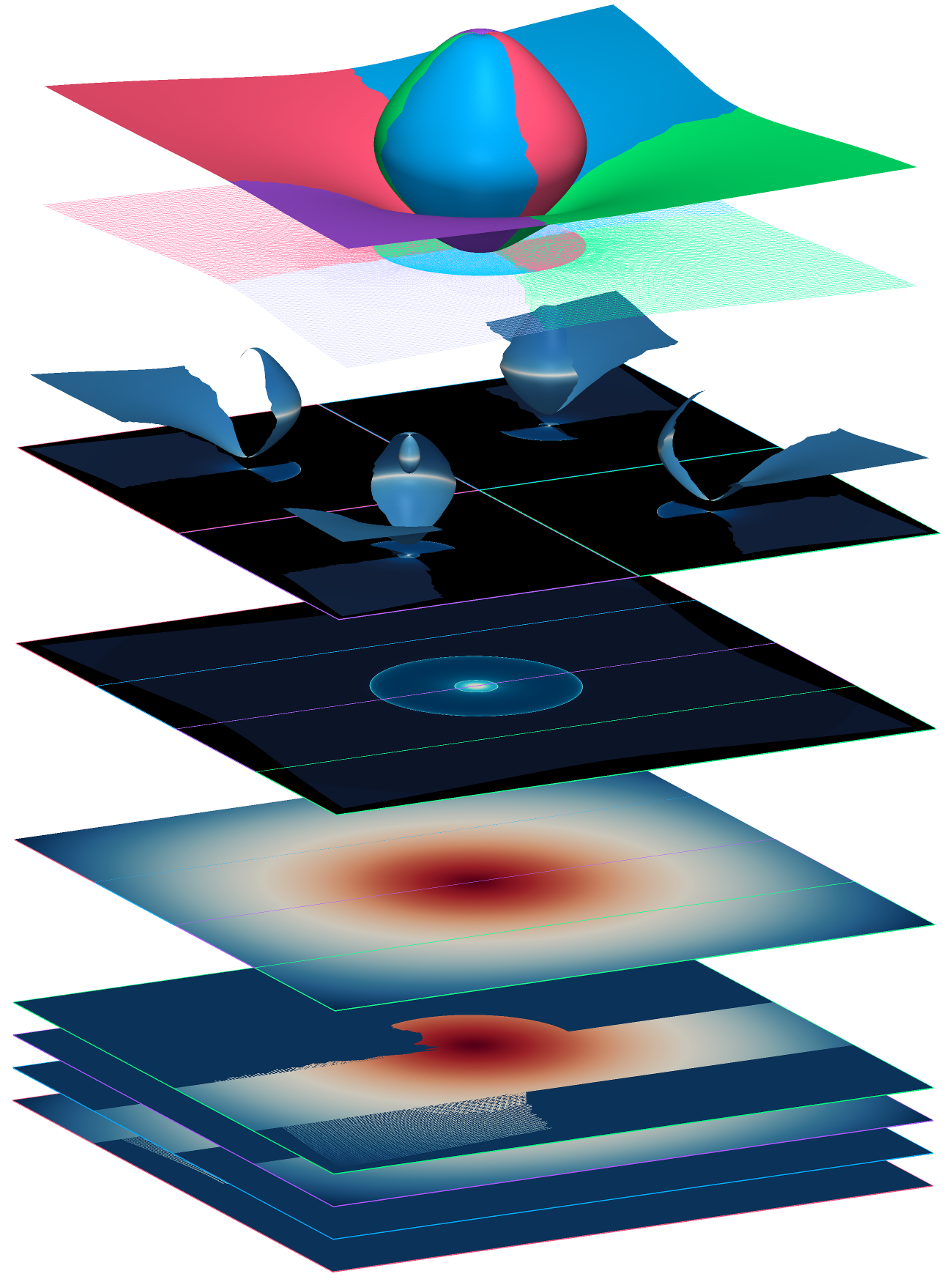}
\hfill
\end{centering}
\caption{Illustration of the different steps involved in the calculation of the acceleration during a single time step of the Vlasov-Poisson solver using $4$ MPI processes, in chronological order from top to bottom. The simplicial mesh is initially partitioned into four Lagrangian patches projected onto local AMR grids. These AMR grids are summed over a global uniform grid at maximum resolution, each process storing only a slab of the global density field. The gravitational potential is obtained by solving Poisson equation in Fourier space via distributed FFT. Finally, required pixels values are redistributed to each process so that the acceleration can be computed at the position of every vertex and tracer of the local mesh patch.\label{fig_timestep}}
\end{figure}

A graphic summary of the operations taking place during this step is given on figure \ref{fig_timestep}.

\subsection{Anisotropic refinement}
\label{sec:aniso}
Here, we implement anisotropic mesh refinement based on the generic method presented in section \ref{sec_meshRefine} (see also \ref{app_meshRefine}). We use measurements of local Poincar\'e invariants to decide when and how to refine: a refinement criterion is checked on a per simplex basis (section \ref{sec_refinementCriterion}) and anisotropic refinement is achieved by splitting a carefully selected edge of the simplex via the introduction of a newly created vertex, resulting in the splitting of all simplices incident to this edge (section \ref{sec_simplexSplitting}).
\subsubsection{Refinement criterion}
\label{sec_refinementCriterion}
Hamiltonian systems preserve symplectic two-forms during motion, or equivalently, in integral form, the Poincar\'e invariants defined by equation (\ref{eq:poinc}) \citep[see, e.g.][]{Meiss92}. This fundamental property can be used to define constraints on the geometrical set-up of the tessellation mesh. The discrete nature of our phase-space sheet representation in terms of a simplicial mesh is indeed a source of non-Hamiltonian perturbations. To control these perturbations, a sufficient sampling of the phase-space sheet has to be maintained during runtime to conserve local Poincar\'e invariants, which serves as a basis to set our per-simplex refinement criterion.

The equivalent of equation (\ref{eq:poinc}) can be defined at the microscopic level for any triangle of the tessellation.   In this case, the Poincar\'e invariant reduces to the symplectic area \citep[see, e.g.][]{Meiss92} defined, for a pair of phase-space vectors  $(\mathbf{\delta z_j},\mathbf{\delta z_k})$ aligned with two sides of the triangle, by
\begin{equation}
\label{eq_poincare}
{\bar I}_{jk} \equiv \frac{1}{2}\mathbf{\delta z_j}^\intercal \omega\, \mathbf{\delta z_k},
\end{equation}
with $\omega$ the symplectic matrix:
\begin{equation}
\omega =\left( \begin{array}{cc}
0 & -\mathbf{I}  \\
\mathbf{I} & 0  \end{array} \right). 
\end{equation}
One can therefore associate to each simplex $M_i$ an invariant which should, in the ideal case, remain null during motion,
\begin{equation}
\label{eq_invariant}
{\cal I}_i=\sup_{(j,k)}{\left|{\bar I}_{jk} - {\bar I}_{\rm ini}\right|},
\end{equation}
where the sum $j,k$ is performed over the pairs of vectors associated to each triangle of the simplex, that is the triangular elements themselves in 2D and the tetrahedra facets in 3D, while ${\bar I}_{\rm ini}$ is a reference value computed in the initial conditions. Note that in the cosmological case we have ${\bar I}_{\rm ini}=0$ by definition (prior to initial displacement of the vertices) so we assume from now on ${\bar I}_{\rm ini}=0$ although our refinement scheme can be easily generalised to the case ${\bar I}_{\rm ini} \neq 0$. 

While one could use directly equation (\ref{eq_invariant}) to decide when refinement of simplex $M_i$ has to be triggered, a more subtle approach taking better account of the local anisotropy of the phase-space sheet consists in limiting the maximum violation of symplecticity that could be obtained after one bisection. To this end, we define a modified invariant ${I}_i$,
\begin{equation}
\label{eq_invariantRefined}
I_i=\sup_j\left[{\cal I}^\prime_i(j), {\cal I}^{\prime\prime}_i(j) \right],
\end{equation}
where ${\cal I}^\prime_i(j)$ and ${\cal I}^{\prime\prime}_i(j)$ are the invariants associated through equation (\ref{eq_invariant}) to the two simplices obtained by splitting simplex $M_i$ along its $j^{\rm th}$ edge. Refinement of simplex $M_i$ is therefore triggered whenever 
\begin{equation}
I_i>\epsilon_I L_{x} L_{u},\quad \epsilon_I \ll 1,
\label{eq:refcritt}
\end{equation} 
where $L_{x}$ and $L_{u}$ are the typical size of the system in configuration and velocity space, respectively, and $\epsilon_I$ is the refinement threshold. This latter should be a very small number, e.g. $\epsilon_I \lesssim 10^{-6}$, and values as tiny as $\epsilon_I \sim 10^{-10}$ are expectable in extreme cases \citep[see, e.g.,][for detailed analyses in the one-dimensional gravitational dynamical case]{Colombi2014}. 

One consequence of our choice of refinement is that the deviation from symplecticity is maintained roughly constant at the simplex level, $I_i \sim \epsilon_i L_x L_u$.  Hence, the cumulated absolute error on the Poincar\'e invariants  due to a piecewise linear approximation of the phase-space sheet scales like $E_I \sim N \epsilon_I$ where $N$ is the number of simplices. The numerical system is thus expected to lose progressively its Hamiltonian nature and to become inaccurate at some point. This is hardly unavoidable, although our second order local description of the phase-space sheet largely compensates for this. Note thus that a proper choice of $\epsilon_I$ is a subtle task depending both on the nature of initial conditions and on the number of dynamical times one aims for to follow the system. 
\subsubsection{Simplex splitting}
\label{sec_simplexSplitting}
Once a simplex $M_s$ has been selected for refinement, the actual splitting has to be carried out. In our implementation (see section \ref{sec_meshRefine}), this means that a procedure has to be designed for the following $3$ tasks:
\begin{itemize}
\item One of the edges must be selected for splitting.
\item A new vertex with carefully computed coordinates has to be introduced in order to break the splitting edge.
\item New edge tracers (see section \ref{sec_quadraticElements}) have to be introduced.
\end{itemize}

Because we chose the conservation of the Poincaré invariant (equation~\ref{eq_poincare}) as the base of our refinement criterion, it certainly makes sense to also use it when choosing the edge to split and we therefore implement splitting edge selection based on the {\em minimisation} of our refinement criterion (equation~\ref{eq_invariantRefined}). This is achieved by simulating the splitting of $M_s$ along all its edges and computing the new value of the invariants $I^\prime_s(j)$ and $I^{\prime\prime}_s(j)$ measured for each of the two resulting simplices $M^\prime_s$ and $M^{\prime\prime}_s$ obtained by splitting along the $j^{\rm th}$ edge of $M_s$. The edge for which ${\rm max}[I^\prime_s(j),I^{\prime\prime}_s(j)]$ is the lowest is then selected for splitting. We note that such a procedure involves simulating two consecutive refinements along all possible edges of $M_s$ as well as along those of $M^\prime_s$ and $M^{\prime\prime}_s$, since equation (\ref{eq_invariantRefined}) is based on the computation of the Poincaré invariant after refinement. This could make refinement slow if the splitting procedure itself requires a lot of computations but this is not problematic in practice as only a small fraction of all simplices in the mesh is expected to be refined at a given time step.

An important issue is that refinement should converge in a finite number of steps (and ideally in a single step), which is not necessarily the case with our splitting scheme: some situations might require two or more consecutive splittings for the invariant to satisfy the refinement criterion. When such cases occur, i.e. whenever a single split does not lead to a sufficient improvement, we choose, for stability purpose, to refine in turn the segments with longest Lagrangian size (i.e. those that were refined the least). This basically amounts to refining isotropically in Lagrangian space until the here-above described method is able to converge in a single refinement. In practice, we observed that a single split is enough to meet refinement criterion in the vast majority of cases.

Once a splitting edge is selected, actual bisection requires the introduction of a new vertex (see figure \ref{fig_tetRefine}), which we naturally chose to be the supplementary tracer that we use for our second order description of the sheet (see section \ref{sec_quadraticElements}). As a consequence, simplices are split into equal Lagrangian (initial) volume children by this procedure and the mass can be partitioned equally between them to a very good approximation as long as the initial mass distribution of the simplices is homogeneous. When initial simplices are allowed to have different masses, the mass gradient in Lagrangian coordinates has to be taken into account to compute the fraction of mass associated to each split simplex.
\begin{figure}
\begin{centering}
\hfill%
\begin{subfigure}[b]{0.49\textwidth}
\centering
\includegraphics[width=.99\linewidth]{figs/tracers_interp_low.png}
\caption{Initial tessellation\label{fig_refineSplit1}}
\end{subfigure}
\hfill%
\end{centering}

\begin{centering}
\hfill%
\begin{subfigure}[b]{0.49\textwidth}
\centering
\includegraphics[width=.99\linewidth]{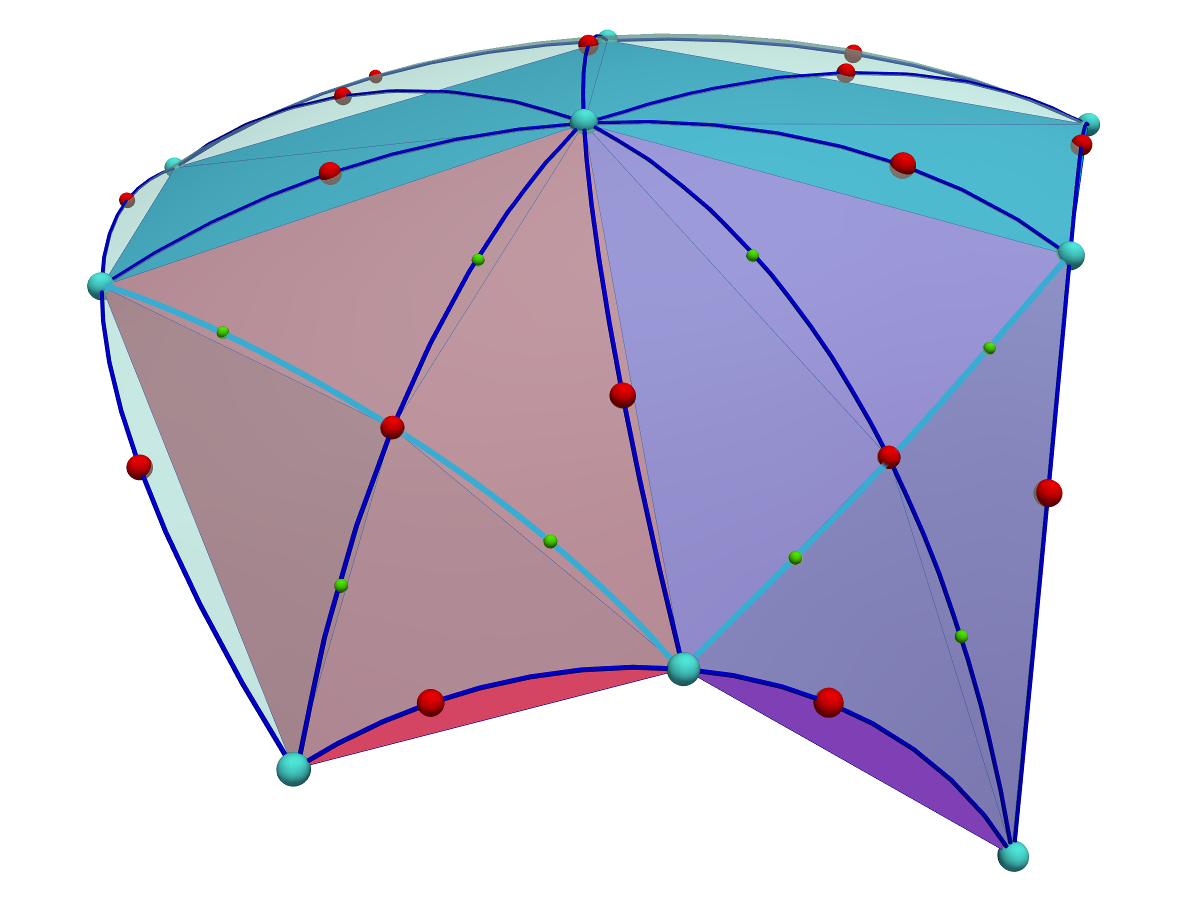}
\caption{First pass\label{fig_refineSplit2}}
\end{subfigure}
\hfill%
\begin{subfigure}[b]{0.49\textwidth}
\centering
\includegraphics[width=.99\linewidth]{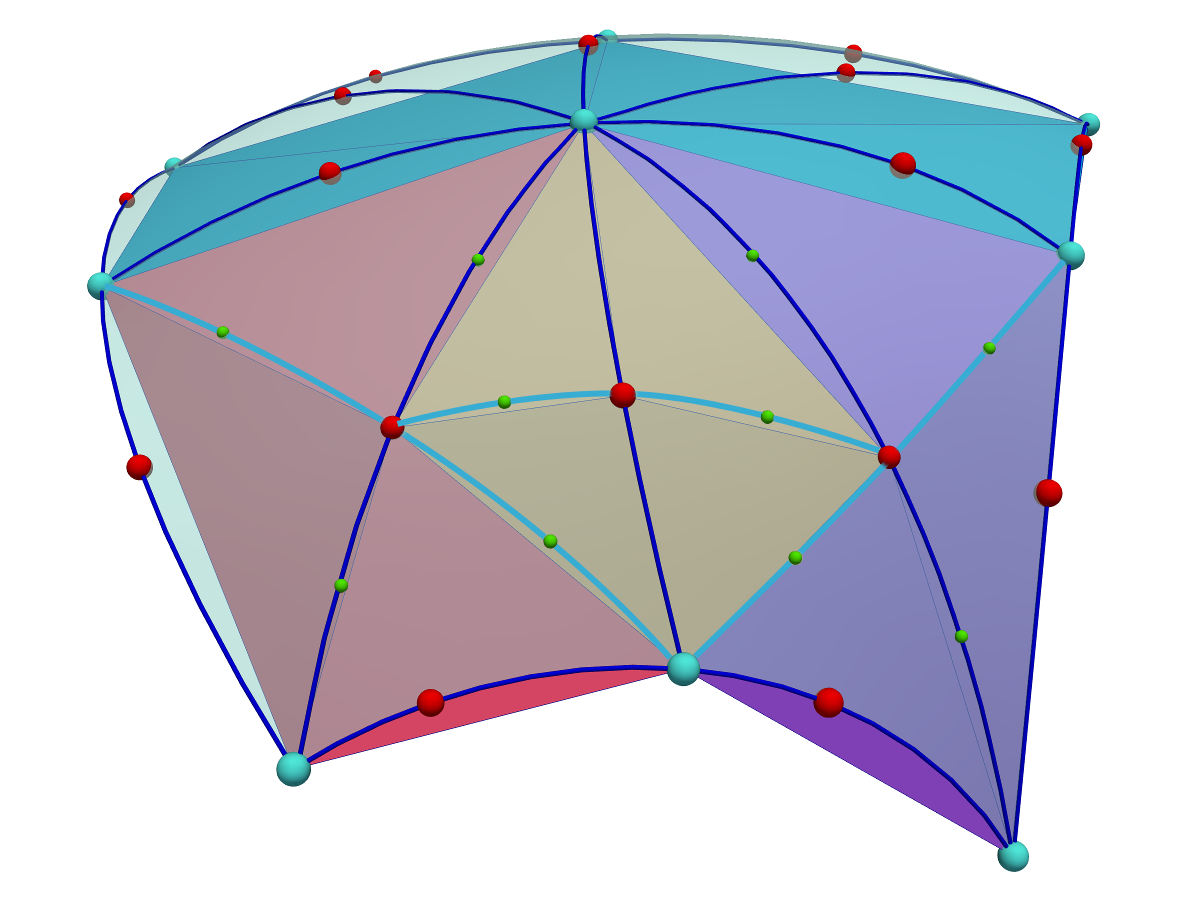}
\caption{Second pass\label{fig_refineSplit3}}
\end{subfigure}
\hfill%
\end{centering}
\caption{Illustration of the phase-space sheet refinement in the 2D case. The initial sheet is described at second order by a linear simplicial tesselation augmented with edge tracers (figure \ref{fig_refineSplit1}). Suppose that the $4$ triangles facing the viewer need refinement, it may be achieved in two passes: two segments that do not share a simplex are first selected for refinement and their edge tracers is used to split them (figure \ref{fig_refineSplit2}). New edge tracers are interpolated at the mid points of the split edge along the quadratic surface (small green balls) to maintain an identical second order description of the surface geometry. Finally, one edge adjacent to the remaining simplices is split in the same fashion to obtain a fully refined sheet. Note that the second order approximation of the sheet remains the same during the procedure. The 3D case is similar, except that a non bounded number of simplices may be incident to the splitting edge. \label{fig_refineSheet}}
\end{figure}

Finally, in order to recover a refined tessellation consistent with the initial one, new supplementary nodes have to be associated to the new edges. These supplementary nodes are simply chosen to be the mid points of theses edges in Lagrangian coordinates and their Eulerian position is computed according to equation (\ref{eq_lagMapping}). As a result, the second order description of the geometry of phase-space sheet is unaffected by the refinement, but rather it is given more degrees of freedom to adapt to future changes of curvature, allowing for a smooth continuous deformation in time. The refinement procedure is illustrated on figure \ref{fig_refineSheet} in the 2D case.

\section{Applications}
\label{sec:exemple}

\subsection{2D chaotic static potential}
\label{sec:2dchaos}
To assess the behavior of our anisotropic refinement in an extreme situation, we simulate the evolution of a small patch in a static chaotic potential of the form \cite{Binney1982}
\begin{equation}
\label{eq_chaosPotential}
\Phi \left(x,y\right)=\frac{1}{2} \log \left({R_{\rm c}}^2 + x^2 +
  \frac{y^2}{q^2} -\frac{x^2 - y^2}{R_{\rm e}}\sqrt{x^2 + y^2}\right),
\end{equation}
where $R_{\rm c}$ is the central radius, $q$ quantifies the level of eccentricity and $R_{\rm e}$ is a tuning parameter controlling the amount of chaos in the orbits, the orbits being regular for $R_{\rm e}=\infty$. The parameters are chosen
such that $R_{\rm c}=0.2$, $q=0.9$, $R_{\rm e}=2$ and the initial patch is set as a $128^2$ elements tessellation of a square of size $0.01$ centered on $(x,y)=(1,0)$ and with uniform initial velocity $\bm{u}=(0,0.4)$. To resolve the equations of motion, we used directly the analytic expression of the force derived from equation (\ref{eq_chaosPotential}) 
with a fixed time step $\Delta t=10^{-3}$. 

\begin{figure}
\begin{centering}
\hfill%
\begin{subfigure}[b]{0.49\textwidth}
\centering
\includegraphics[width=.99\linewidth]{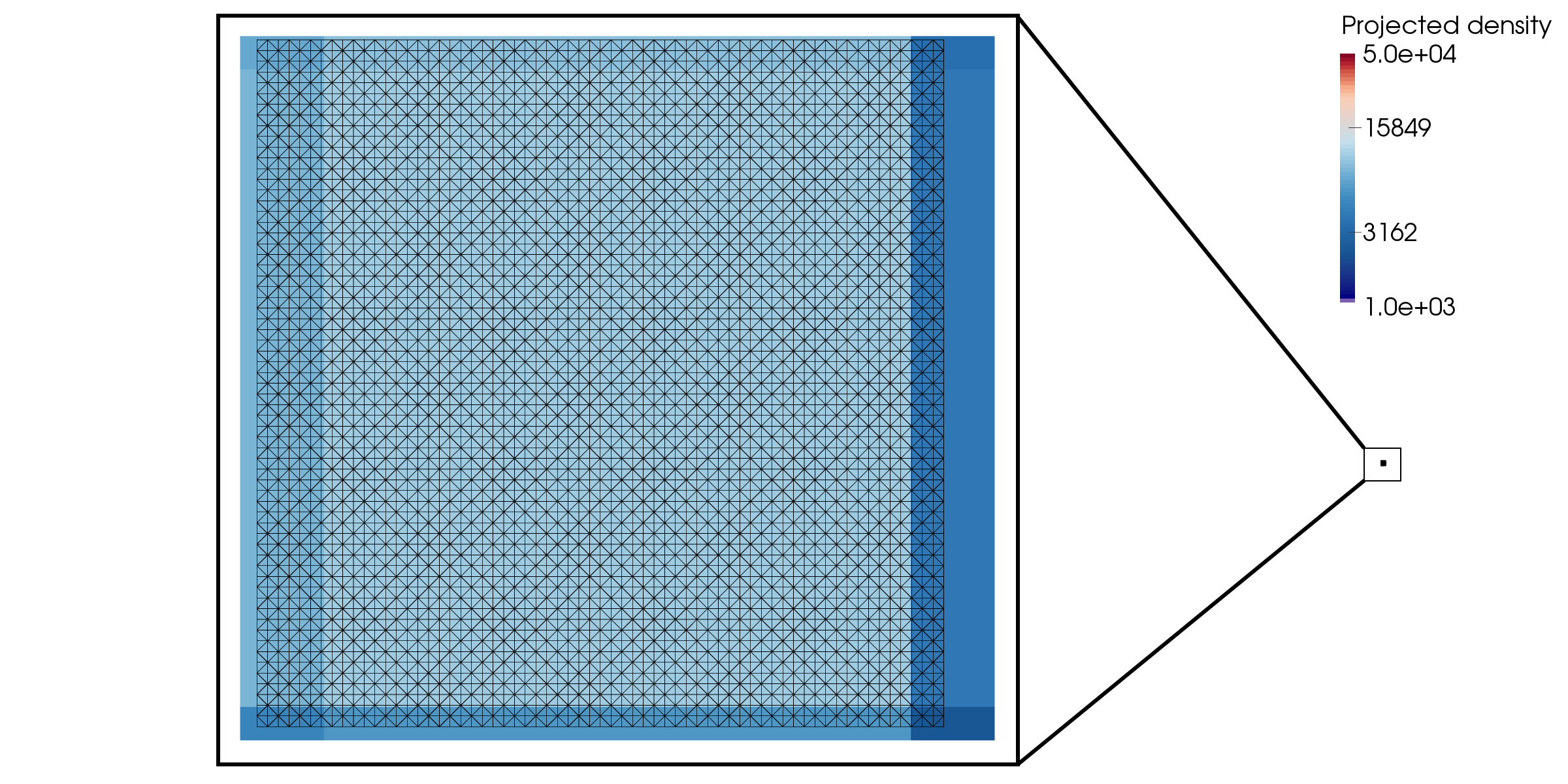}
\caption{Initial square patch (zoomed), t=0\label{fig_chaosA}}
\end{subfigure}
\hfill%
\begin{subfigure}[b]{0.49\textwidth}
\centering
\includegraphics[width=.99\linewidth]{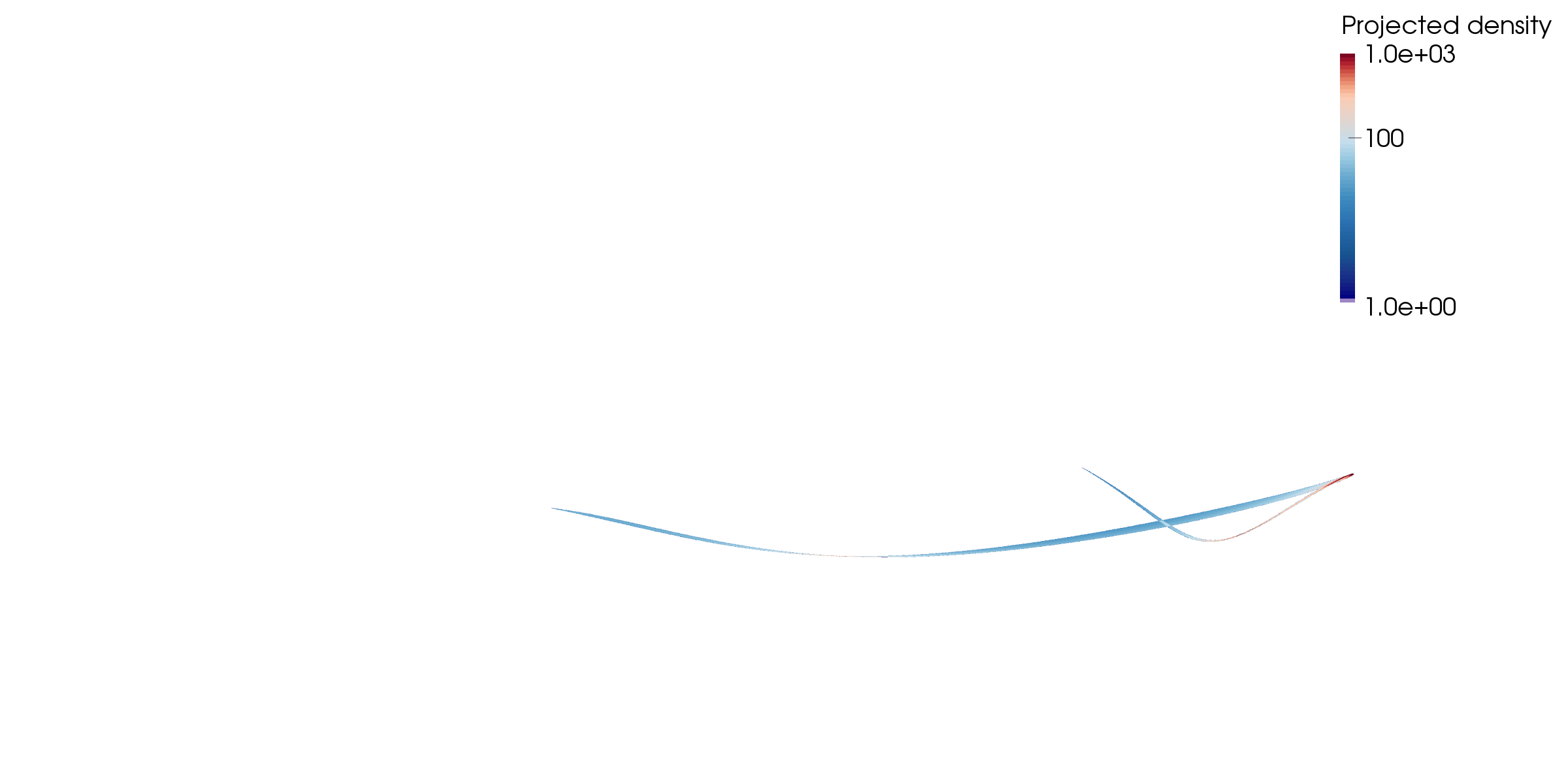}
\caption{t=15\label{fig_chaosB}}
\end{subfigure}
\hfill%
\end{centering}

\begin{centering}
\begin{subfigure}[b]{0.90\textwidth}
\centering
\includegraphics[width=.99\linewidth]{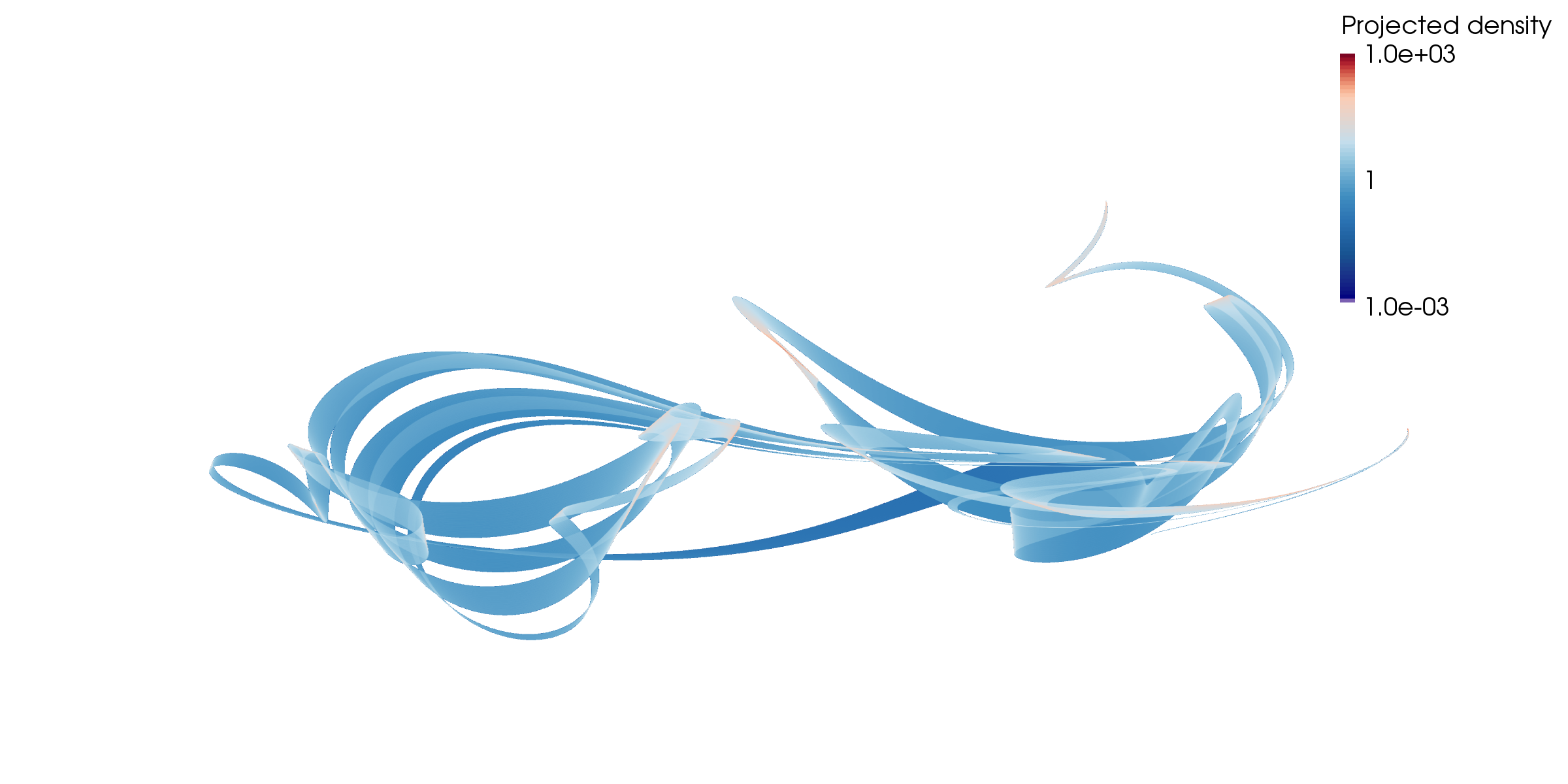}
\caption{t=25\label{fig_chaosC}}
\end{subfigure}
\hfill%
\end{centering}

\begin{centering}
\hfill%
\begin{subfigure}[b]{0.90\textwidth}
\centering
\includegraphics[width=.99\linewidth]{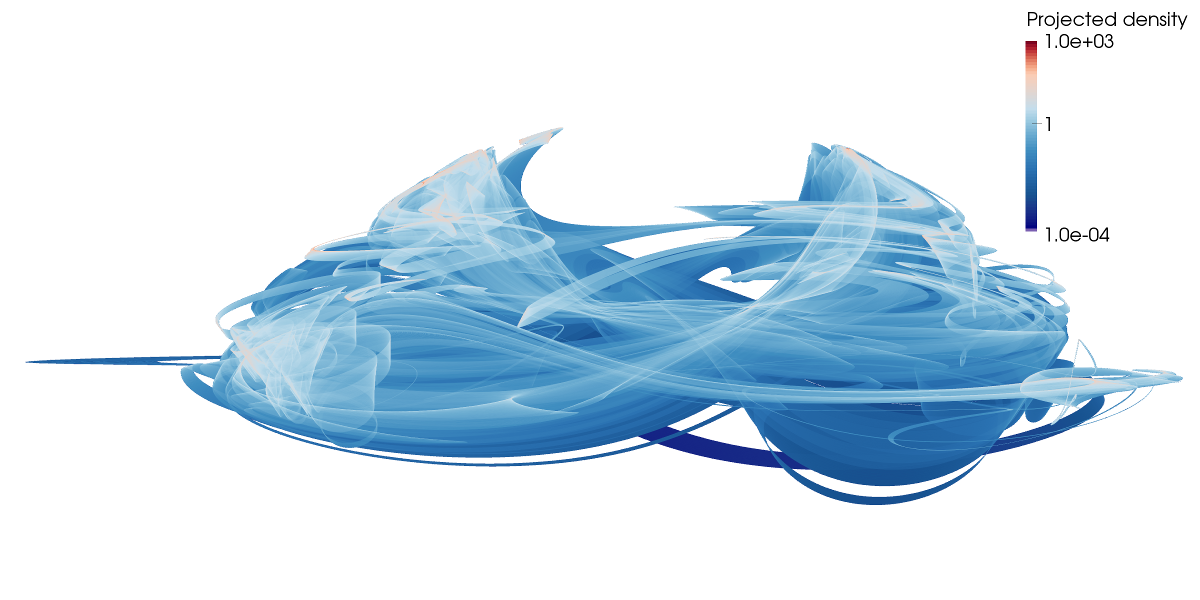}
\caption{t=35\label{fig_chaosD}}
\end{subfigure}
\hfill%
\end{centering}
\caption{Evolution in time of the projected density produced by an initially small square patch of size $0.01$ in a chaotic potential. On each panel, the density is obtained by projecting the phase-space sheet onto a $2048^2$ pixels grid of extent $[-2.5,2.5]$ along the x-axis and $[-1,1]$ along the y-axis. The initial tessellation mesh of the patch is shown plotted over the projected density as a zoom on figure \ref{fig_chaosA}. A movie is also available at the following address: \url{http://www.vlasix.org/index.php?n=Main.ColDICE}. \label{fig_chaos}}
\end{figure}

\begin{figure}
\begin{centering}
\hfill
\includegraphics[width=0.9\linewidth]{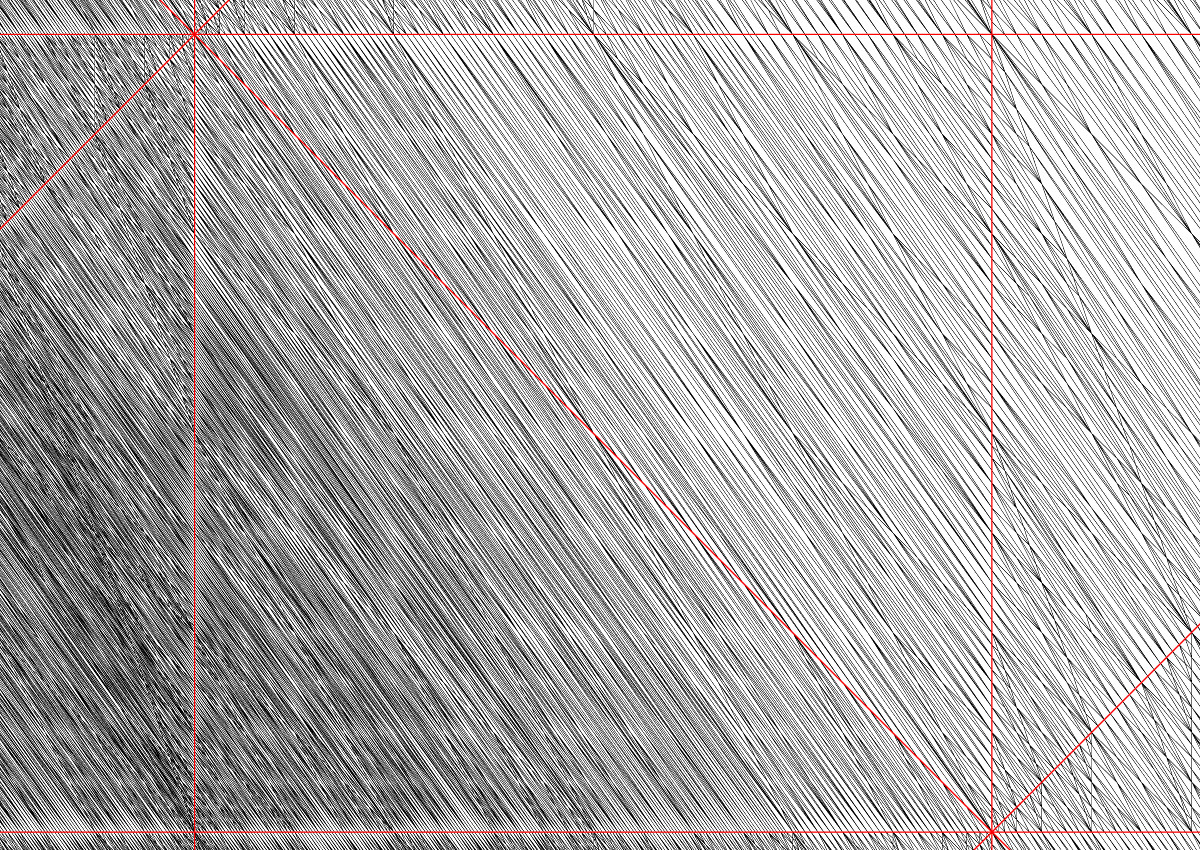}
\hfill
\end{centering}
\caption{Refinement of two initial tessellation mesh elements in Lagrangian space. The red lines show a zoom on two simplices of the initial mesh in Lagrangian space at $t=0$. The black lines delineate the refined elements at $t=28$, showing how our anisotropic refinement implementation is able to track correctly the anisotropy of the deformation and increase effective resolution only in the direction where it is really needed. Note that this is not a particularly refined region of the tessellation, which at $t=28$ has been refined up to level $13$.\label{fig_chaosRefinement}}
\end{figure}

\begin{figure}
\begin{centering}
\hfill%
\begin{subfigure}[b]{0.49\textwidth}
\centering
\includegraphics[width=.99\linewidth]{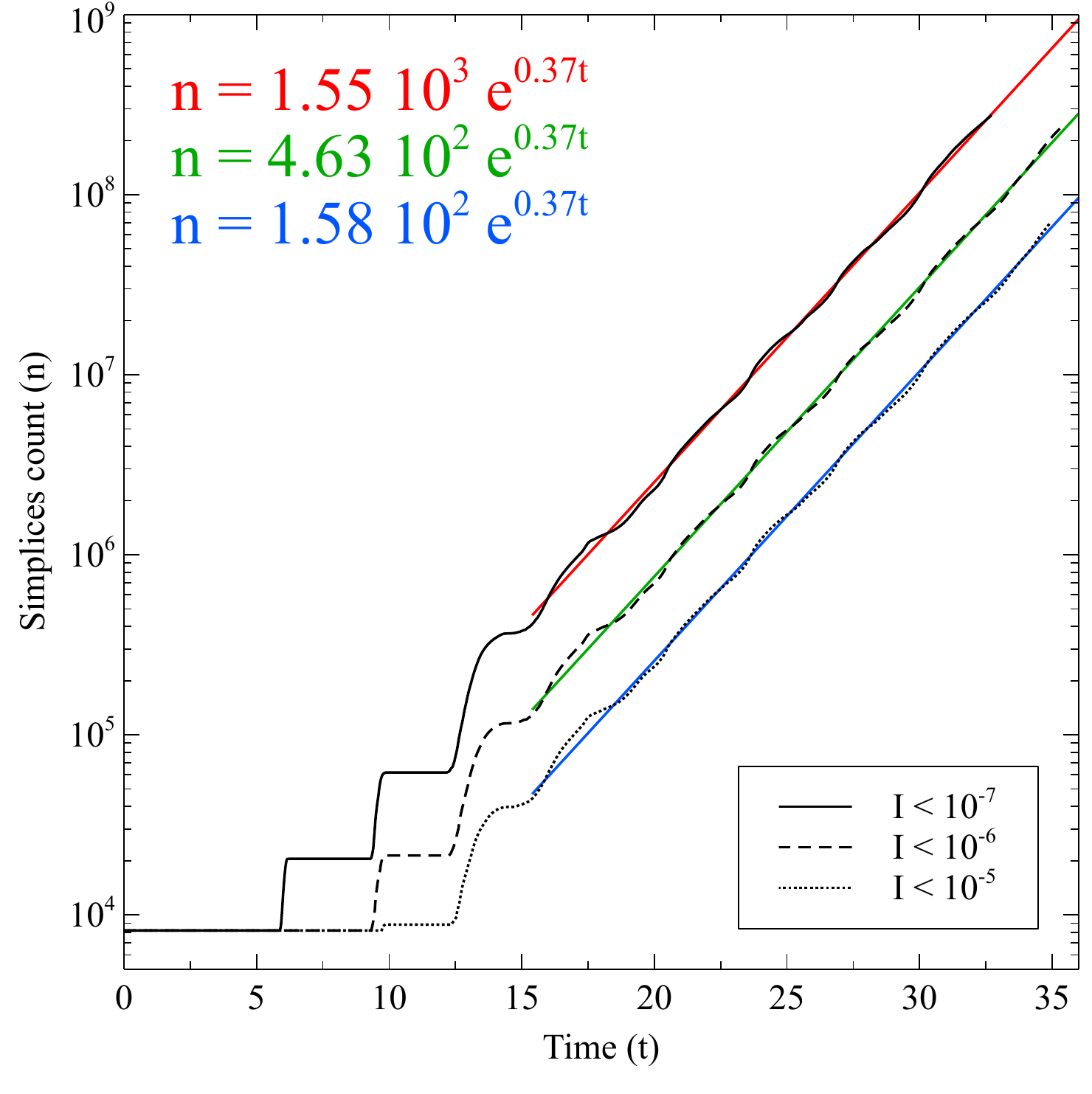}
\end{subfigure}
\hfill%
\begin{subfigure}[b]{0.49\textwidth}
\centering
\includegraphics[width=.99\linewidth]{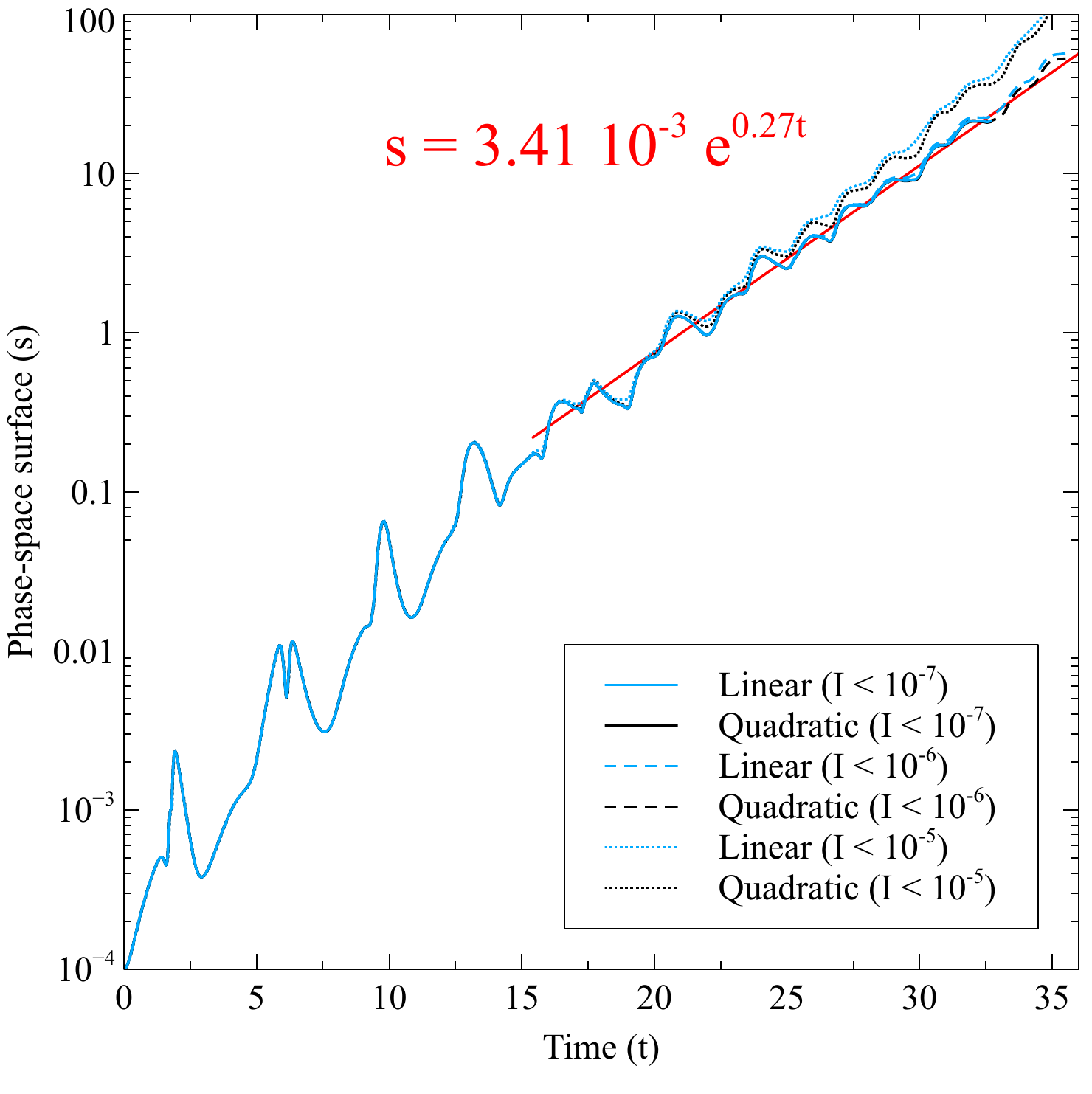}
\end{subfigure}
\hfill%
\end{centering}
\caption{Exponential growth of the number of simplices in the mesh (left) and of the phase-space surface of the patch (right) in the chaotic potential (\ref{eq_chaosPotential}). The solid, dashed and dotted lines correspond to runs with a Poincaré invariant refinement threshold of $I < \epsilon_I=10^{-7}$, $10^{-6}$ and $10^{-5}$ respectively. The blue and black curves on the right panel correspond to the surface of the phase-space sheet using linear elements (i.e. total surface of the simplices) and quadratic elements (i.e. total surface of the quadratic simplices) respectively.\label{fig_chaosSimplices}}
\end{figure}

Figure \ref{fig_chaos} shows various snapshots of the patch in configuration space. It gets considerably stretched with time as a consequence of chaotic orbit instability, triggering strongly anisotropic refinement as illustrated on figure \ref{fig_chaosRefinement}, where a small piece of the simplicial mesh is displayed in Lagrangian space at time $t=28$. 
The number of simplices in the tessellation of the phase-space sheet as well as its surface are shown on figure \ref{fig_chaosSimplices} as functions of time for various values of the Poincar\'e threshold $\epsilon_I=10^{-5}$, $10^{-6}$ and $10^{-7}$. As expected, both quantities display an exponential growth behaviour for $t\gtrsim 15$ , with respective Lyapunov exponents $\alpha_{\rm L}=0.37$
and $\alpha_{\rm L}=0.27$. 

The measured phase-space surface should not depend on the value of the refinement parameter $\epsilon_I$ if numerical convergence is achieved. Examination right panel of Fig.~\ref{fig_chaosSimplices} suggests that, for the number of dynamical times considered, $\epsilon_I  \gtrsim 10^{-6}$ is sufficient to achieve numerical convergence while $\epsilon_I=10^{-5}$ is too loose a threshold. This figure also brings out the difference between the first and second order descriptions of the phase-space sheet: interestingly, this difference becomes noticeable just before deviations related to the choice of $\epsilon_I$ can be observed. Estimating this difference can thus help determining the time range during which the simulation can be trusted.

We now try to understand, in this very anisotropic setting, how refinement relates to the number of simplices $n$ and the phase-space sheet surface. This will allow us to explain certain properties observed in Fig.~\ref{fig_chaosSimplices}.  Firstly, we notice from the left panel of Fig.~\ref{fig_chaosSimplices} that $n$ increases approximately by a factor $3.1$ when $\epsilon_I$ is increased by a factor 10. To understand this scaling, we estimate in \ref{app:refanddis} the constraint induced on the local inter-vertex distance $d$ imposed by our refinement criterion, and we obtain:
\begin{equation}
d \sim \beta   (\epsilon_I R)^{1/3},
\label{eq:dsimeps}
\end{equation}
where $R$ is the local curvature radius along the direction of refinement and $\beta$ some number depending on the orientation of the phase-space sheet in phase-space. This equation, which is true both in 2D and 3D, implies a certain scaling of the number of simplices $n$ with $\epsilon_I$. Two extreme cases can be considered: fully isotropic refinement and extremely anisotropic refinement. 

In the isotropic case, $\beta R^{1/3}$ does not depend on the orientation of the simplex in phase-space and there is no preferred direction in the dynamics. As a result, the simplices count scales like $n \propto d^{-D}$ where $D=2$ or 3 is the dimension of configuration space, so 
\begin{equation}
n \propto \epsilon_I^{-D/3}.
\end{equation}
Anisotropic refinement takes place when local curvature presents a strong angular dependence or when the phase-space sheet becomes very elongated along one direction, as observed on Fig.~\ref{fig_chaos}. In this case, estimating the scaling of $n$ with $\epsilon_I$ is not simple because the result depends on the statistical properties of the geometric set-up. We perform a rough estimate in \ref{sec:refstrans} and find
\begin{equation}
n \propto \epsilon_I^{-\eta_{D}},\label{eq:nsce2}
\end{equation}
with 
\begin{equation}
\eta_{2} \simeq 0.46, \quad \eta_{3} \simeq 0.58, \label{eq:nscea2}
\end{equation}
respectively in two and three dimensions. That is, for $D=2$, $n$ increases by a factor of $2.9$ when $\epsilon_I$  decreases by a factor $10$. This result is to be compared to the ratio $3.3$ and $2.9$ between the red and green line, and between the green and the blue line, respectively, on left panel of Fig.~\ref{fig_chaosSimplices}. Given the very approximate nature of our calculations, this level of agreement is quite respectable. 

It is also possible to relate the Lyapunov exponents obtained respectively for the simplices count and the phase-space sheet surface. If stretching along a single direction is the dominant behaviour, one can use the following similarity argument: an elongation of the phase-space sheet by a factor 2 is roughly equivalent, from equation (\ref{eq:dsimeps}), to a decrease of the refinement criterion $\epsilon_I$ by a factor of 8 in equation (\ref{eq:nsce2}). Then the simplices count is related to the phase-space surface $s$ of the sheet as follows:
\begin{equation}
n \propto  s^{3 \eta_D }.
\label{eq:nspred}
\end{equation}
This means, in two dimensions, that if $n \propto \exp(0.37 t)$,  then $s \propto \exp(0.27 t)$ which provides an excellent prediction for the overall behaviour of $s(t)$ in the regime $t \gtrsim 15$, as shown by the straight line on the right panel of figure \ref{fig_chaosSimplices}. 
\subsection{Double sinusoidal wave collapse}
\label{sec:doublesine}
We now turn to a classical test consisting in studying the evolution of perturbations induced by sinusoidal waves across each axis of the simulation box \cite{Melott1983b,Melott1983,Alimi1990,Moutarde1991,Moutarde1995}. Here, we simulate, similarly to \cite{Melott1983,Alimi1990}, the cosmological gravitational collapse of two orthogonal sinusoidal density waves in $2+2$ dimensional phase-space. The initial configuration of the phase-space sheet corresponds to the tessellation of a $256^2$ pixels grid with two simplices per pixel. Its vertices/tracers positions are perturbed according to the following displacement field:
\begin{equation}
P_x(\bm{q})=\frac{0.4 L}{2\pi} \sin\left( \frac{2\pi}{L} q_x \right), \quad
P_y(\bm{q})=\frac{0.3 L}{2\pi} \sin\left( \frac{2\pi}{L} q_y \right),
\label{eq_2sineDisp}
\end{equation}
where $L \equiv 1$ is the size of the periodic simulation box covering the range $[0,L]$ in each dimension. Such an initial setup is illustrated on figure \ref{fig_2sineInit} at a lower resolution. The initial velocity of each vertex/tracer is set according to the fastest Lagrangian linear growing mode (\ref{app:inic}). The cosmology is chosen to be of Eistein-de Sitter type with a matter density parameter $\Omega_0=1$ and the starting expansion factor is $a=a_{\rm ini}=0.01$. The CFL and dynamical time step parameters are given by $C_{\rm CFL}=0.25$ and $C_{\rm dyn}=0.01$. We performed several simulations with various values of  the refinement parameter, $\epsilon_I=\{10^{-6}, 10^{-7},10^{-8}\}$ and a FFT grid resolution $N_{\rm g}=1024$ to compute the force.  In addition to these, we also performed one additional simulation with $N_{\rm g}=2048$ and $\epsilon_I=10^{-7}$ to test for the effects of spatial resolution. Note that, due to computational cost issues related to very different simplices counts between various configurations, the simulations were not all stopped at the same final times. 

\begin{figure}
\begin{center}
\begin{subfigure}[b]{0.4\textwidth}
\centering
\includegraphics[width=.99\linewidth]{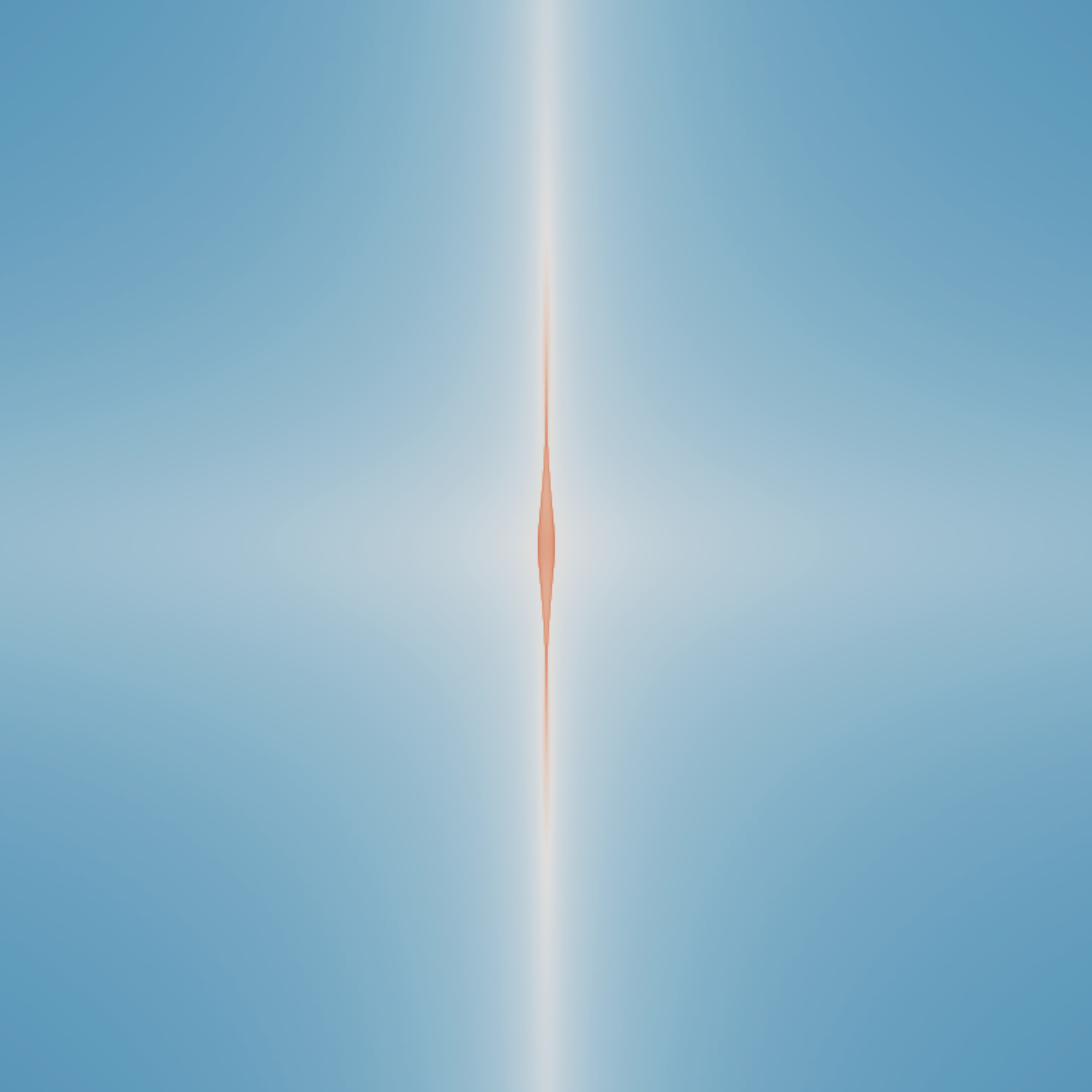}
\end{subfigure}
\hskip 0.1cm
\begin{subfigure}[b]{0.4\textwidth}
\centering
\includegraphics[width=.99\linewidth]{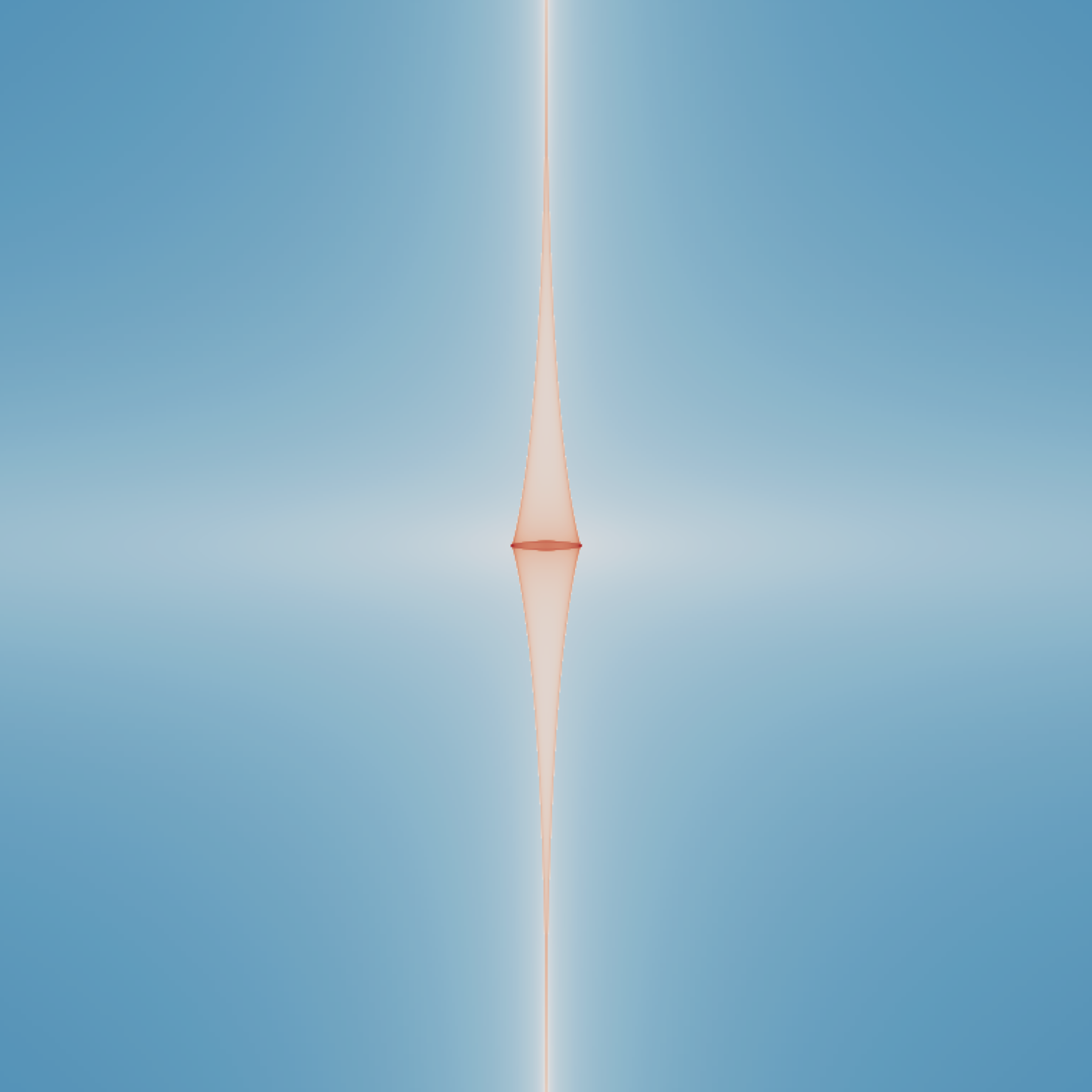}
\end{subfigure}
\end{center}
\begin{center}
\begin{subfigure}[b]{0.4\textwidth}
\centering
\includegraphics[width=.99\linewidth]{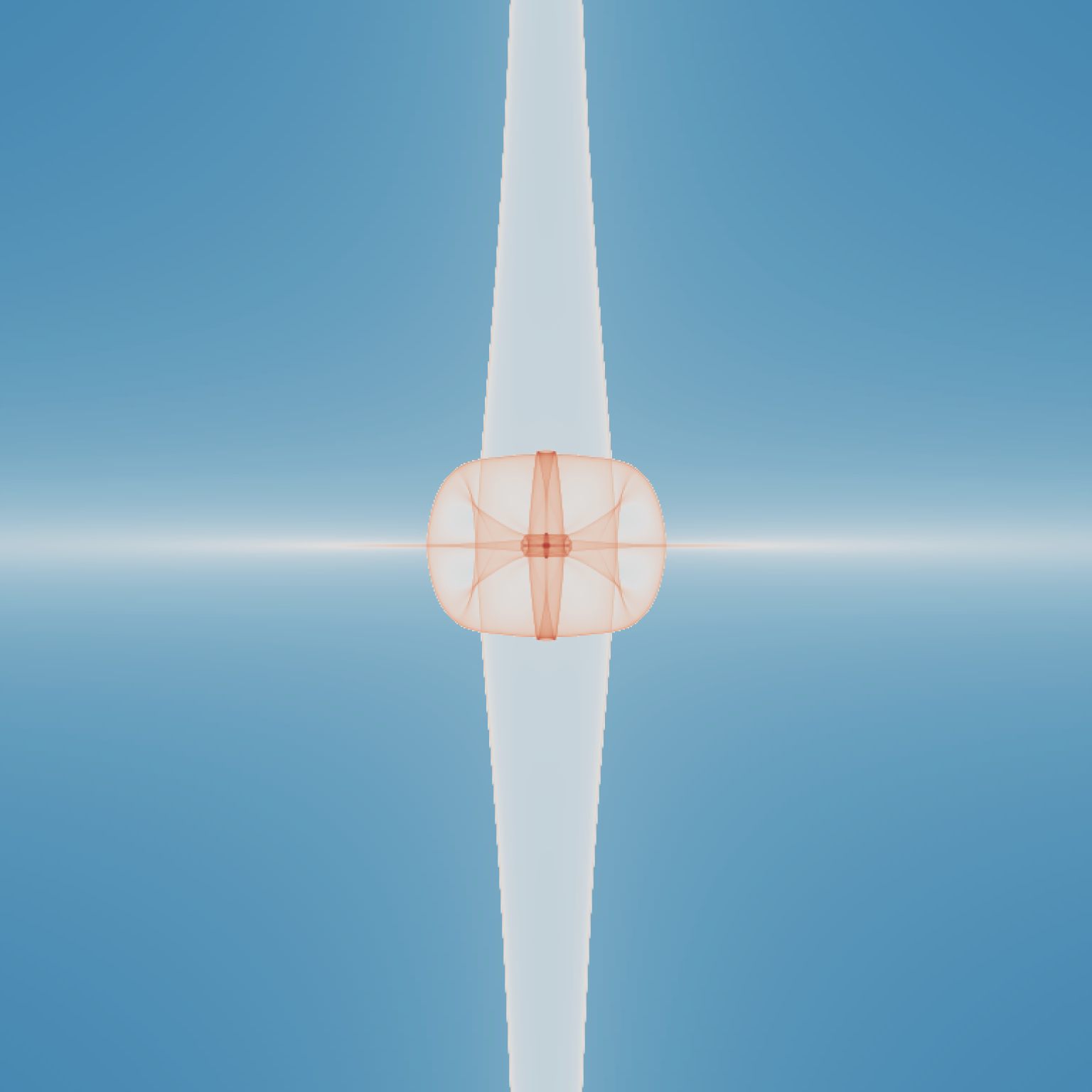}
\end{subfigure}
\hskip 0.1cm
\begin{subfigure}[b]{0.4\textwidth}
\centering
\includegraphics[width=.99\linewidth]{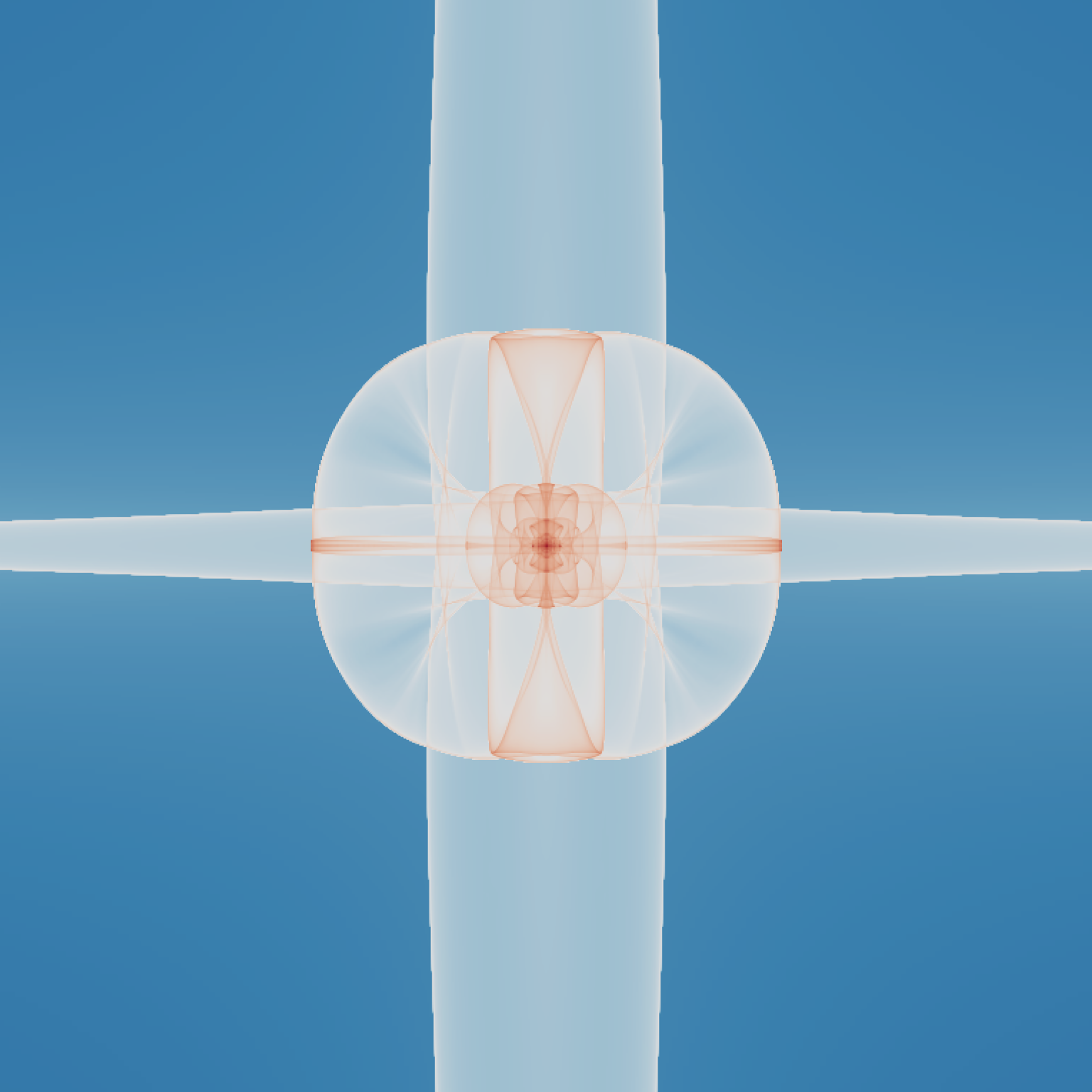}
\end{subfigure}
\end{center}
\begin{center}
\begin{subfigure}[b]{0.4\textwidth}
\centering
\includegraphics[width=.99\linewidth]{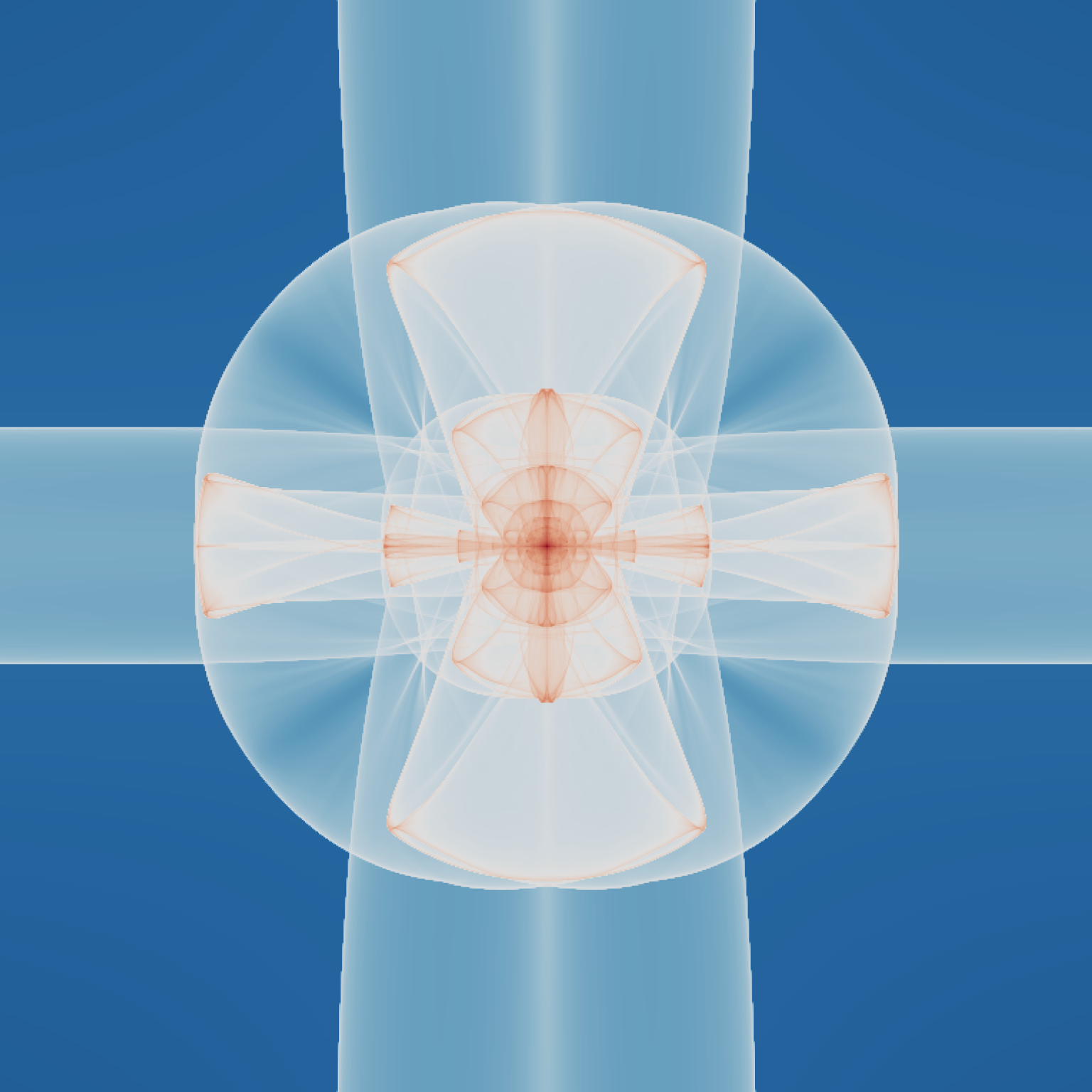}
\end{subfigure}
\hskip 0.1cm
\begin{subfigure}[b]{0.4\textwidth}
\centering
\includegraphics[width=.99\linewidth]{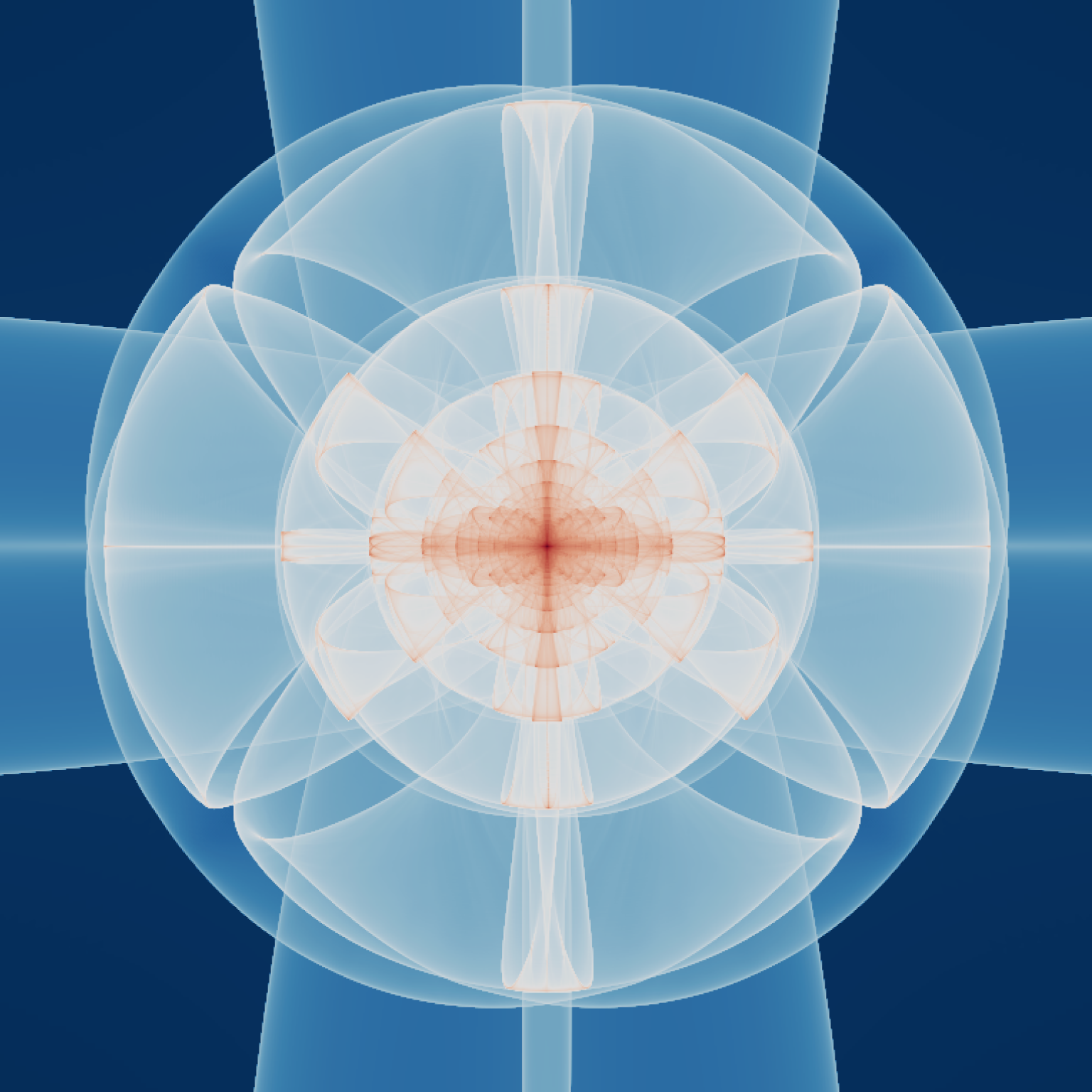}
\end{subfigure}
\end{center}
\caption{Projected density at various times of the two sinusoidal waves simulation with $\epsilon_I=10^{-7}$ and a force resolution of $2048^2$ pixels. A zoom in the region $[L/4+L/16, 3L/4-L/16]$ where $L$ is the simulation box size has been performed. From top to bottom and left to right, the value of the expansion factor is $a=0.023$, $0.025$, $0.033$, $0.048$, $0.088$ and $0.255$. The image is obtained directly from the projected density field used to solve Poisson equation. \label{sin_time_evolv}}
\end{figure}

Figure \ref{sin_time_evolv} shows various snapshots of the projected density of the $N_{\rm g}=2048$ simulation in a region of size one quarter that of the simulation box. The system first experiences a collapse in the $x$ axis direction and becomes dominated by a thin vertical structure (top left panel), followed by a collapse in the $y$ axis direction, which creates a thin horizontal bar inside the previously formed pattern (top right panel).  Studying the physics of successives crossing along each axis and nonlinear couplings between various directions of collapse is probably one of the keys to understanding the formation of dark matter halos.  Anisotropy in the dynamics due to the different amplitudes given to the initial displacement field in each direction is then progressively erased and the system becomes nearly circular at advanced times of the simulation, with a clear self-similar pattern \cite{Alimi1990}. 

\begin{figure}
\begin{centering}
\hfill
\includegraphics[width=0.98\linewidth]{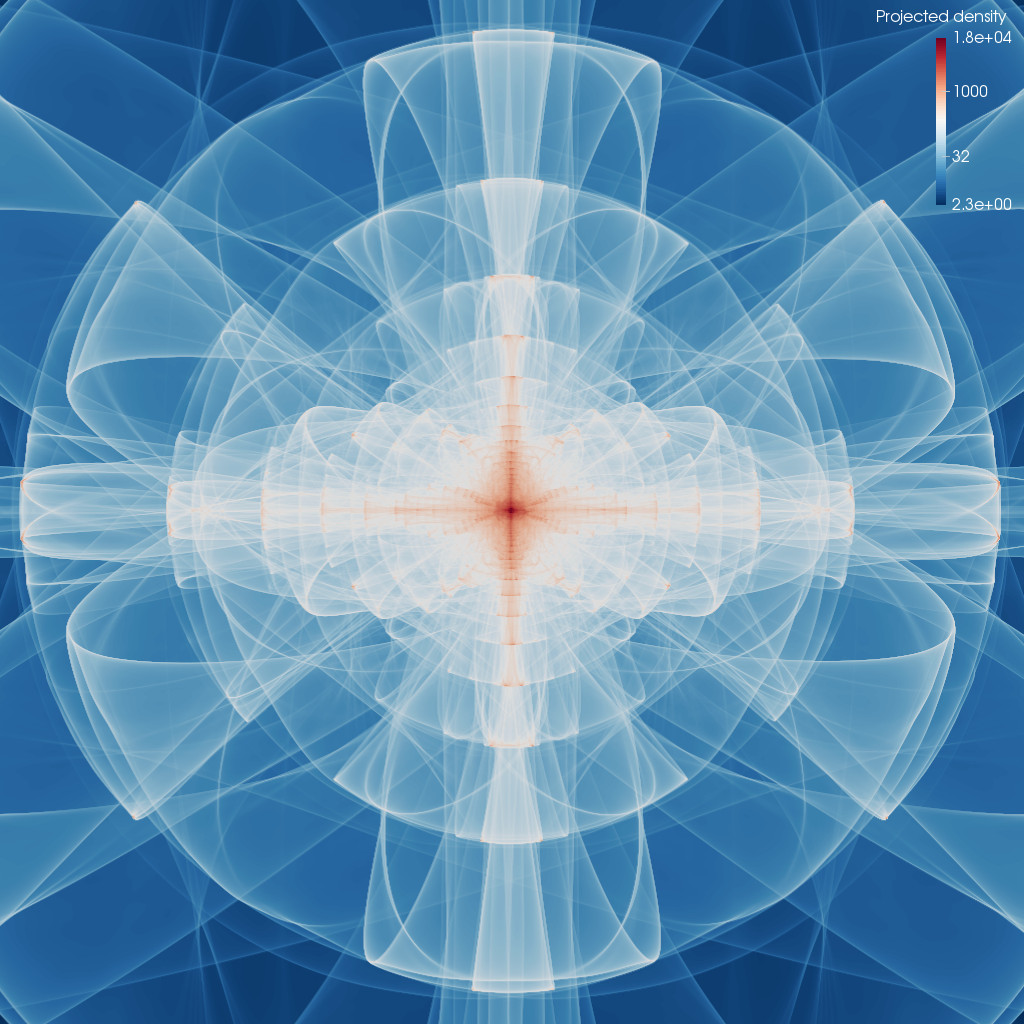}
\hfill
\end{centering}
\caption{Zoom on the central region of the projected density field of the two sine simulation with $\epsilon_I=10^{-8}$ and a force resolution of $1024^2$ pixels, at the same time as bottom right panel of Fig.~\ref{sin_time_evolv} ($a=0.255$). This image was obtained by reprojecting the phase-space sheet at a resolution of $8192^2$. A patch of size exactly $1/8^{\rm th}$ of the actual bounding box in each dimension is represented on the figure.}
\label{fig_2sineDensityZoom}
\end{figure}

\begin{figure}
\begin{centering}
\hfill
\includegraphics[width=0.98\linewidth]{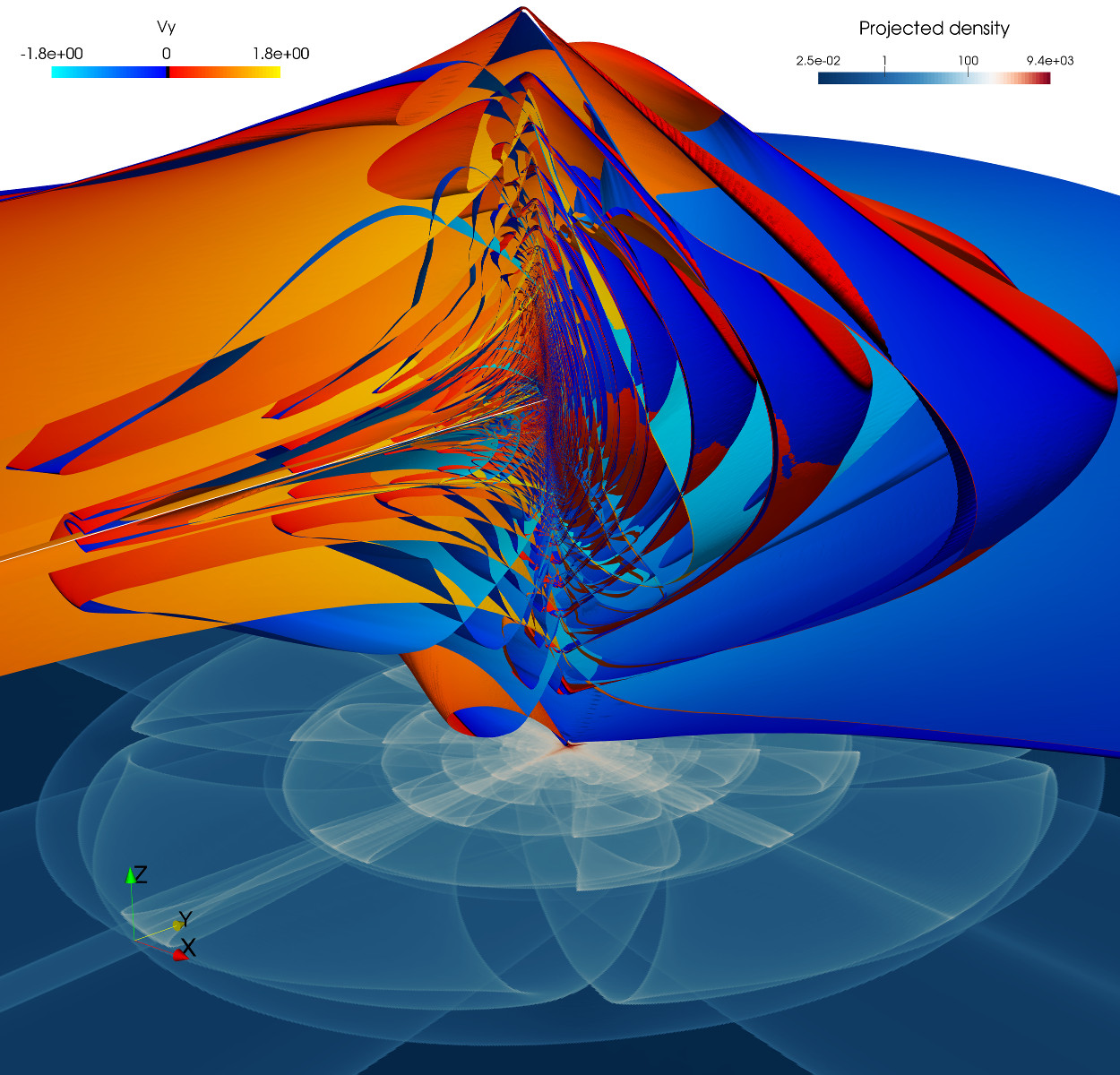}
\hfill
\end{centering}
\caption{Structure of the phase-space sheet of the two sine waves simulation with $\epsilon_I=10^{-6}$ and a force resolution of $1024^2$ pixels, at the same time as Fig.~\ref{fig_2sineDensityZoom} and bottom right panel of Fig.~\ref{sin_time_evolv}.  The $x$ and $y$ axes represent the 2D position space while the velocity $u_x$ is represented along the $z$-axis, compressed by a factor of $8$ for clarity (the higher the $z$ coordinate, the higher the value of $u_x$, with $u_x=0$ at about middle altitude). The velocity $u_y$ is color coded in blue and red, with a black stripe corresponding to $u_y=0$. On the bottom part of the picture, the projected density used for force integration (as extracted directly from the simulation) is represented on a plane. Note that the small defects that can be seen on the image come from the fact that the phase-space sheet is only represented at order $1$, whereas a smoother order $2$ approximation is actually available in the code.\label{fig_2sinePS}}
\end{figure}

Figure~\ref{sin_time_evolv}  also illustrates the exquisite accuracy of the numerical approximation of the phase-space sheet by the adaptive tessellation, with both a smooth and a sharp description of the projected density with well defined caustics. This state of facts is even better exemplified by Figs.~\ref{fig_2sineDensityZoom} and \ref{fig_2sinePS}. Figure \ref{fig_2sineDensityZoom} shows the density in the center of the $(N_{\rm g},\epsilon_I)=(1024,10^{-7})$ simulation at time $a=0.255$ (similar to that of image on the bottom right panel of figure \ref{sin_time_evolv}), but re-projected at a much higher resolution of $8192^2$ pixels. The quality and smoothness of the phase-space sheet can be clearly seen on this image through the details of the complex caustic patterns. Note that this figure can thus be compared directly to the bottom right panel of Fig.~\ref{sin_time_evolv}, as long as one stays aware of its better resolution. Of course, some small discrepancies are expectable between both simulations, particularly at late times, since they do not have the same spatial force resolution. On the other hand, effects due to the choice of $\epsilon_I$ are not noticeable for the time considered. Figure~\ref{fig_2sinePS} shows, still for $a=0.255$, a phase-space diagram of the simulation with $(N_{\rm g},\epsilon_I)=(1024,10^{-6})$. It evidences again the smoothness of the representation of the system, which is only displayed at order $1$ on the figure whereas an even smoother order $2$ approximation is actually available in the code. It also brings out the very rich pattern of the phase-space sheet, which is the object of many windings. This reflects the fact that the system has already evolved during a significant number of dynamical times. 

\begin{figure}
\begin{centering}
\hfill%
\begin{subfigure}[b]{0.49\textwidth}
\centering
\includegraphics[width=.99\linewidth]{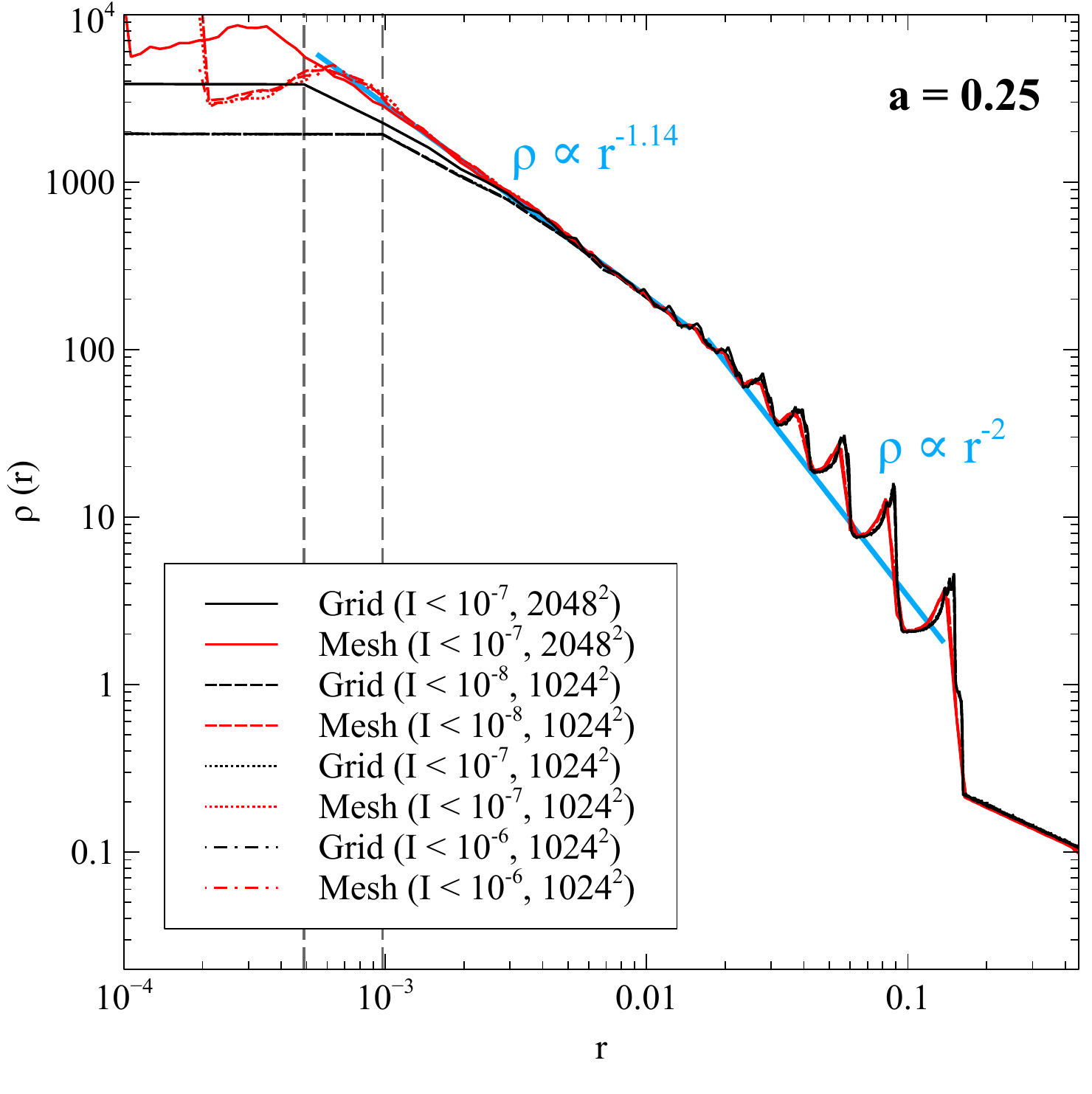}
\end{subfigure}
\hfill%
\begin{subfigure}[b]{0.49\textwidth}
\centering
\includegraphics[width=.99\linewidth]{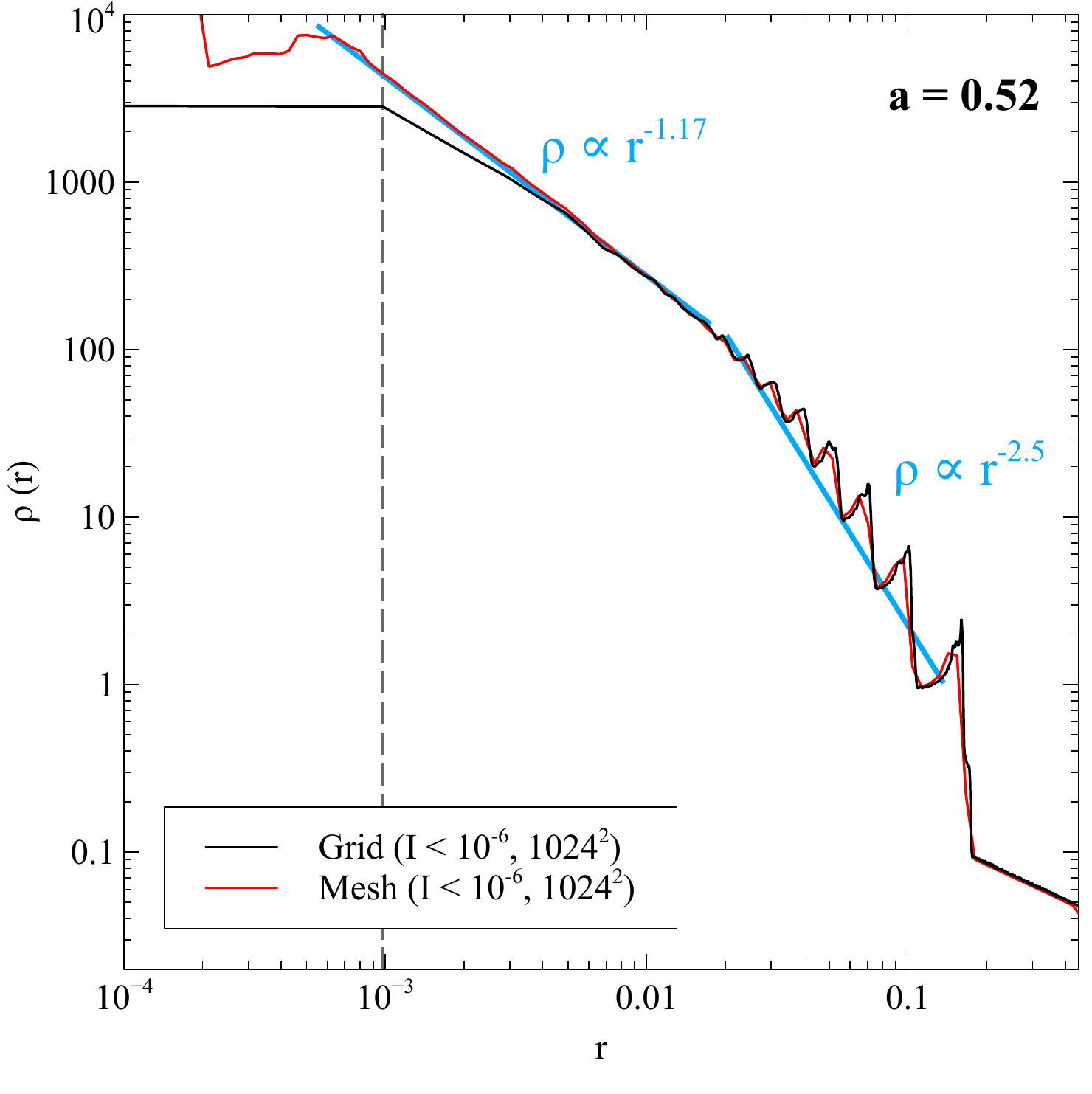}
\end{subfigure}
\hfill%
\end{centering}
\caption{Density profiles of the two sine waves simulations at times corresponding to expansion factor $a=0.25$ (left panel) and $a=0.52$ (right panel). For each simulation, the density profile was computed from the projected density grid used to solve Poisson equation (black curves) and directly from the phase-space sheet tessellation (red curve) by randomly sampling each simplex with $100$ particles. Note that the profiles on the left plot were computed at the same time as the distributions depicted in bottom right panel of  Fig.~\ref{sin_time_evolv}, in Fig.~\ref{fig_2sineDensityZoom} and Fig.~\ref{fig_2sinePS}.\label{fig_2sineProfiles}}
\end{figure}

To complete visual inspection, figure~\ref{fig_2sineProfiles} displays the radial density profile $\rho(r)$ measured in each simulation for $a=0.255$ (left panel) and $a=0.52$ [right panel and only for the $(N_{\rm g},\epsilon_I)=(1024,10^{-6})$ simulation], using two methods to estimate $\rho(r)$: (a) directly from the computational grid used to compute the force and (b) by randomly sampling each simplex of the phase-space sheet with 100 particles. Interestingly, the phase-space sheet seems to carry information slightly beyond the grid resolution so measurements using method (b) are clearly the best. This is also an effect of binning, on which we have better control when using method (b). 

The left panel of Fig.~\ref{fig_2sineProfiles} demonstrates the excellent match between all the runs, with a central power-law behaviour of the density profile $\rho(r) \propto r^{\alpha_\rho}$ of index $\alpha_\rho \simeq -1.14$. This result agrees with measurements of \cite{Alimi1990} performed using PM simulations in a case where the initial amplitudes of the sinusoidal waves are equal. The slope does not seem to change significantly with time, although it might show a small increasing trend as already noticed by \cite{Alimi1990}, since we measure $\alpha_\rho=-1.17$ at $a=0.52$. 

\begin{figure}
\begin{center}
\begin{subfigure}[b]{0.49\textwidth}
\centering
\includegraphics[width=.99\linewidth]{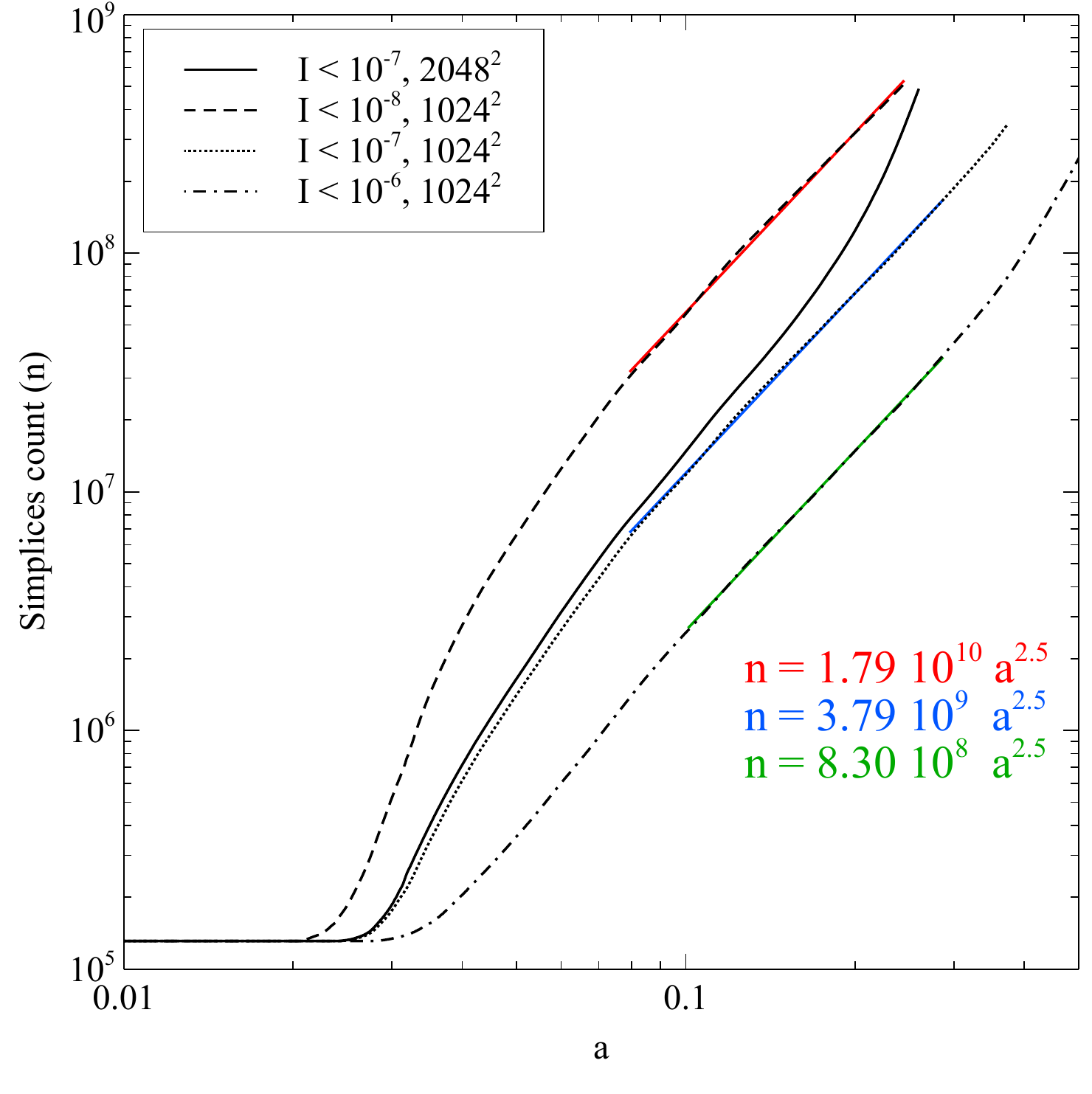}
\end{subfigure}
\hskip 0.2cm%
\begin{subfigure}[b]{0.49\textwidth}
\centering
\includegraphics[width=.988\linewidth]{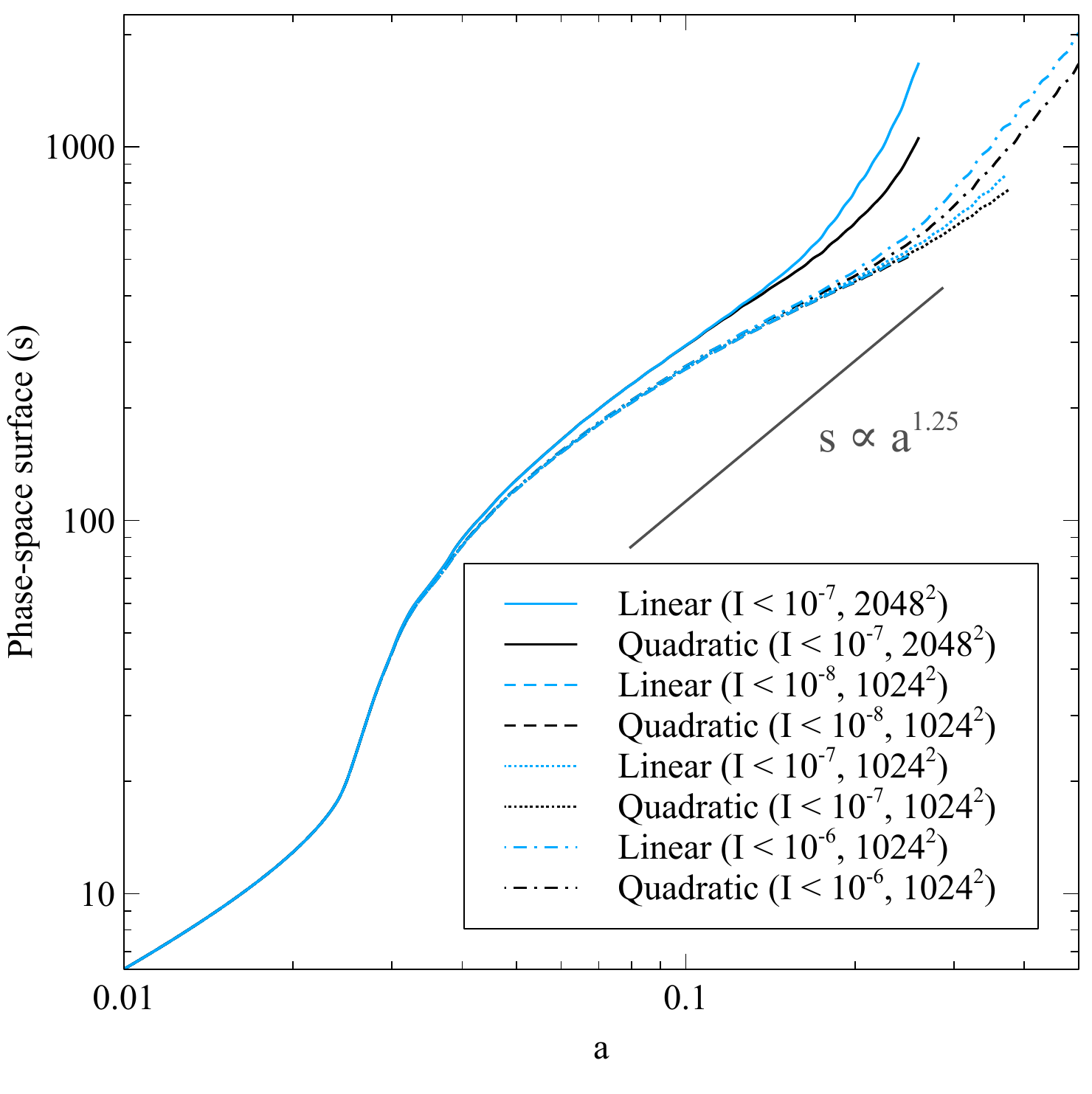}
\end{subfigure}
\end{center}
\caption{Number of simplices in the mesh (left) and phase-space surface of the mesh (right) as a function of the expansion factor for a two sine wave initial condition. The dash, dotted, and dash-dotted lines correspond to runs with a Poincaré invariant refinement threshold $\epsilon_I=10^{-8}$, $10^{-7}$ and $10^{-6}$ respectively and a force resolution of $1024^2$ pixels. The plain lines correspond to $\epsilon_I=10^{-7}$ and a force resolution of $2048^2$ pixels. The blue and black curves on the right panel correspond to the surface of the phase-space sheet using linear elements (i.e. total surface of the simplices) and quadratic elements (i.e. total surface of the quadratic simplices) respectively. \label{fig_2sineSimplices}}
\end{figure}

Figure \ref{fig_2sineSimplices} shows the simplices count $n$ (left panel) and the surface of the phase-space sheet $s$ (right panel) as functions of the expansion factor. Self-similarity is evidenced again with a power-law behaviour, $n \propto a^{\alpha_n}$ with $\alpha_n=2.5$, at least in some regime for the $N_{\rm g}=1024$ simulations, as illustrated by the coloured lines on the left panel of Fig.~\ref{fig_2sineSimplices}. The value of $\alpha_n$ is probably related to the slope $\alpha_\rho$ of the radial density. A detailed analysis of the topology of self-similar solutions, as in e.g.~\cite{Alard2013}, would be needed to predict this relation but this is beyond the scope of the present work. The normalisation of the coloured lines on the left panel of Fig.~\ref{fig_2sineSimplices} scales about as $\epsilon_I^{-2/3}$, as expected for a predominantly isotropic refinement. Hence the advantage of having the possibility of performing anisotropic refinement does not seems to be obvious for this particular setup.  Purely isotropic refinement implies $s\propto \sqrt{n}$ as represented by the grey line in right panel of Fig.~\ref{fig_2sineSimplices} which is clearly a poor fit for the $N_{\rm g}=1024$ simulations. In fact, $s$ does not really behave convincingly as a power-law of the expansion factor. Note however that scaling arguments relating $s$ to the vertex number count are valid only if they apply everywhere on the phase-space sheet, which is not necessarily the case here. For instance, visual inspection of Fig.~\ref{sin_time_evolv} shows that at early times, dynamics is predominantly strongly anisotropic between two crossing times. This might induce anisotropic refinement during these lapses of time. However, because collapse happens in successive directions which are orthogonal to each other, one might not be surprised that refinement becomes isotropic in the end, but finding how the number of vertices $n$ actually relates to the phase-space sheet surface in this case is not a trivial task. 

The $N_{\rm g}=2048$ simulation starts diverging from the other simulations at some point, both for vertices number count and phase-space sheet surface. This was of course expectable since force resolution is twice better in this simulation. However, both $n(a)$ and $s(a)$ seem to be the subject of a dramatic increase when $a \gtrsim 0.15$, which corresponds also to the time when first order approximation and second order approximations of the surface start diverging from each other. While we do not have sufficient dynamical time at our disposal to make any claim, this potentially exponential increase is probably the signature of some numerical resonant instability. It is also visible at later time, $a \gtrsim 0.3$, in the $(N_{\rm g},\epsilon_I)=(1024,10^{-6})$ simulation. In the latter case, diminishing $\epsilon_I$ seems to resolve, at least partly, the problem. Clearly, as already discussed in section \ref{sec:2dchaos}, this is a sign that $\epsilon_I$ is not small enough to correctly follow the system beyond $a \simeq 0.15$ and $a \simeq 0.3$ respectively for $(N_{\rm g},\epsilon)=(2048,10^{-7})$ and $(N_{\rm g},\epsilon)=(1024,10^{-6})$. 

\begin{figure}
\begin{centering}
\hfill
\includegraphics[width=0.5\linewidth]{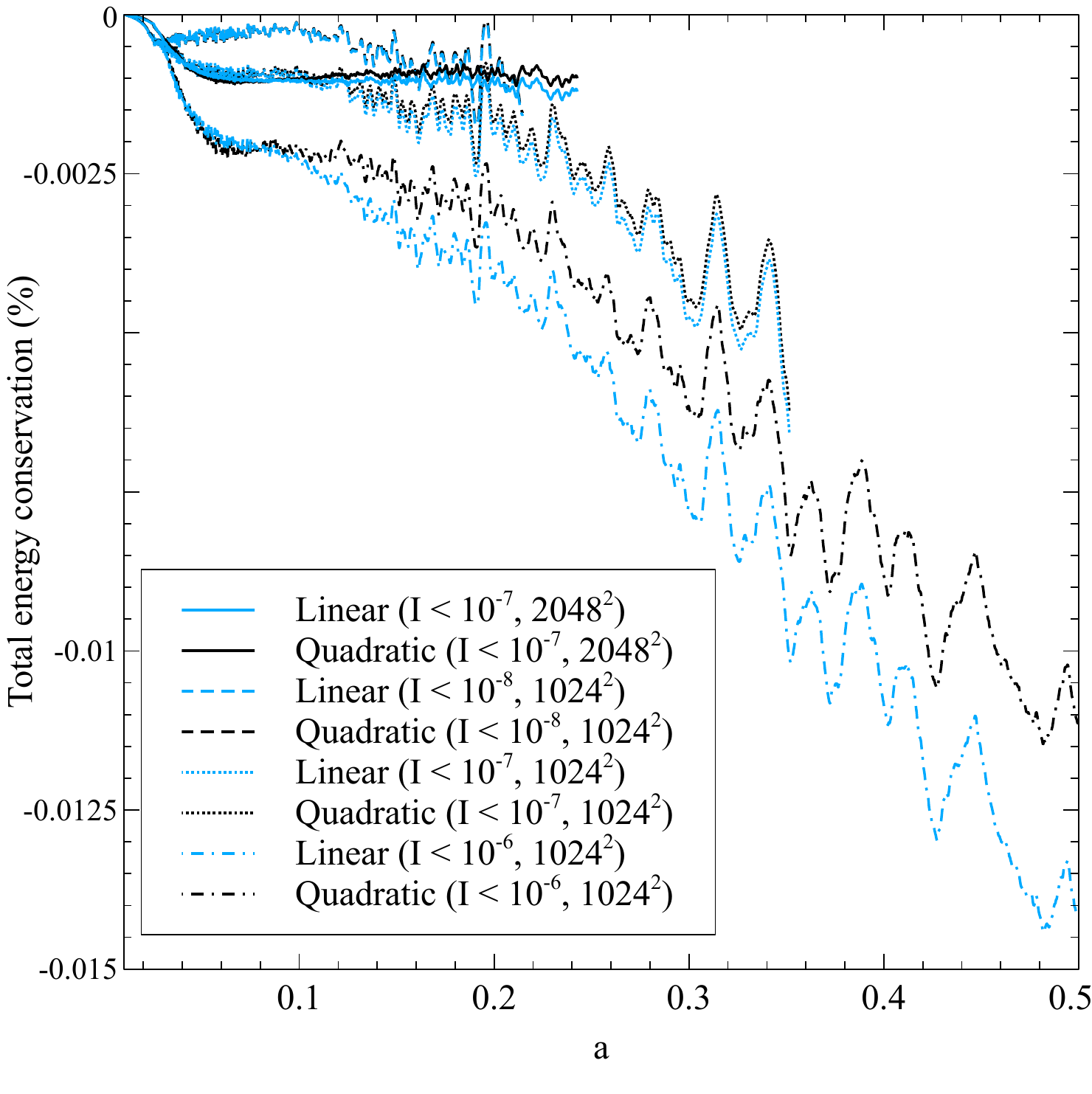}
\hfill
\end{centering}
\caption{Conservation of the total energy as a function of the expansion factor for a two sine wave initial condition. The total energy variation is normalised to the sum of the total of the final kinetic energy and the magnitude of the potential energy during the run with $\epsilon_I=10^{-7}$ and $1024^2$ pixels force resolution. The dash, dotted, and dash-dotted lines correspond to runs with a Poincaré invariant refinement threshold $\epsilon_I=10^{-8}$, $10^{-7}$ and $10^{-6}$ respectively and a force resolution of $1024^2$ pixels. The plain lines correspond to a threshold $\epsilon_I=10^{-7}$ and a force resolution of $2048^2$ pixels. The blue and black curves correspond to a computation of the kinetic energy using linear and quadratic elements respectively.\label{fig_2sineEnergy}}
\end{figure}

In figure \ref{fig_2sineEnergy}, we test energy conservation for the various simulations. To compute the total kinetic energy, we integrate directly the square of the velocity over the phase-space sheet: this allows us to perform  two estimates, one valid at linear order, by simply using equation (\ref{eq:barycentricinterp}) inside each simplex to interpolate the projected density of the phase-space sheet, and one valid at quadratic order by exploiting equation (\ref{eq:quadintbar}). The total potential energy is estimated directly from the FFT grid, which means that we do not take into account the convolution effects related to the finite difference differentiation of the potential followed by dual TSC interpolation to compute the force at the vertice positions. Note also that in the cosmological case, calculation of total energy involves an additional term compared to the standard physical case, due to the expansion of the Universe (see \ref{app:cosener}). 

The choice of the constraints on the time step are on the conservative side, so they do not represent the dominant factor in energy conservation: we indeed checked that increasing the time step by a factor of two does not significantly impact the energy balance. The parameters that matter, as illustrated by Fig.~\ref{fig_2sineEnergy}, are the grid resolution $N_{\rm g}$ used to compute the force and the refinement control parameter $\epsilon_I$. As expected, increasing $N_{\rm g}$ or decreasing $\epsilon_I$  improves energy conservation, although $\epsilon_I$ seems to be the most determinant parameter. In principle, because of the softening of the force below the grid pixel size, curvature of the phase-space sheet should be somewhat related to the size of a pixel: $L/N_{\rm g}$. This is particularly true at sufficiently late time, when the phase-space sheet literally wraps around the central pixels of the simulation, as illustrated by Fig.~\ref{fig_2sinePS}, which explains in part the difference in the late time phase-space surface measured in the $N_{\rm g}=2048$ and the $N_{\rm g}=1024$ simulations (right panel of Fig.~\ref{fig_2sineSimplices}). In order to maintain energy conservation at a given level, one might thus intuitively expect the choice of the refinement parameter $\epsilon_I$ to correlate with the spatial resolution of the simulation.  However, this intuition is difficult to exploit in practice because such a correlation depends on the initial conditions and on the dynamical state considered.  For instance, during a significant amount of time, there is no big difference in the energy check up between the $(N_{\rm g},\epsilon_I)=(2048,10^{-7})$ and the $(N_{\rm g},\epsilon_I)=(1024,10^{-7})$ simulations. Note that the insufficiently small value of $\epsilon_I$ does not seem to affect the energy balance of the $N_{\rm g}=2048$ simulation, while we know that something wrong is happening for $a \gtrsim 0.15$ when considering the phase-space surface. 

Examining each curve of Fig.~\ref{fig_2sineEnergy} more in detail, one notices a decrease of total energy, followed by a plateau and then another regime where energy decreases again. The first decrease is not of real concern. The plateau, of which the size seems to be mainly conditioned by the value of $N_{\rm g}$, is reassuring, because it is a signature of a well behaved symplectic behaviour. The second decrease, on the other hand, does not seem to have an end and is potentially dramatic but remains small in the simulations considered here. Indeed, energy conservation is excellent in all the runs: in the worse case, corresponding to $(N_{\rm g},\epsilon_I)=(1024,10^{-6})$, the variation of total energy relative to the sum of total kinetic energy and magnitude of total potential energy is less than $0.015$ percent. In the best cases, $(N_{\rm g},\epsilon_I)=(1024,10^{-8})$ and $(2048,10^{-7})$, energy conservation is preserved at the $0.001$ percent level! 
\subsection{Cosmological simulation}
\label{sec_WDM}
Our six-dimensional phase-space test consists in simulating a Warm Dark Matter (WDM) universe, similarly to \cite{Hahn15}. The cosmology considered here follows closely the WMAP7 data release measurements \cite{Komatsu2011} and corresponds to a matter density parameter $\Omega_0=0.276$, a cosmological constant $\Omega_{\rm L}=0.724$, a Hubble constant $H_0=70.3$ km/s/Mpc, a spectral index $n_{\rm s}=0.96$  and a r.m.s. of density fluctuations in a sphere of radius $8 h^{-1}$ Mpc linearly extrapolated to present time equal to $\sigma_8=0.811$. This cosmology thus slightly differs from more recent constraints provided by Planck satellite \cite{Planck2015}. Furthermore, the warm dark matter particle mass considered here, $m_{\rm WDM}=250$ eV, is unrealistically low as high redshift Lyman-$\alpha$ forest suggest $m_{\rm WDM} \gtrsim 3.3$ KeV \cite{Viel2013}. However, this simulation is just meant for illustration and tests in terms of parallel performances (performed in section \ref{sec:parallelscaling}), so these choices should not have any impact on the analyses performed below. 

\begin{figure}
\begin{centering}
\hfill%
\begin{subfigure}[b]{0.49\textwidth}
\centering
\includegraphics[width=.99\linewidth]{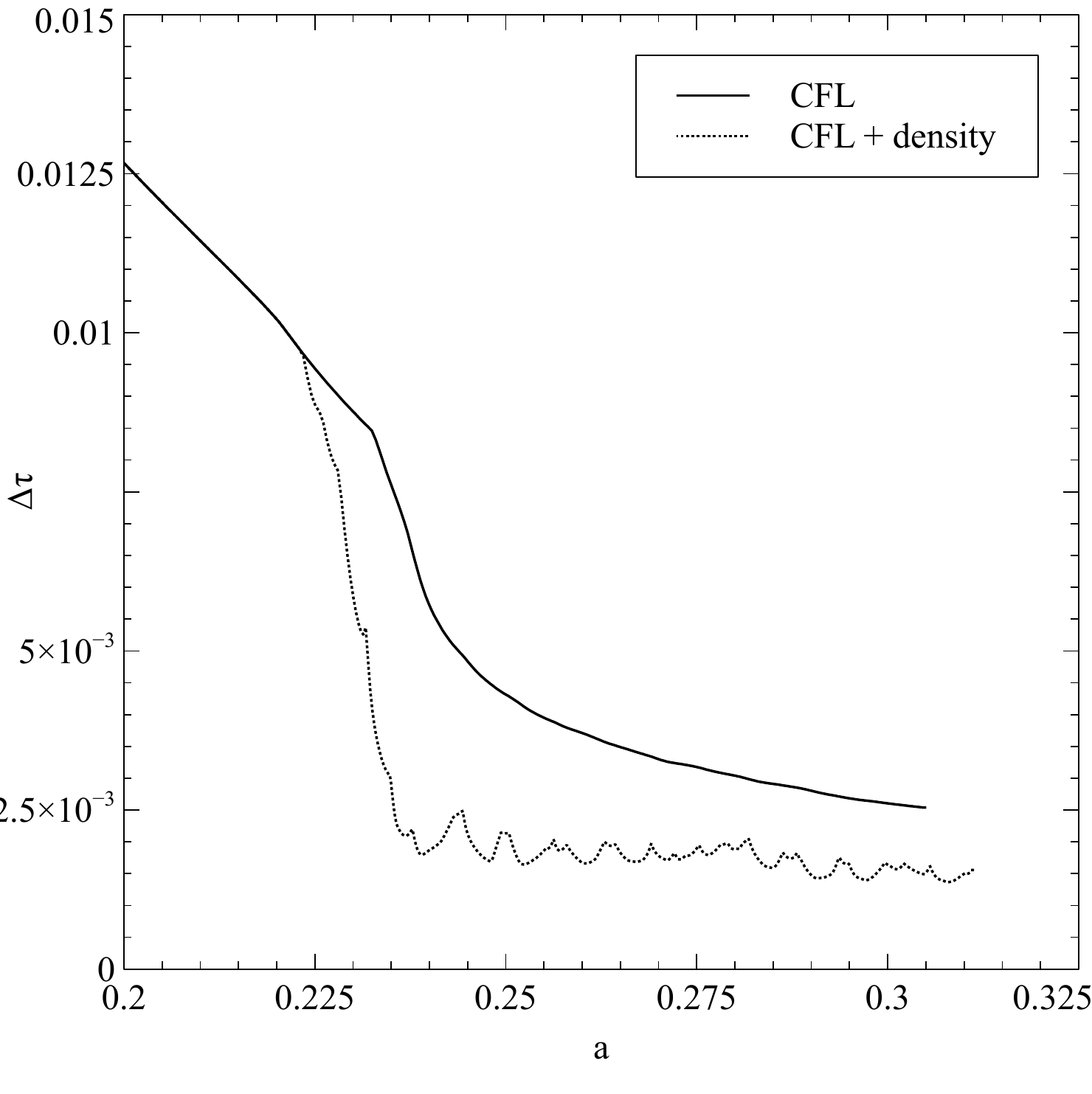}
\end{subfigure}
\hfill%
\begin{subfigure}[b]{0.49\textwidth}
\centering
\includegraphics[width=.99\linewidth]{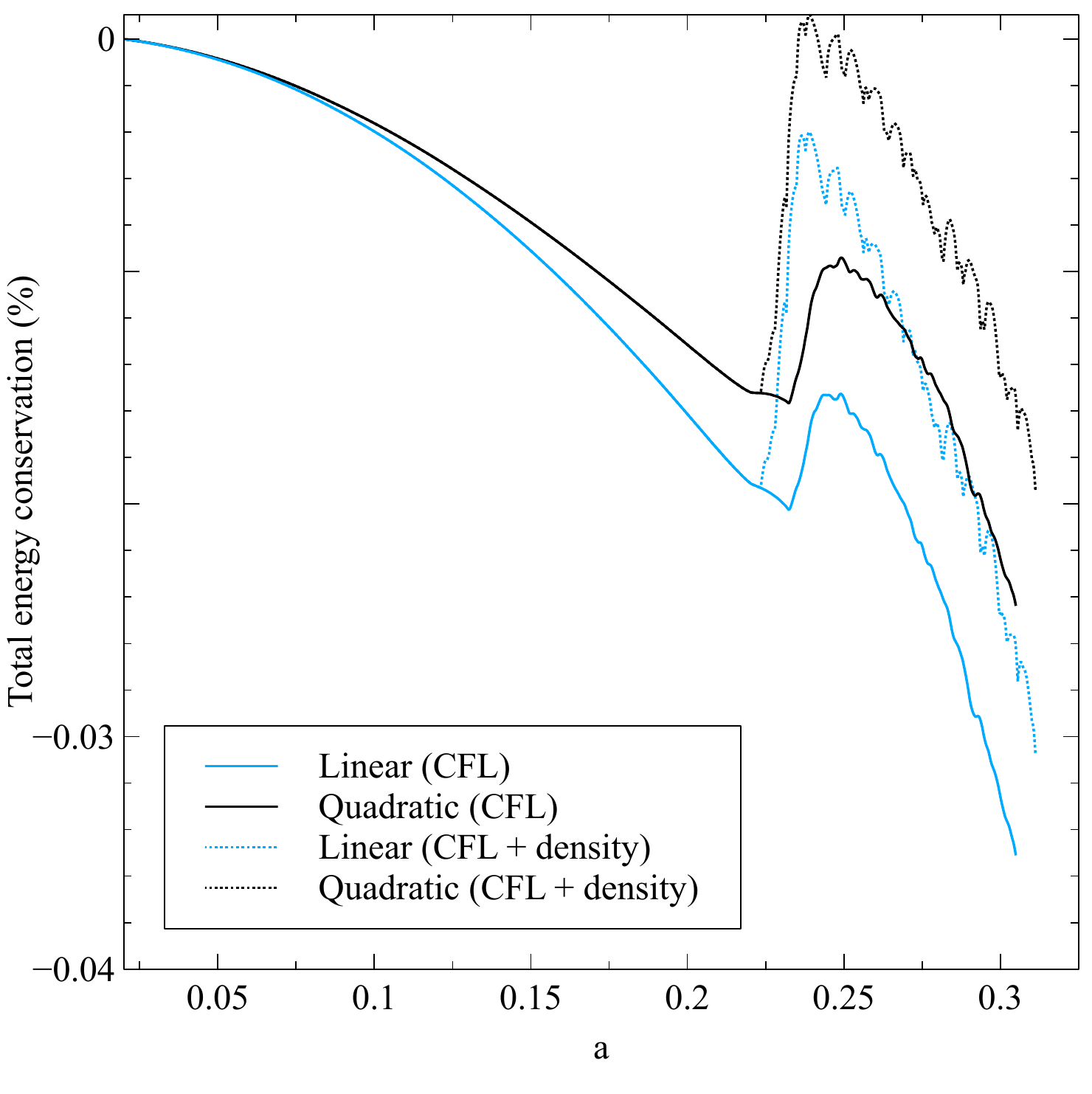}
\end{subfigure}
\hfill%
\end{centering}
\caption{Size of the time step in conformal time $\tau$ (left) and conservation of the total energy (right) as a function of the expansion factor for the cosmological simulation of a WDM universe. The total energy variation is normalised to the sum of the total variation of the kinetic energy and absolute potential energy during the run with a CFL condition only. The blue and black curves correspond to a computation of the kinetic energy using linear and quadratic elements respectively.\label{fig_WDMEnergy}}
\end{figure}

We use a set of $512^3$ particles to create initial conditions in a periodic box of size $L=28.4$ Mpc. The particles positions and velocities are generated using the public code {\tt MUSIC} \cite{Hahn2011} at a cosmic time corresponding to initial expansion factor $a_{\rm ini}=1/51$. These particles provide the initial coordinates of $256^3$ vertices defining the $100,663,296$ simplices of the initial tessellation, the rest of the particles being used to initialise the additional tracers. The grid employed to resolve Poisson equation has $1024^3$ voxels and the simplicial mesh refinement control parameter is $\epsilon_I=10^{-7}$. We performed two simulations, ${\rm WDM}_{\rm CFL64}$ and ${\rm WDM}_{64}$, which differ from each other only by the way time step constraints were set. The first one uses a time step criterion based solely on the CFL condition (\ref{eq:CFL}) with $C_{\rm CFL}=0.25$, while for the second one, the additional dynamical constraint (\ref{eq:DYN}) was used with $C_{\rm dyn}=0.01$.  Figure \ref{fig_WDMEnergy} shows the time step value $\Delta \tau$ as a function of expansion factor, as well as the energy balance for both these simulations. In both simulations, energy conservation is very good at a level better than $0.04$ percent, but seem to be subject to a systematic drift at late times, as already noticed for the sine wave simulations in 2 dimensions. Note however that the simulations do not cover a sufficient dynamical range to be absolutely certain that this drift is going to stand at later time, but our analyses of the two sine waves simulations suggest that it is probably the case. Using the dynamical condition on the time step provides, after collapse, a more constant time step value of $\Delta \tau$ (see the plateau on the dotted curve of left panel of Fig.~\ref{fig_WDMEnergy}) albeit not as smooth as when using the courant condition alone. 

Both simulations were run on 768 cores of the supercomputer Occigen of CINES, using 64 MPI processes with 12 cores each. Each compute node has 24 cores distributed over 2 sockets with 128 Go shared memory. We assigned one MPI process per socket, while local parallel processing on the socket was performed using OpenMP. The ${\rm WDM}_{\rm CFL64}$ and ${\rm WDM}_{64}$ ran respectively for 11 and 27 hours in elapsed time. We stopped the ${\rm WDM}_{64}$ at expansion factor $a=0.31$, corresponding to the time when the number of simplices had just reached $10^9$. 

\begin{figure}
\begin{centering}
\hfill%
\includegraphics[width=0.95\linewidth]{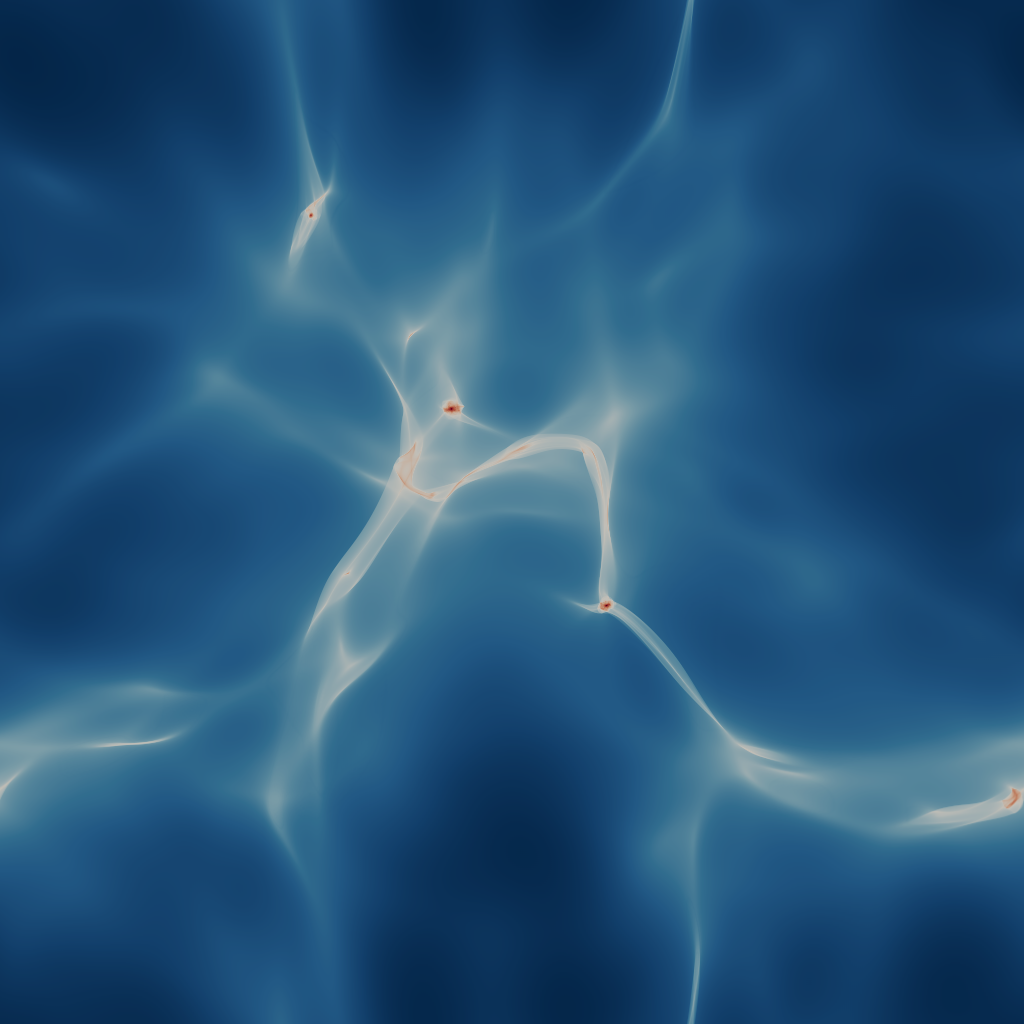}
\hfill%
\end{centering}
\caption{Projected density at the end of the ${\rm WDM}_{64}$ simulation, integrated along the $x$-axis. The image is produced directly from the grid used to compute the force in the simulation and the color scale is logarithmic. The phase-space structure of the second largest halo in the lower central region of the image is shown on figures \ref{fig_halo}  and \ref{fig_haloSubsetsZoom}.\label{fig_WDMprojZ}}
\end{figure}

\begin{figure}
\begin{centering}
\hfill%
\begin{subfigure}[b]{0.8\textwidth}
\centering
\includegraphics[width=.99\linewidth]{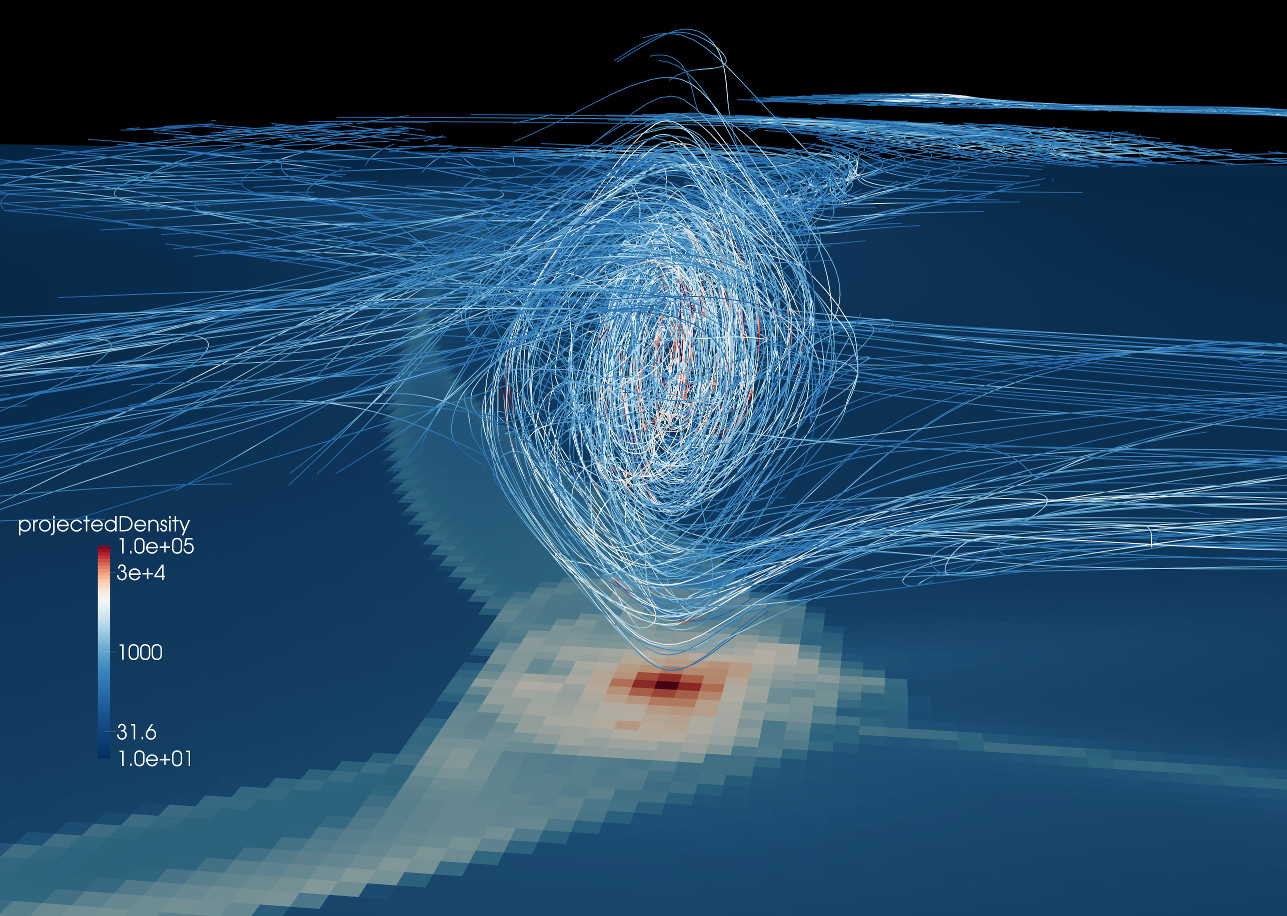}
\end{subfigure}
\hfill%
\end{centering}

\begin{centering}
\hfill%
\begin{subfigure}[b]{0.8\textwidth}
\centering
\includegraphics[width=.99\linewidth]{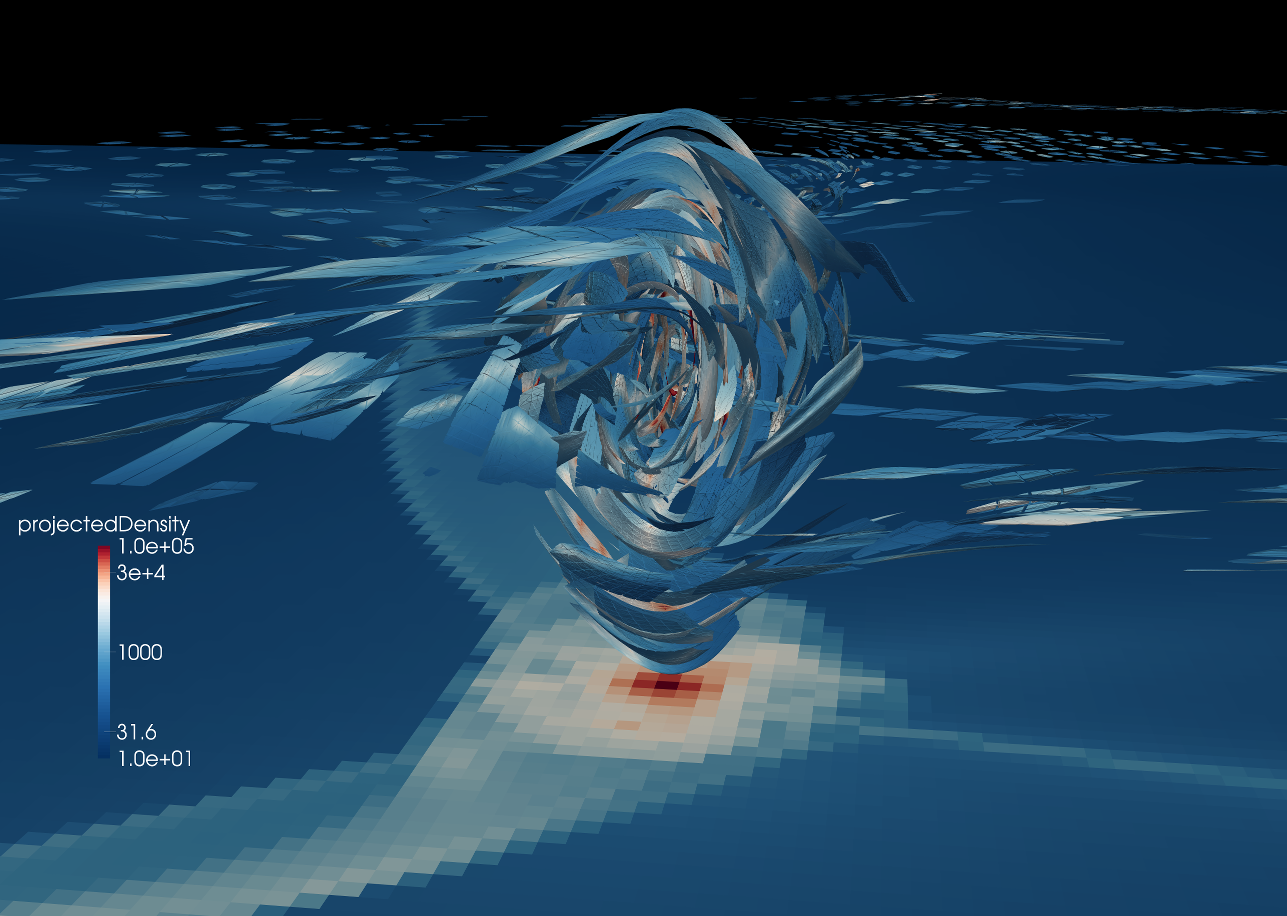}
\end{subfigure}
\hfill%
\end{centering}
\caption{Phase-space structure of a halo at the end of the ${\rm WDM}_{64}$ simulation. The image on the bottom of each picture represents a $1$ voxel thick slice perpendicular to the $z$ axis and going through the center of the halo of the projected density field used in the simulation. The halo itself is represented in the $(x,y,v_x)$ sub-space, as the projection of a wire corresponding to a uniform grid in Lagrangian space (top panel) and Lagrangian subsets initially cubical (bottom panel). In the later case, each subset was originally tessellated into $6$ tetrahedra in the initial conditions, before refinement occurs. In both panels, only elements for which the projected density contrast is higher than $0.3$ are shown. Note also that the simplices in the Lagrangian cubes have been shrunk down to $80\%$ of their actual size for illustration purpose. The color scale on the panels corresponds to the projected density contribution of the phase-space sheet over the wire (upper panel) or the Lagrangian cube (lower panel). It is given in units of $10^{10}M_\odot h^2 {\rm Mpc}^{-3}$, where $M_{\odot}$ is the mass of the sun. \label{fig_halo}}
\end{figure}

\begin{figure}
\begin{centering}
\hfill%
\includegraphics[width=0.9\linewidth]{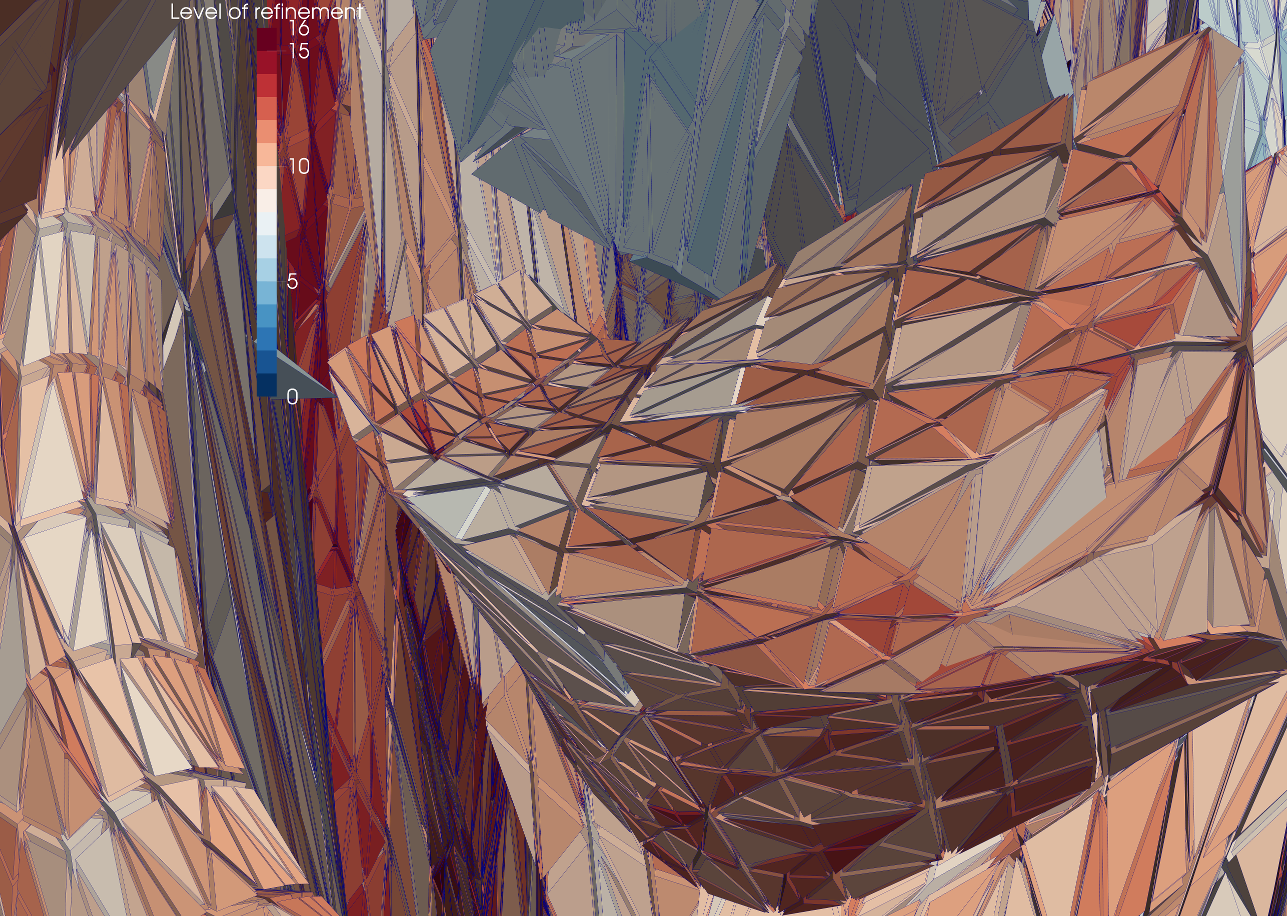}
\hfill%
\end{centering}
\caption{Zoom on a few Lagrangian cubes of the bottom panel of Fig.~\ref{fig_halo}. The color corresponds now to the level of refinement. The maximum level of refinement in the entire simulation is $\ell=22$ but for clarity we display here refinement levels only up to $\ell=16$. Note also that the simplices in the Lagrangian cubes have been shrunk down to $80\%$ of their actual size for illustration purpose.\label{fig_haloSubsetsZoom}}
\end{figure}

Figure \ref{fig_WDMprojZ} shows the projected density over all the simulation box at the end of the ${\rm WDM}_{64}$ simulation, bringing up filamentary structures and the presence of three halos. The structure of one of these halos (the second largest one on the central lower half of figure \ref{fig_WDMprojZ}) is shown in more details on Figs.~\ref{fig_halo}  and \ref{fig_haloSubsetsZoom} where a set of evolved Lagrangian lines and Lagrangian cubes selected from the initial conditions is represented in the $(x,y,v_x)$ sub-phase-space. These figures illustrate again the smoothness of the representation of the phase-space sheet and the good behaviour of refinement. Despite the fact that the halo is only resolved by a dozen of cells in each dimension of the FFT grid, it is quite evolved from the dynamical point of view, with a significant amount of windings of the phase-space sheet. 

\begin{figure}
\begin{centering}
\hfill%
\begin{subfigure}[b]{0.49\textwidth}
\centering
\includegraphics[width=.99\linewidth]{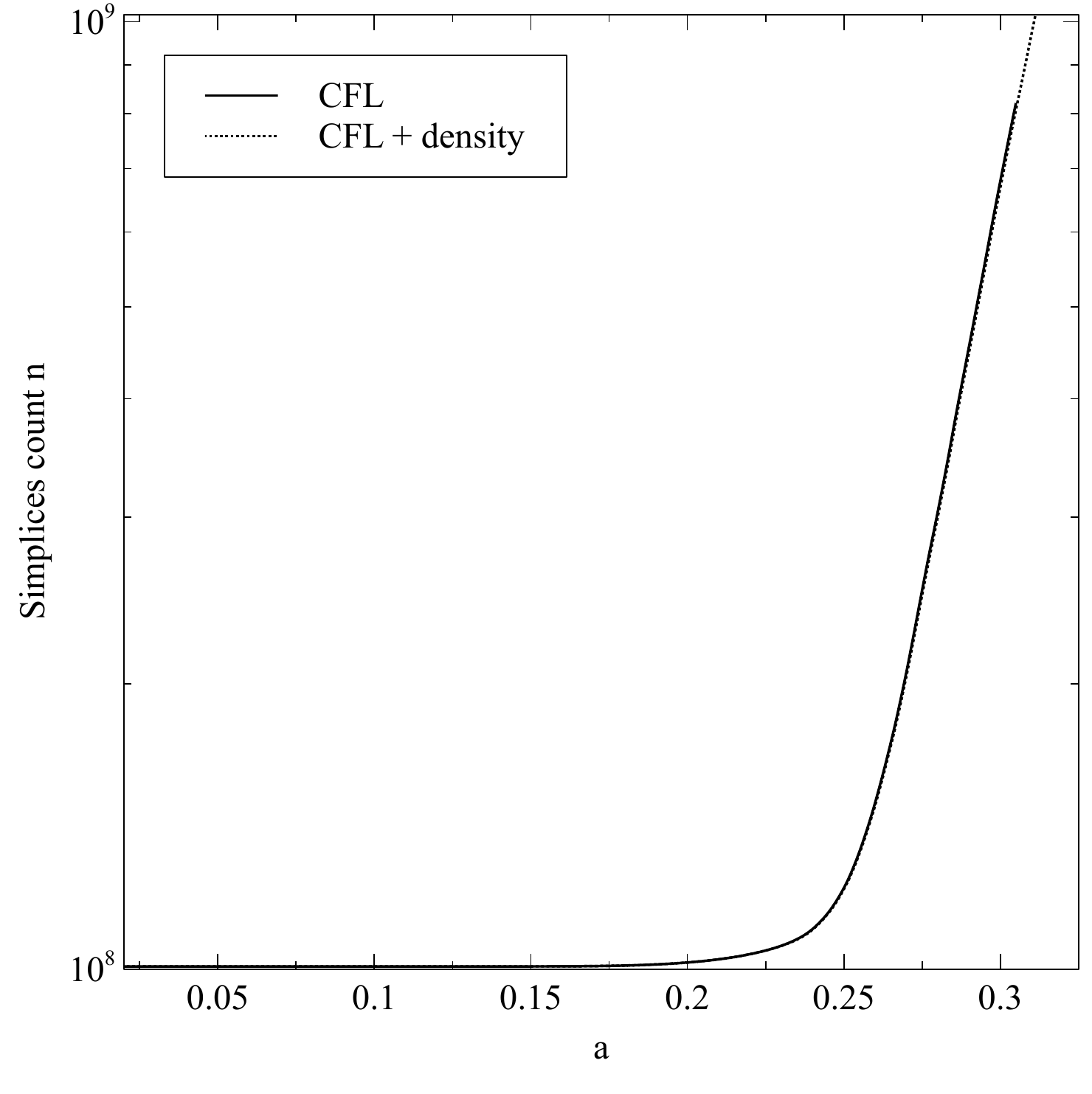}
\end{subfigure}
\hfill%
\begin{subfigure}[b]{0.49\textwidth}
\centering
\includegraphics[width=.99\linewidth]{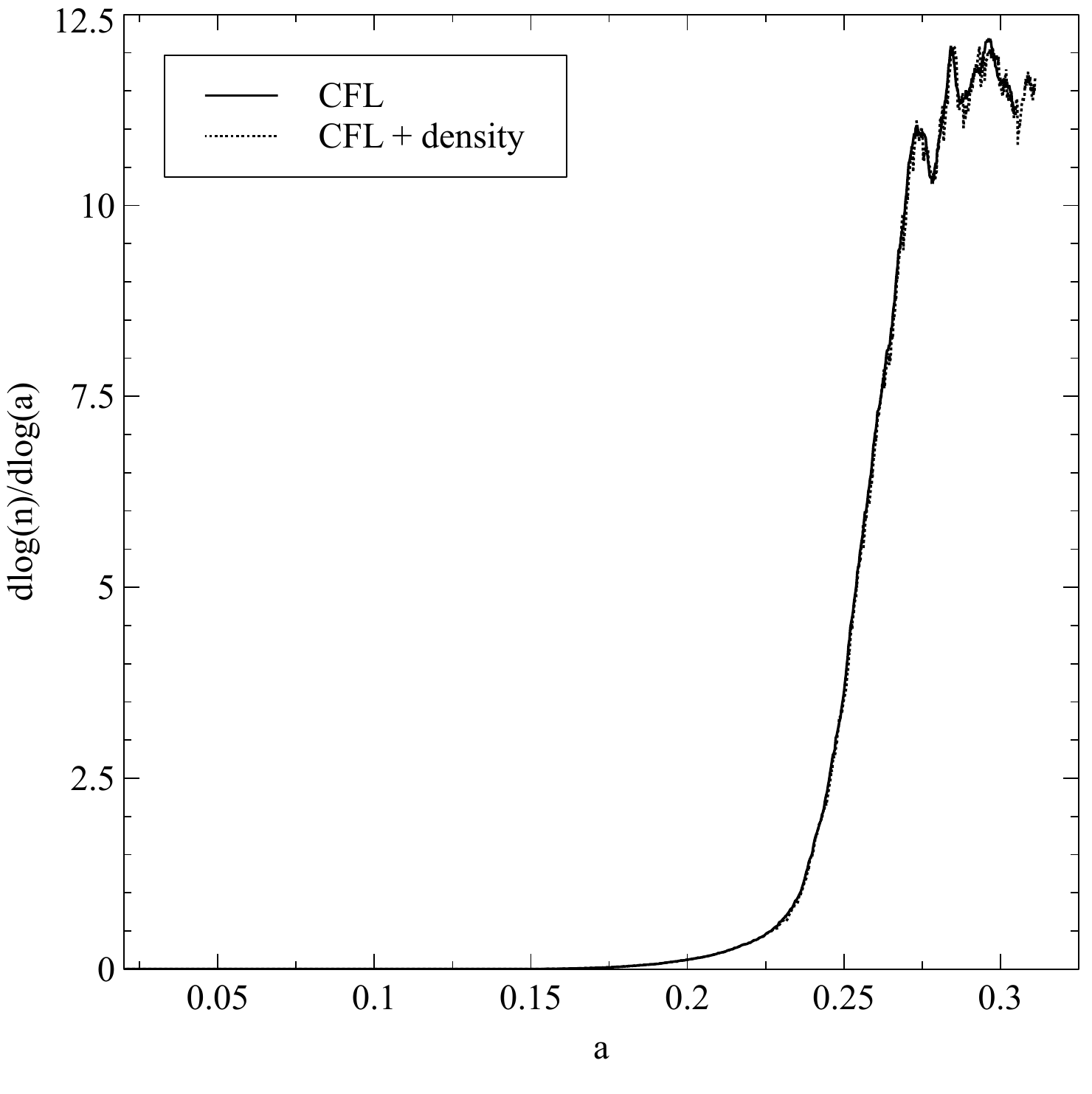}
\end{subfigure}
\hfill%
\end{centering}
\caption{Number of simplices in the mesh (left) as a function of expansion factor and its logarithmic slope (right) for the cosmological WDM simulations.\label{fig_WDMSimplices}}
\end{figure}

In fact, the cost of these windings is rather large in terms of simplices count $n$, as illustrated by the left panel of Fig.~\ref{fig_WDMSimplices}, which displays $n$ as a function of expansion factor. By measuring the logarithmic slope of this curve, at late times (right panel), we find that the simplices count scales roughly as $n \propto a^\alpha$, with $\alpha \simeq 12 \pm 1/2$! Having a power-law simplex count is good news as it means that the system is not chaotic. However, such a quickly increasing count brings out the limits of the tessellation method in the three-dimensional case. For instance if the curves of Fig.~\ref{fig_WDMSimplices} can be extrapolated to later times, just pushing this simulation to a twice large expansion factor of $a=0.62$ would require approximately $2^{12}=4096$ times more simplices, i.e. $4\times 10^{12}$ simplices at the end of the simulation! 
\section{Parallel scaling}
\label{sec:parallelscaling}
\subsection{Multi-threading with OpenMP}
\label{sec_ompScaling}
\begin{figure}
\begin{centering}
\hfill%
\begin{subfigure}[b]{0.49\textwidth}
\centering
\includegraphics[width=.99\linewidth]{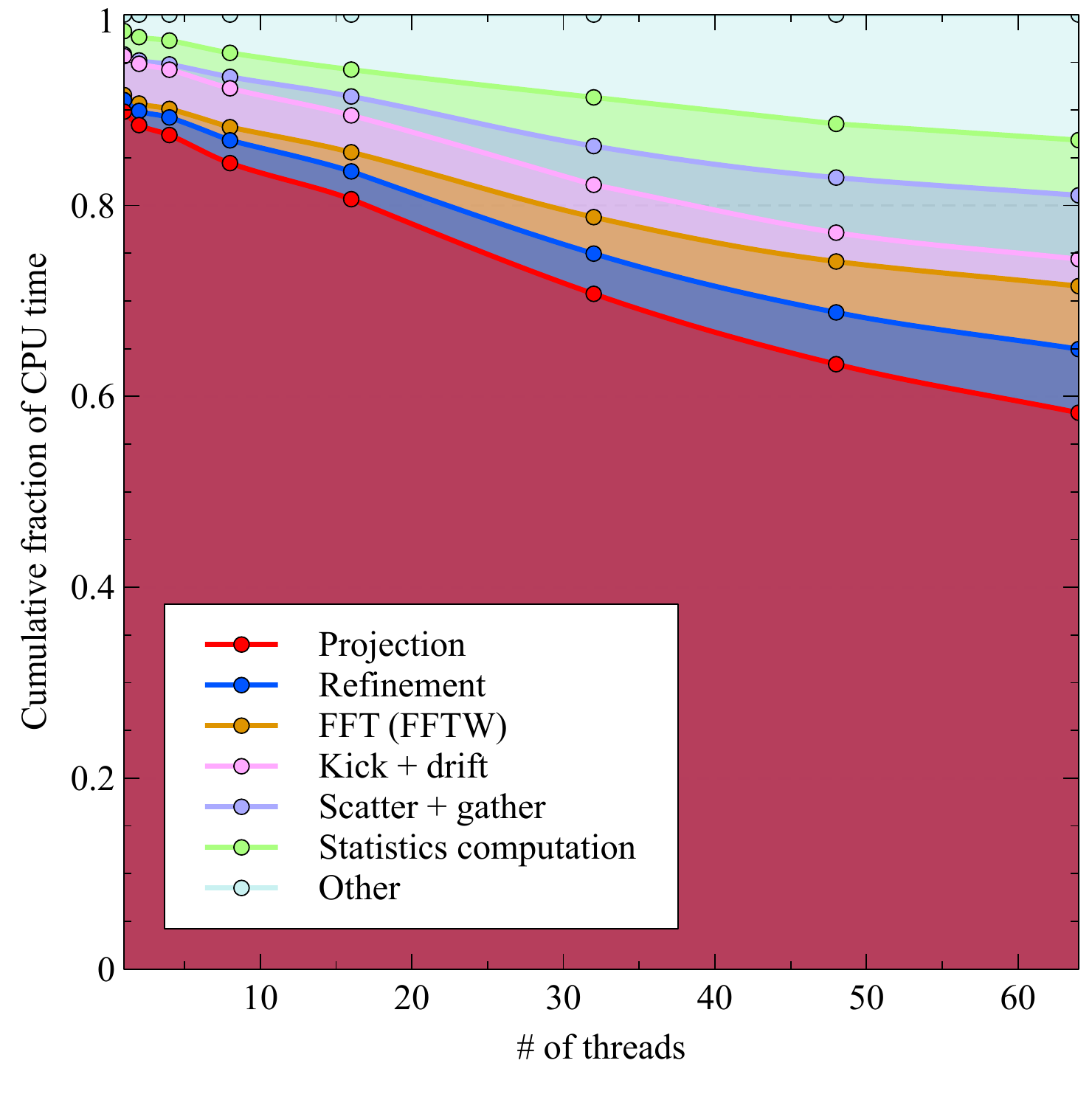}
\end{subfigure}
\hfill%
\begin{subfigure}[b]{0.49\textwidth}
\centering
\includegraphics[width=.99\linewidth]{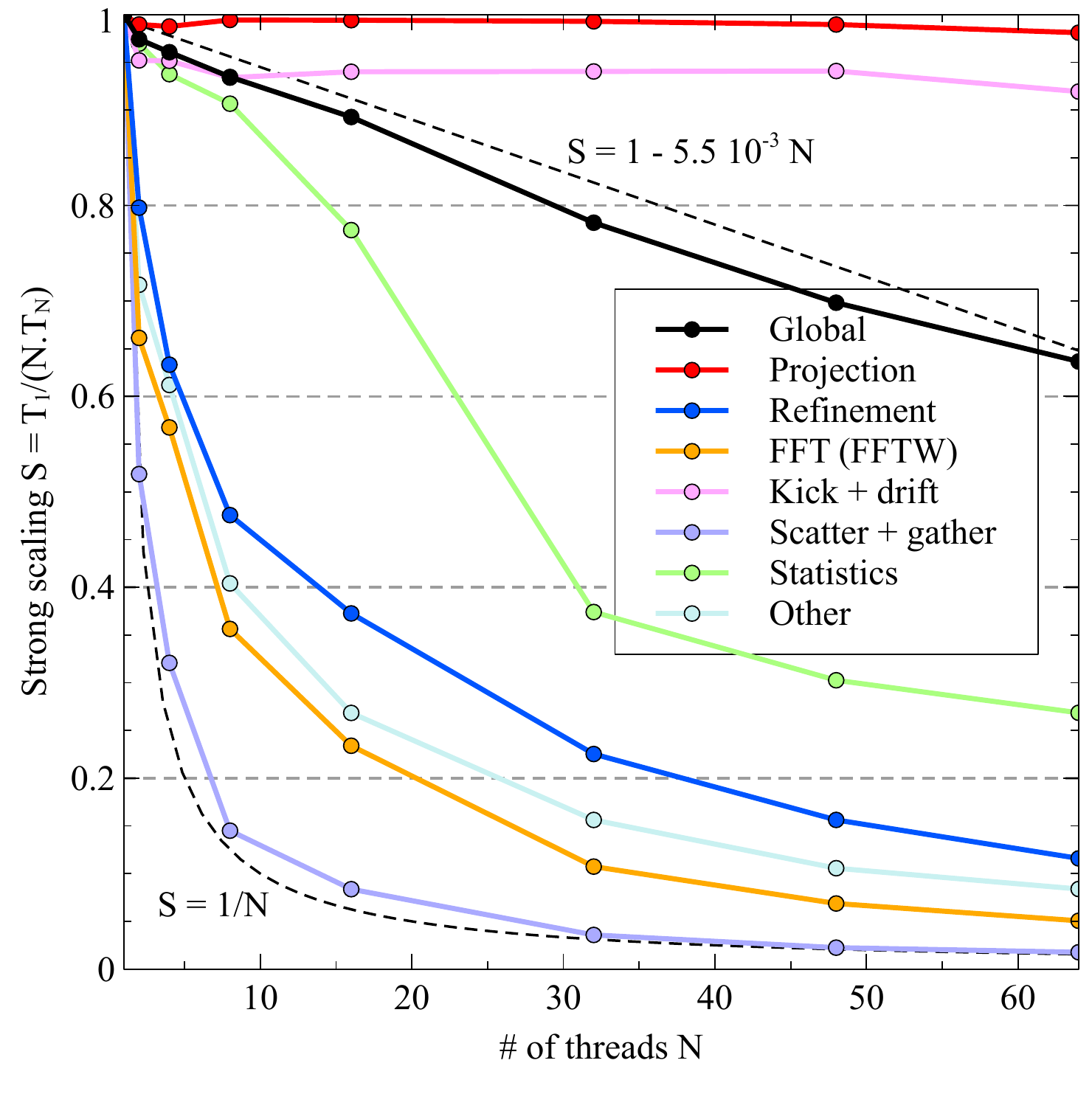}
\end{subfigure}
\hfill%
\end{centering}
\caption{Scaling of the last time step of the WDM$_{\rm OMP}$ simulation as a function of the number of threads. The cumulative fraction of the total time step used for the main tasks is displayed on the left panel while the strong scaling of these tasks is shown on the right panel. The duration of the time step was $17471$ seconds  in the single threaded case and $428$ seconds in the $64$ threads case. The different tasks represented are described on each panel. They correspond to the projection of the phase-space sheet tessellation onto the AMR grid, the refinement of $~0.1\%$ of the simplices, the resolution of Poisson equation in Fourier space using {\em in-place} FFT, the kick and drifts on the vertices/tracers, the scattering and gathering of the projected density and potential before and after solving Poisson equation, the calculations of statistics such as surface and energy and the rest of the tasks, that include building the AMR grid, computing the projected density associated to each vertex of the sheet, and diverse management related operations. A more precise account of each task is given in section \ref{sec:vlapoi}.\label{fig_scalingOMP}}
\end{figure}

In order to assess the pure OpenMP multi-threading performances of the code, we performed a ${\rm WDM}_{\rm OMP}$ simulation with the same parameters as ${\rm WDM}_{\rm CFL64}$, but using a single MPI process and $64$ OpenMP threads. The simulation was ran over the $64$ cores of a dedicated shared memory computer featuring $8$ Intel Xeon E7-8837 $2.67$ GHz octo-core processors with 24 Mb of L3 cache memory and it took $28$ hours to reach time step $300$. We then restarted the simulation from step $300$ and measured the timings of the different sub-tasks during a single time-step using a number of threads ranging from $1$ to $64$. The results are reported on figure \ref{fig_scalingOMP}. 

On the left panel, the duration of a time-step appears to be dominated by the phase-space sheet projection, which occupies from up to $90\%$ of the time-step in the single threaded case down to $60\%$ in the $64$ threads case. This decrease is explained by the near-perfect scaling of the projection algorithm (see section \ref{sec_exactProj}) compared to most other tasks, as can be seen on the right panel. In this particular example, projection was performed using double-double precision floating point numbers everywhere\footnote{i.e. using a combination of two double precision numbers to achieve $S=104$ bits precision arithmetic with the help of the {\tt libQD} library.} with an average rate of $6560$ tetrahedra treated per second and per thread for a total of about $10^8$ tetrahedra and a $1024^3$ grid to compute the gravitational potential.

The calculation of the force from the potential, the kick and the drift operations also scale very well with the number of threads, but as previously noted, this is not the case for all sub-tasks. In the worse cases, the efficiency on $64$ threads ranges from $12\%$ for refinement down to $2\%$ for the scatter and gather operation (see section \ref{sec:projdenscal}), making this last task roughly as slow in the multi-threaded case as in the single-threaded one. It can also be noted that the FFT operation using the {\tt FFTW} library scales rather poorly, which may be due to the fact that we are using cache inefficient in-place transforms. 

These issues do not however seem to be critical due to the rather small fraction of time spent in these poorly scaling sub-tasks. This is clear when examining the global strong scaling black curve on the right panel of figure \ref{fig_scalingOMP}, which slowly decreases linearly with the number of threads, reaching a worst efficiency of roughly $64\%$ for $64$ threads. We therefore reserve further optimisation for the future, noticing that the refinement task should probably be our main priority. Indeed, its cost and its impact on scaling is expected to increase quickly with time as suggested by the vertice count measurements of section \ref{sec:exemple},  while the contribution of other tasks with poor scaling, proportional to the FFT grid size, should remain constant. 

\subsection{Distributed processing with MPI}
\label{sec_mpiScaling}
We now proceed to measuring distributed processing scaling performances, when more than one MPI process is used. To this end, we take advantage of the ${\rm WDM}_{\rm CFL64}$ and ${\rm WDM}_{\rm 64}$ simulations introduced in section \ref{sec_WDM} as well as a ${\rm WDM}_{\rm CFL8}$ simulation identical to ${\rm WDM}_{\rm CFL64}$, except that only $8$ MPI processes were used instead of $64$ in the original run. All these simulations were performed on the Occigen supercomputer of CINES, using $4$ and $32$ compute nodes respectively, each node featuring $128$ GB shared memory and $2$ Intel Xeon E5-2690v3 $12$-core processors running at $2.67$ GHz. For all runs, each MPI process was bound to a single processor, taking advantage of the $12$ cores via $12$ OpenMP threads. The properties of the simulations are summarised in table \ref{tab_mpi}.

\begin{table}[!htbp]
\centering
\begin{tabular}{|c|c|c|c|c|c|}
\hline
 Name &  MPI processes &  OpenMP threads &  Cores &  CPU time (h)& Wallclock time (h)\\
\hline
${\rm WDM}_{\rm CFL8}$  & 8 & 12 & 96 & 2208 & 23 \\
${\rm WDM}_{\rm CFL64}$ & 64 &12  &768  & 8448 & 11 \\
${\rm WDM}_{\rm 64}$ &  64  & 12 & 768 & 20736  & 27 \\
\hline
\end{tabular}
\caption{Characteristics of the MPI simulations.\label{tab_mpi}}
\end{table}

\begin{figure}
\begin{centering}
\hfill%
\includegraphics[width=0.5\linewidth]{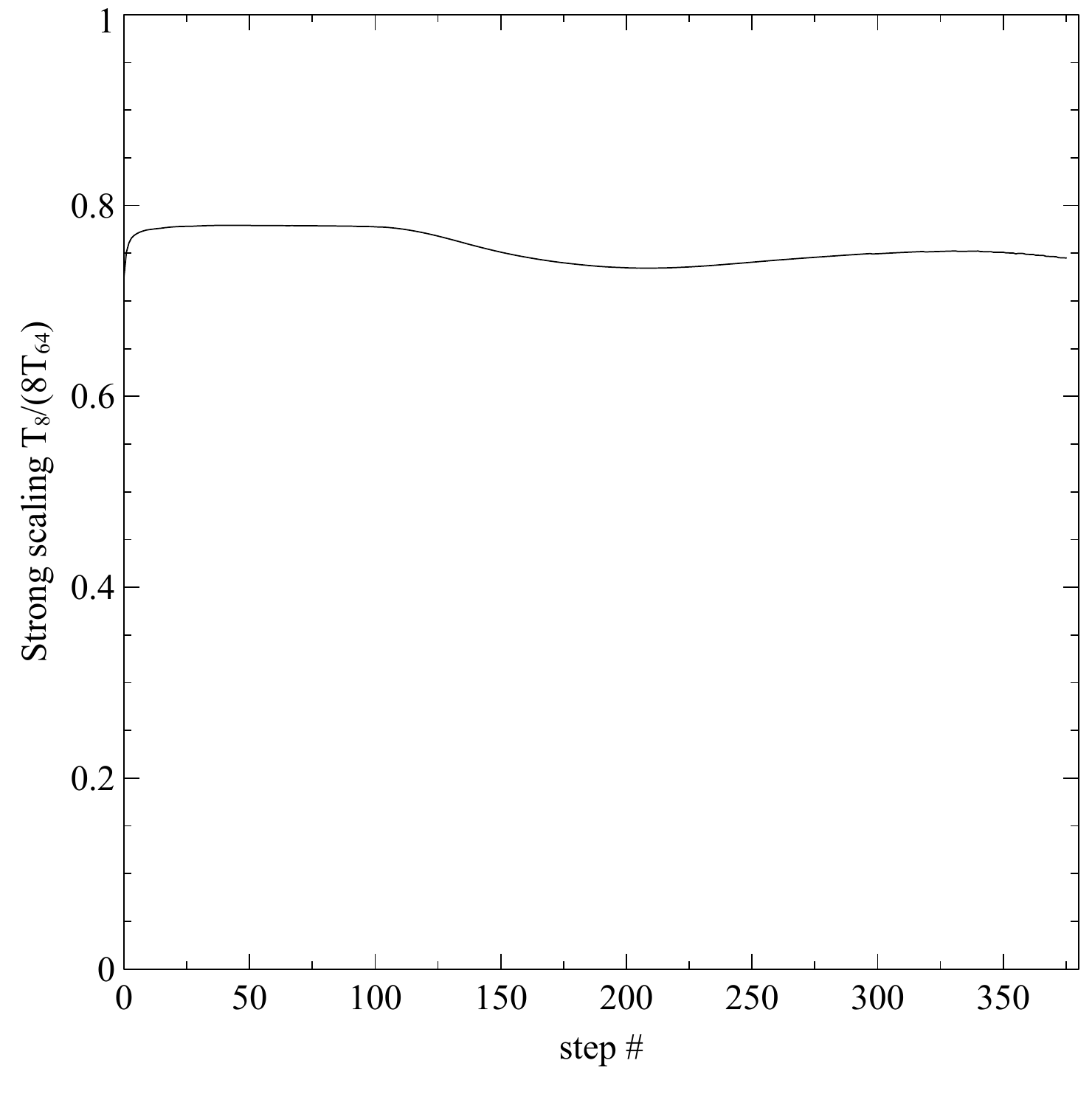}
\hfill%
\end{centering}
\caption{Strong scaling of the WDM simulation as a function of time step index for MPI tasks. The scaling is measured as ${T_8}/{8T_{64}}$, with $T_{8}$ and $T_{64}$ representing the cumulative wallclock time spent to reach a given time in the ${\rm WDM}_{\rm CFL8}$ and ${\rm WDM}_{\rm CFL64}$ simulations using $8$ MPI tasks ($96$ cores) and $64$ MPI tasks ($768$ cores) respectively.\label{fig_scalingMPI_strong}}
\end{figure}

We start by noting that the ${\rm WDM}_{\rm CFL8}$ simulation reached time step $300$ in $16$ hours and $45$ minutes while the similar ${\rm WDM}_{\rm OMP}$ simulation introduced in section \ref{sec_ompScaling} took $28$ hours, making the MPI run roughly $10\%$ faster per core than the pure OpenMP one. Such a comparison is however very limited due to the different nature of the machines used for both runs. A more accurate account of the strong scaling properties is given by the comparison of the ${\rm WDM}_{\rm CFL8}$ and ${\rm WDM}_{\rm CFL64}$ runs as a function of the time step index over the $375$ time steps of ${\rm WDM}_{\rm CFL8}$, as shown on figure \ref{fig_scalingMPI_strong}. Inspection of this figure indicates that the strong scaling efficiency oscillates around $75\%$ for the whole duration of the run when the number of MPI processes is multiplied by $8$, therefore exhibiting reasonably good and stable scaling properties. This good but imperfect scaling can be mostly explained by the fact that the resolution of the AMR grids used for the projection of the phase-space sheet differs between the two runs, as mentioned in section \ref{sec:paralfrank}. Indeed, the refinement level of the AMR grid in ${\rm WDM}_{\rm CFL8}$ and ${\rm WDM}_{\rm CFL64}$ is allowed to range from level $8$ to level $10$, but resolution must be fixed at maximum level $10$ on the processes boundaries. As a result, projection is slower in ${\rm WDM}_{\rm CFL64}$ as it features more boundaries, and measurements show that this constraint alone accounts for between $50\%$ and $75\%$ of the lack of scaling, the rest being due to the increased computations and communication requirements stemming from the larger number of processes, in particular for the scatter and gather operations (see section \ref{sec:projdenscal}) as well as for the simplices refinement (see section \ref{sec:aniso}).

\begin{figure}
\begin{centering}
\hfill%
\begin{subfigure}[b]{0.49\textwidth}
\centering
\includegraphics[width=.99\linewidth]{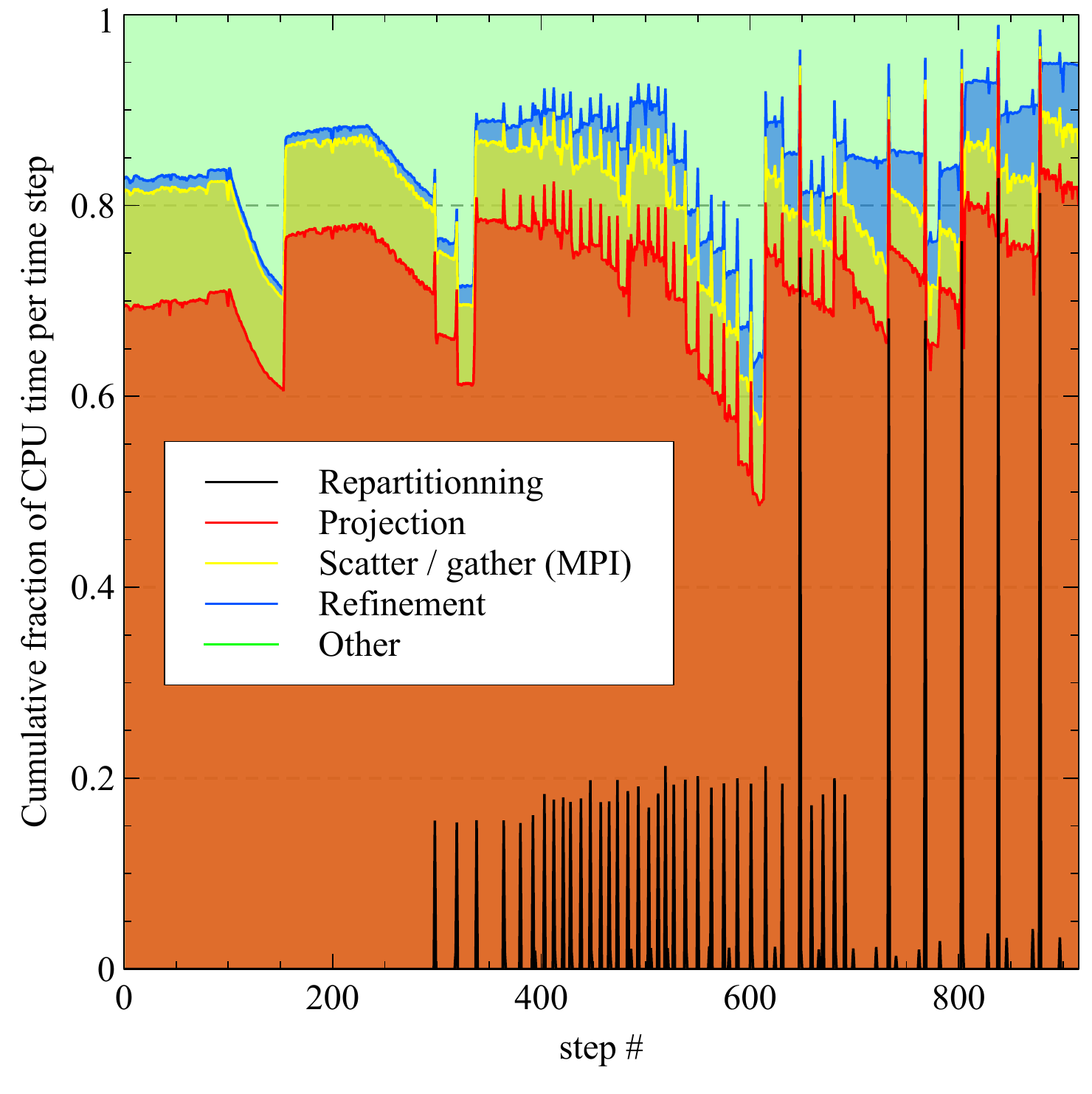}
\end{subfigure}
\hfill%
\begin{subfigure}[b]{0.49\textwidth}
\centering
\includegraphics[width=.99\linewidth]{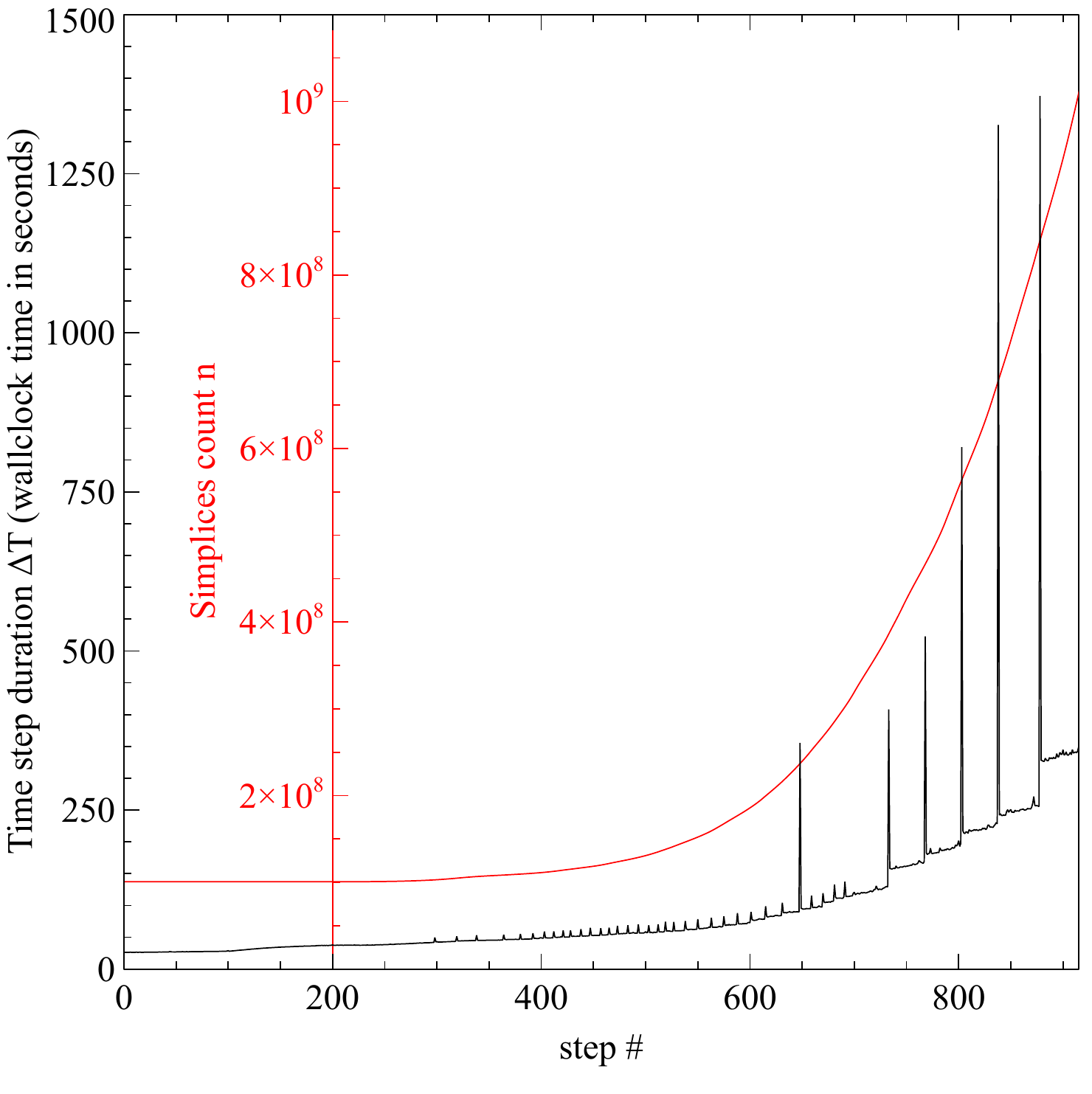}
\end{subfigure}
\hfill%
\end{centering}
\caption{Cumulative fraction of CPU time taken by specific sub-tasks as a function of time step index in ${\rm WDM}_{64}$ (left panel) and total wallclock duration of individual time steps (right panel). The simplices count was over-plotted in red on the right panel to emphasise the fact that time step duration is growing slower than the simplices count as the simulation advances. The spikes on this figure correspond to time steps where re-partitioning was triggered. The different tasks represented on the left panel correspond,  in reverse order, to re-partitioning of the tessellation over the MPI processes when load imbalance becomes too high, projection of the phase-space sheet tessellation onto the AMR grid, scattering and gathering of the projected density and potential before and after solving Poisson equation, refinement of the phase-space sheet tessellation and the remaining tasks. Note that the imbalance threshold that triggers re-partitioning was changed from $15\%$ to $50\%$ around time step $700$.\label{fig_scalingMPI_fraction}}
\end{figure}

The cumulative fraction of time spent in each sub-task of individual time steps in ${\rm WDM}_{64}$ is shown as a function of time step index on the left panel of figure \ref{fig_scalingMPI_fraction}. Compared to the pure OpenMP case (left panel of figure \ref{fig_scalingOMP}), one can see that, as expected, the scatter and gather operations take more time in the MPI case due to inter node communications. The duration of these operations is however still small compared to the time needed for projection and one can actually see that it is significantly shrinking during the course of dynamics in comparison to the duration of projection and refinement. The increasing fraction of CPU time spent refining can be easily understood by looking at figure \ref{fig_WDMSimplices}, which illustrates the power law growth of the simplices count as the simulation advances in time. Refinement, however, is not uniform and the rapid increase in the number of simplices is not balanced among the MPI processes. Load balancing is therefore often required once refinement has started. In the ideal case, it is achieved by exchanging simplices among processes treating neighbouring patches of the phase-space sheet, so that the simplices distribution among processes can be made uniform with relatively low communication requirements. This is however not always possible, as illustrated by the black curve on the left panel of figure \ref{fig_scalingMPI_fraction}, showing the fraction of time spent re-partitioning. Re-partitioning first happens at around time step $300$ and is required more and more frequently as the rate of refinement increases. Although it seems to take only a small fraction of time up to step $600$, important spikes appear at later times. These spikes correspond to complete re-orderings of the partitions, triggered whenever it is not possible to balance efficiently by direct exchange among neighbours. As shown on the right panel of figure \ref{fig_scalingMPI_fraction}, these complete re-orderings can take a significant amount of time and the duration of time steps for which they occur may be multiplied by a factor of $3$ or $4$.

\begin{figure}
\begin{centering}
\hfill%
\begin{subfigure}[b]{0.49\textwidth}
\centering
\includegraphics[width=.99\linewidth]{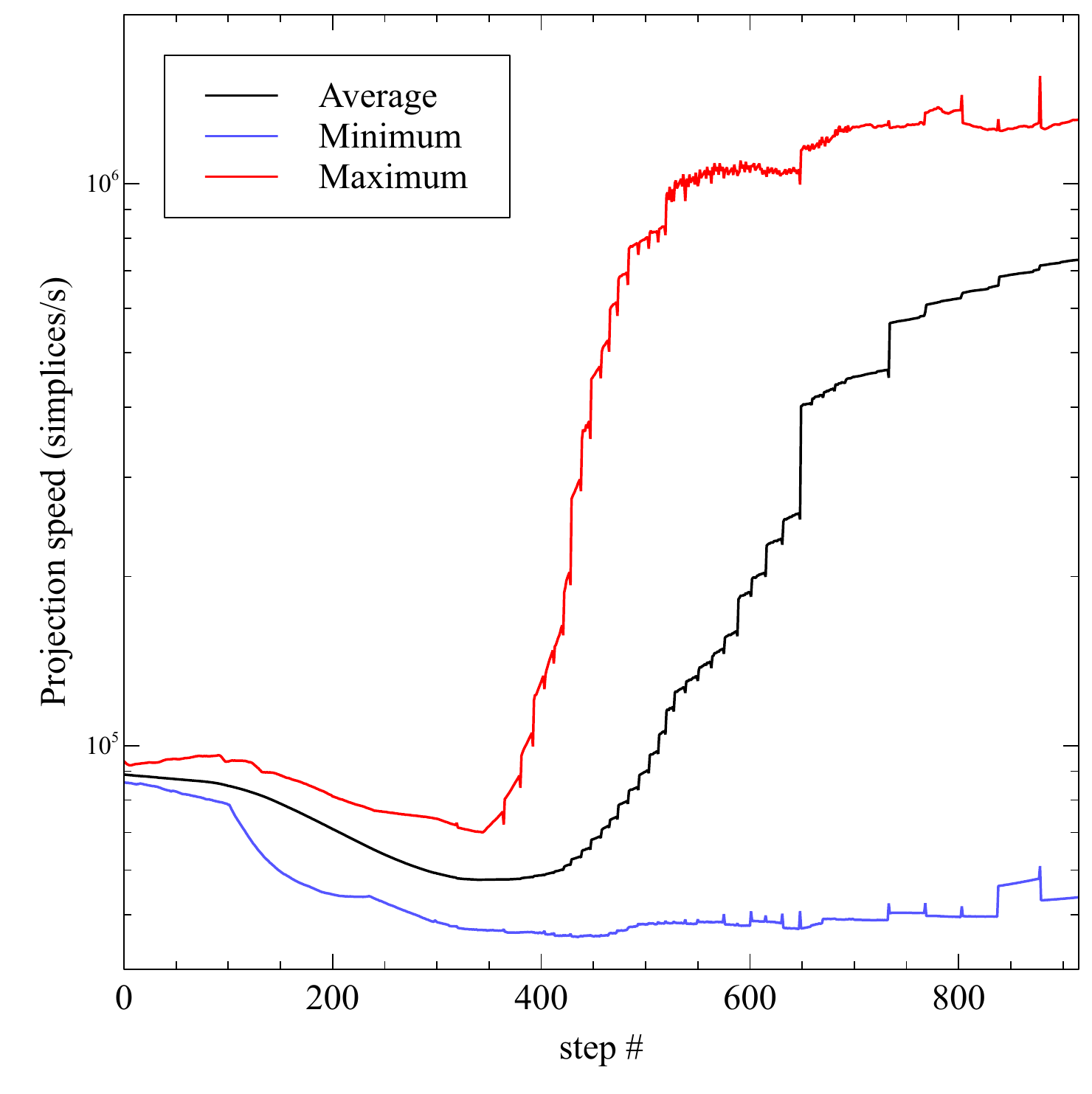}
\end{subfigure}
\hfill%
\end{centering}
\caption{Average phase-space sheet projection speed in simplices per second as a function of the time step index for a single MPI process. The speed of the fastest and slowest process is shown in red and blue respectively while the average speed is shown in black. When the simplices are well balanced over MPI processes, the global duration of projection is that of the slowest process due to the necessity of synchronising. Note that $12$ OpenMP threads were used for each MPI process.\label{fig_scalingMPI_timeStep}}
\end{figure}

Comparing the growth in simplices count (red curve on right panel of Fig.~\ref{fig_scalingMPI_fraction}) to the increase in time step duration, one can see that the former is growing faster than the later on average as the simulation advances. This is because the speed of the projection is roughly proportional to the average number of intersections simplices have with the AMR grid voxels, and in extreme cases where simplices become much smaller than AMR grid voxels, the speed is actually drastically increased since projecting such simplices simply amounts to adding their mass to the AMR voxel they fall in. As the simulation evolves, refinement is mainly triggered in over-dense regions of the simulation, where simplices tend to become smaller and smaller, while their size remains the same or even grows slightly in under-dense regions. As a result, the projection speed largely increases in dense and highly refined regions together with the number of local simplices. As long as the simplices distribution remains imbalanced, the larger number of local simplices somewhat compensates for the increase in speed and the time step duration only slowly increases. But whenever re-partitioning is triggered, the local number of simplices on each process becomes equal and the time step duration suddenly jumps as the total speed of projection is equal to that of its slowest process. This explanation is clearly confirmed by inspection of figure \ref{fig_scalingMPI_timeStep}, where the projection speed of the slowest and fastest MPI process is shown together with the average speed: as the simulation evolves and refinement is required more and more often, the projection speed of the slowest and fastest MPI process tend to largely diverge from each other. A relatively simple way of dampening the impact of re-partitioning on performances consists in loosening the criterion for triggering re-partitioning, which in our case was actually done around time step $700$, where the maximum imbalance threshold was increased from  $15\%$ up to $50\%$.

As previously mentioned, the duration of the global projection is equal to that of the slowest MPI process. A closer examination of the blue curve on figure \ref{fig_scalingMPI_timeStep} is therefore reassuring since it clearly shows a stable or slowly increasing projection speed as a function of time. As a result, the total wall-clock duration of a given run should be bounded from below and worst case scenario predictions for the duration of a given run are possible. But a comparison to the black curve on the same figure, showing the average projection speed, also indicates that performances could be significantly improved. Indeed, the average speed corresponds to the theoretically optimal speed one could reach if the simplices distribution was perfectly homogeneous and the projection speed was equal among all processes. The fact that it is about an order of magnitude higher than that of the slowest process therefore indicates that there is opportunity for significant optimisation, which could be achieved by modifying the re-partitioning procedure to take into account the projection speed of each individual simplex. This is however a relatively complex problem and we leave such optimisation for future work.

\section{Discussion: possible improvements of the code}
\label{sec:conclusion}
The applications studied above show that {\tt ColDICE} is able to follow very accurately the evolution of the phase-space sheet. Its design allows one to run large configurations in parallel on supercomputers, but there is still room for improvement and we conclude here by listing several modifications that can come to mind. 
\begin{itemize}
 \item{\em Poisson equation: adaptive mesh refinement.} In the current implementation, Poisson equation is solved using FFT method on a fixed resolution grid, which is suboptimal in the gravitational case, because of the very high dynamical contrasts the system builds during the course of the dynamics. This is illustrated by Fig.~\ref{fig_WDMprojZ}, where only very small regions of space experience highly nonlinear evolution. A more optimal approach would thus consist in using an adaptive grid to compute the force and employ for instance relaxation methods to solve Poisson equation, as in AMR particle codes \citep[see, e.g.][]{Teyssier2002}. This would represent a very natural --albeit non trivial from the algorithmic point of view--  extension of our code since an AMR grid structure is already implemented on a per MPI process basis. However, two major issues remain. The first one is that having a better local resolution implies having a smaller global time step. In standard AMR $N$-body codes, this is dealt with by having the time step depend on the level of refinement, which helps reducing computer cost. In our code however, this would be problematic as the phase-space sheet would become discontinuous except at times when all the levels of the AMR grid are synchrone. The second issue is that symplecticity is necessarily broken when using adaptive mesh refinement as the effective softening of the potential varies spatially and with time. However, despite these potential problems, adaptive mesh refinement is worth testing because it probably corresponds to the most natural extension of our code.
\item{\em Poisson equation: treecode technique.} An alternative approach would simply consist in forgetting the grid approach and instead solve Poisson equation directly from the tessellation, by combining exact calculation of the force from individual simplices close to the vertex/tracer position under treatment \citep[see, e.g.][and references therein]{Waldvogel1979} and standard treecode techniques for groups of remote simplices. Using such a method, it would however be technically difficult to warrant a sufficiently smooth force field and smoothness of the force is mandatory to avoid inducing catastrophic irregular patterns over the tessellation on the long term. 
\item{\em Hybrid approach.} Because of mixing, the tessellation needs to be constantly enriched with additional vertices. This can become prohibitive and could be a real issue in the three-dimensional case, where we found that the number of vertices was increasing like the twelfth power of the expansion factor in our cosmological WDM simulation. An alternative would be to find a way to coarse grain the numerical solution. The simplest idea is probably to go back to the $N$-body approach when refinement exceeds some threshold, i.e. by replacing simplices with an ensemble of particles when refinement level exceeds some threshold, similarly to the technique employed by \cite{Hahn15}. But doing so would unfortunately reintroduce the defects of the $N$-body approach we aim to avoid. An alternative that is more in the spirit of the Vlasov approach, but also much more complex to implement, would be to create a six-dimensional domain bounded by a five dimensional waterbag inside which a semi-Lagrangian approach would be employed to evolve a coarse grained version of the phase-space distribution function. This follows naturally from the fact that in strongly mixed regions, the phase-space sheet tends to fully fill-up the local six-dimensional volume at the coarse level.  
\item{\em Load balancing}. In section \ref{sec_mpiScaling}, we noticed that projection of the phase-space sheet onto the FFT grid takes much more time for large simplices than for small ones, which makes optimisation of load balancing both in terms of memory and CPU impossible in the general case in practice. This opens the window for improvements of the code, consisting for instance in providing to each MPI process several carefully chosen patches of the phase-space grid so that the overall memory and computational cost per MPI process would be approximately the same. To do so, it is necessary to perform a domain decomposition of the phase-space sheet on smaller patches containing approximately the same number of simplices; then compute an appropriate CPU global cumulative cost function associated to each of these patches as the sum of an elementary cost function computed on a per-simplex basis; finally repartition the patches on each MPI process so that the total CPU cost per process is as uniform as possible. The cost function remains to be found by combining analytical and experimental means, which should not be very difficult. Note however that performing a domain decomposition with smaller patches will decrease the strong scaling efficiency because of boundary effects.
\end{itemize}

\section*{Acknowledgements}
We thank J. Touma for suggesting using Poincar\'e invariants to constrain refinement and J. Barnes for mentioning the test with the chaotic potential.  We also thank T. Abel, R. Angulo, A. Bak, N. Besse, O. Hahn, S. Shandarin and Y. Suto for fruitful discussions. TS acknowledges the hospitality of Y. Suto and the University of Tokyo and financial support by the Research Center for the Early Universe (RESCEU). This work has been funded in part by ANR grant ANR-13-MONU-0003. It has made use of the HPC resources of CINES (Occigen supercomputer) under the GENCI allocation 2015047012. Three-dimensional data visualisation was performed with {\tt ParaView} \cite{Ahrens2005,Ayachit2015} (\url{www.paraview.org}).
\appendix
\section{Data structure for storing adaptive mesh elements}
\label{app_mesh}
\begin{figure}
\centering
\includegraphics[width=0.95\linewidth]{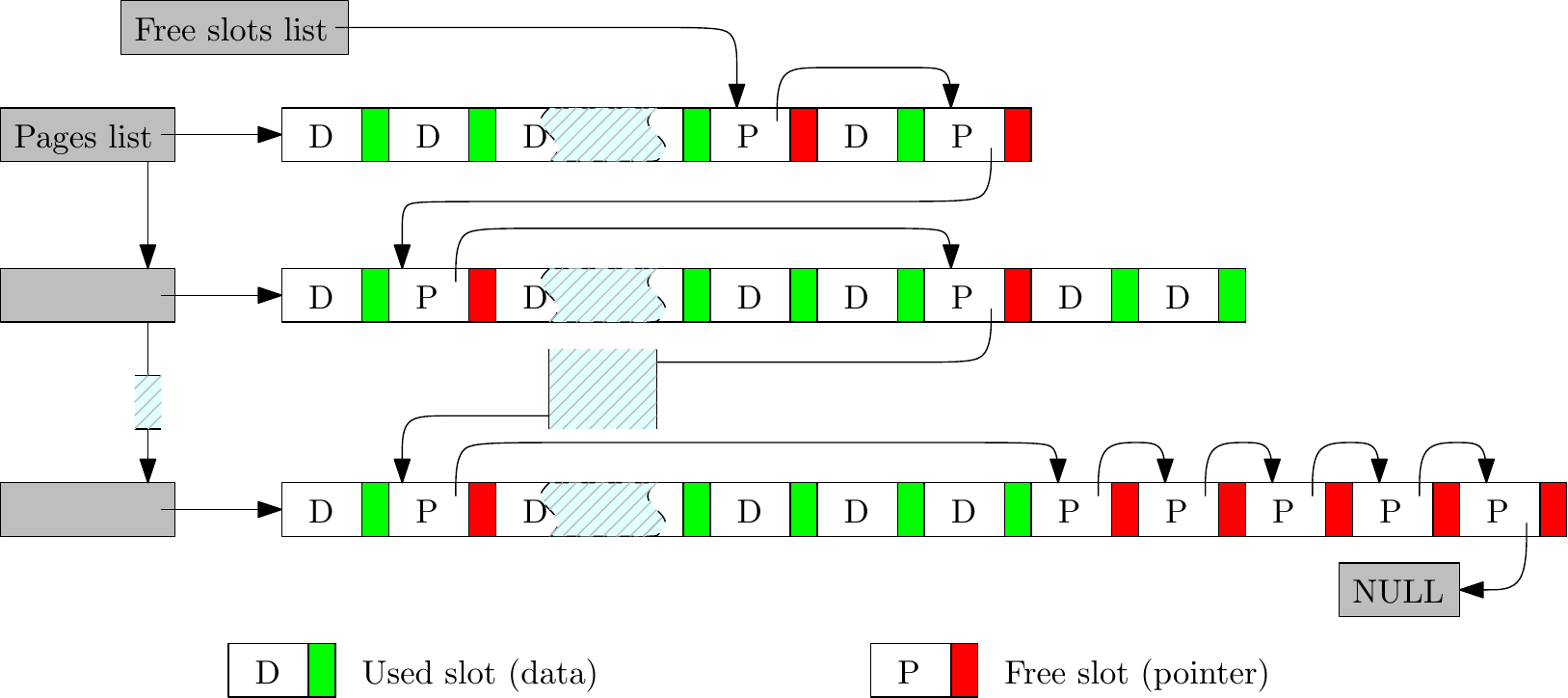}
\caption{Representation of the data structure of the memory pool used to manage simplices and vertices of the adaptive mesh presented in section \ref{sec_mesh}. A trailing state bit (represented as a green or red rectangle) is added to the data structure of the elements to store in order to identify used and free memory slots. Memory pages are allocated on demand and stored as lists of increasing size arrays while a free slots linked list is implemented using directly the memory available in the free slots. Whenever an element is destroyed, its state bit is updated and it is added to the free slots list. New elements can then be assigned to free slots and old elements cleared from it in constant time using the free slots list. It is also easy to iterate through the elements by skipping them according to the value of their state bit. Note that elements whose data structure requires padding will not grow in size due to the addition of the state bit.\label{fig_memoryPool}}
\end{figure}

An adaptive mesh is a dynamically evolving data structure composed of elements of different types (such as vertices, simplices, ...). Using the right design to store and manage these elements is a critical issue in terms of performances and flexibility. In our implementation, we use a dedicated memory pool structure to store each type of element, each element being internally stored together with a trailing state bit\footnote{This is easily implemented in C++ using inheritance, allowing the addition of that state bit in a way completely transparent to the user.} that allows for fast insertion and removal, as illustrated on figure \ref{fig_memoryPool}. The state bit is used to mark pre-allocated memory slots that are allocated but currently not in use. The pool itself is composed of a linked list of memory pages used to store data. Initially, a single page is allocated such that it can store a fixed number of elements and the state bit of each slot it contains is set to indicate that they are currently available. The linked list of memory pages (``Pages list'' on the diagram) is then initialised in such a way that the previously allocated page is its unique element. Now, because allocated slots of the first memory page are not yet in use, it is possible to use their dedicated memory space for other purposes. Interpreting the first 64 bits of each slot as a pointer, we set each of them to the address of the next free slot in the page. The last free slot is set to the null pointer and a ``free slots'' pointer is initialised to the start of the page. In practice, the ``free slots'' pointer therefore acts as the head of a linked list that does not store anything, but whose nodes are all stored in memory at the address of an available slot.

Once such a structure is in place, a new element can be easily inserted in the pool at the slot pointed to by the ``free slots'' pointer by setting its state bit to ``used'' and updating the ``free slot'' pointer to the next element in the ``free slot'' list (i.e. the element it points to). This process will work until their is no more slot available in the first memory page. At this point, a new page can be allocated, initialised as described above, and stored as the next element of the ``memory pages'' list. In practice, we set the size of this new array to a fraction\footnote{A fraction of $100\%$ is used in our implementation, which amounts to doubling the pool size every times it saturates.} of the total number of elements in the pool so that the number of required allocations is only proportional to the log of the pool's size. With this structure, removing an element is also a simple constant time operation that consists in setting the bit flag of the memory slot that stores it to ``free'', setting its first 64 bits to the null pointer, and making the last element of the ``free slots'' list point it. Thanks to the presence of the state bit, it is also easy iterate in a quick and flexible way over the elements stored in the pool as it suffice to scan all slots in the memory pages while skipping the ones tagged as ``free''. Multi-threaded iteration is also easily achieved by assigning equally sized subsets of the ``used'' slots to each thread.

In practice, the way these subsets are created is very important as it may have a great influence on the processor cache hit to miss ratio of the algorithms that use multi-threaded iteration. In our adaptive mesh implementation (section \ref{sec_mesh}), processor cache usage is improved by sorting vertices and simplices along a Peano-Hilbert curve (PH hereafter). This allows for a significant acceleration of most algorithms and a gain of $\approx 25\%$ in speed can be observed on average. Such an optimisation is even more profitable on multi-core architectures as the size of the cache available per thread shrinks and memory bandwidth becomes a critical limiting factor for good scaling. Newly inserted elements cannot however be PH sorted on the fly and only part of the elements stored in the pool benefit from the ``fast'' access while others remain slower on average. It is therefore important to distribute elements subsets to the different threads in a smart way when iterating so that the ratio of fast to slow access elements is well balanced among the threads. We achieve this by keeping track of how many elements are sorted and dividing the pool into $N$ subsets of contiguous elements from the ``sorted'' fast part and $N$ from the ``unsorted'' slow part, where $N$ is the number of threads used for iteration. Each thread is then assigned one of the sorted and one of the unsorted subset so that all of them exhibit comparable performances on average.

\section{Multi-threaded distributed mesh refinement}
\label{app_meshRefine}
In this appendix, we give a detailed version of the algorithm we use for the distributed mesh refinement introduced in section \ref{sec_mesh}. Note that the ``cache'' variable that we mention in the algorithm description is included in the simplices data structure (see section \ref{sec_meshImpl}) and can often be used to conveniently associate arbitrary data to simplices while alleviating multi-threading issues.

\begin{myalgo}{Distributed mesh refinement}{
    A distributed mesh $M$ and two user defined functions:
  \begin{itemize}
    \item {\em bool} checkRefine$( M_i)$: a function taking a simplex $M_i$ as argument and returning {\em true} if refinement is required, {\em false} otherwise.
    \item {\em (edge,score)} getEdge$(M_i)$: A function taking a simplex $M_i$ requiring refinement as argument and returning the preferred bisection edge and a corresponding score corresponding to how much refinement is needed.
    \item {\em vertex} splitEdge$(e)$: A function taking an edge $e$ and returning a vertex used to bisect $e$.
  \end{itemize}}{A refined version of $M$ such that calling checkRefine$(M_i)$ returns false for every $M_i$ in $M$.}
\label{algo_meshRefine}

\item Iterate in parallel over each simplex and ghost simplex $M_i$ on the local process, calling checkRefine$(M_i)$. Whenever {\em true} is returned, tag simplex $M_i$ for refinement, call getEdge$(M_i)$ as a different task, and store the returned $(e,${\em score}$)$ pair with simplex $M_i$ (using the ``cache'' variable defined in the simplices data structure). Edge $e$ is said to be refined by simplex $M_i$.

\item Iterate over each refinement edge generated by {\em local} simplices in parallel, recover for each of them the incident simplices and check if they are tagged for refinement. If so, compare their respective score and un-tag all the simplices but that with highest score.

\item Create a list $L_e$ of edges to refine with all edges refined by a {\em local} simplex.

\item Synchronise the ghost simplices refinement tags with neighbouring regions and add edges refined by ghost simplices to $L_e$.

\item Iterate with a single thread over each refinement edge $e_i$ in $L_e$. For each simplex incident to $e_i$, tag it as refined by $e_i$ if it is not already tagged. If it is already tagged as refined by another edge $e_j$, cancel refinement for the edge with lowest score and remove it from $L_e$. This step ensures that two neighbours of a simplex won't refine two of its edges simultaneously.

\item Synchronise ghost simplices refinement tags only for simplices whose refinement was canceled at previous step. At this point, refinements tags should already be consistent except for very rare unfavourable configurations. If inconsistencies appear, remove corresponding refinement edges from $L_e$.

\item Iterate in parallel over each edge $e_i$ to refine and insert vertex $V$ returned by calling function splitEdge$(e_i)$ in the mesh. Lets $v_0$ and $v_1$ be the two vertices of $e_i$. For each simplex $M_k$ incident to $e_i$:
\begin{itemize}
\item Insert a copy $M^\prime_k$ of $M_k$ in the mesh, and update its local index as appropriate.
\item Replace vertex $v_0$ in $M_k$ by $V$.
\item Replace vertex $v_1$ in $M^\prime_k$ by $V$.
\item Update user defined data associated to $M_k$ and $M^\prime_k$ as appropriate.
\end{itemize}

\item Synchronise the updated data and global identities of the refined ghost and shadow simplices and vertices and add newly created shadow simplices and vertices.

\item Fix simplices neighbourhood by updating the ``neighbour'' pointers in the simplices data structures as appropriate.

\item If at least one simplex was refined during this pass, repeat from start.
\end{myalgo}

\section{Derivation of the exact projection formula at first order}
\label{app_order1}
In this appendix, we show how to calculate equations (\ref{eq_firstOrder}) to (\ref{eq:E13D}). Any convex polygonal surface (2D) or volume (3D) ${\cal P}$ can be easily decomposed into a tessellation of simplices by  simply adding a new vertex $O$ inside it and associating it to each facet of the polygonal volume/surface. Then one can use the fact that the integral of an arbitrary polynomial over a triangle or a tetrahedron is well known and easily computed by expressing the polynomial in terms of barycentric coordinates \cite{Eisenberg1973}: the integral over ${\cal P}$ is simply given as the sum of the integrals over each triangle/tetrahedron. The point here is to rewrite this result at linear order in a form compatible with the algorithm of \cite{Franklin93}, that is in terms of quantities calculated at the vertices positions only, $\bm{P}.\bm{T}$, $\bm{P}.\bm{N}$, $\bm{P}.\bm{B}$, following the notations of section \ref{sec_franklinForm}. While we perform here the calculations up to linear order, the reader will be convinced that the procedure described below can be easily generalized to arbitrary order and, iteratively, to arbitrary number of dimensions. 

\begin{figure}
\centering
\includegraphics[width=0.6\linewidth]{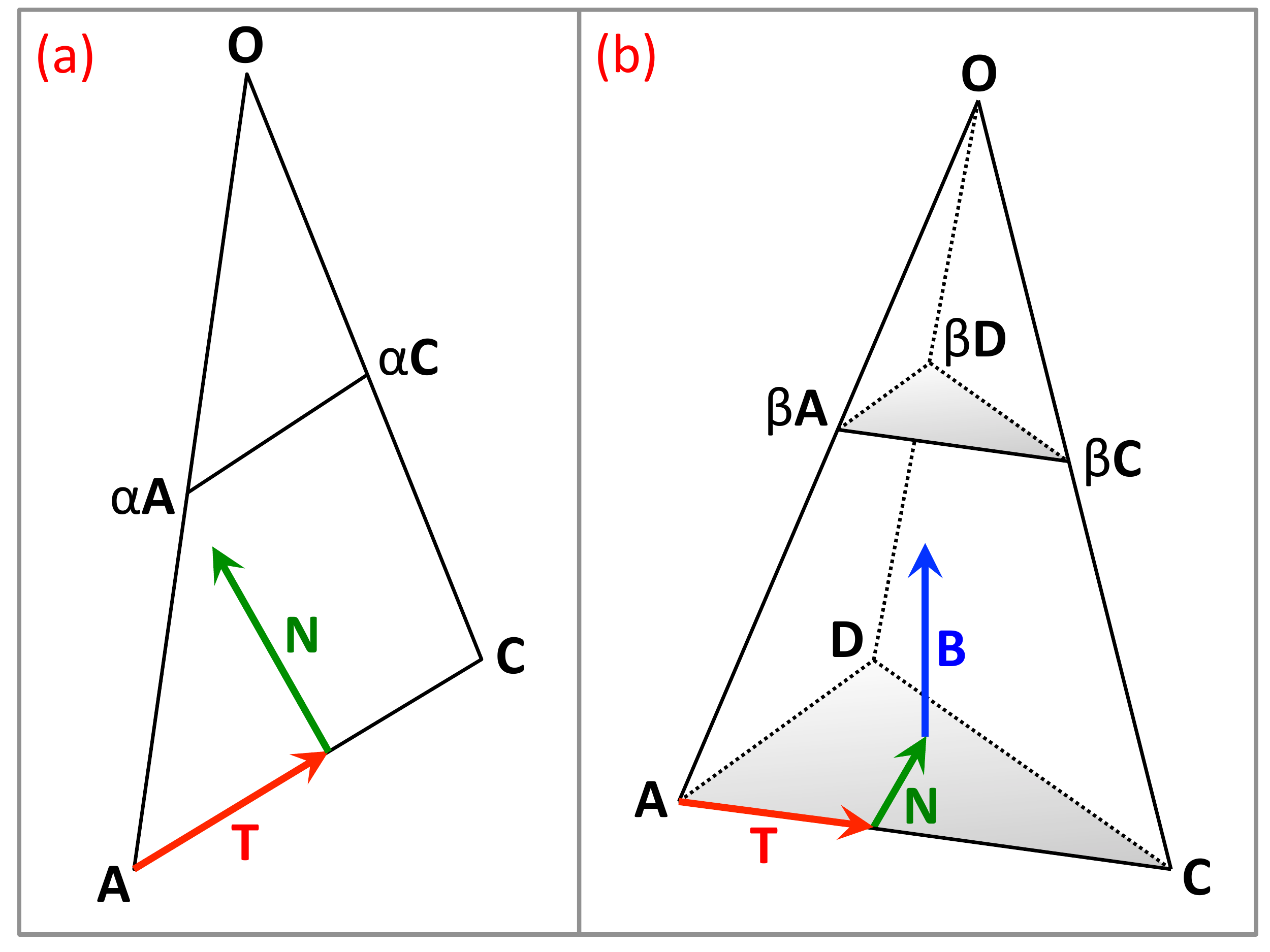}
\caption{Illustration on how to compute the integral of the projected density up to linear order inside a simplex of which one point $O$ is at the origin of coordinates, as discussed in the text. The left and the right panel correspond to the 2- and 3-dimensional cases, respectively. }
\label{fig:monbotriangle}
\end{figure}

We assume without any loss of generality that the expression for the projected density at linear order at point $\bm{X}$ is given as
\begin{equation}
\rho(\bm{X})={\tilde \rho}^0 + \nabla \rho^1.\bm{X}.
\end{equation}
We first start with the two-dimensional case. Consider a convex polygon ${\cal P}=AC\ldots$ and assume that the origin $O$ of the system of coordinates is inside it. Let us focus on triangle ${\cal T}=ACO$ for a start. We define a new system of coordinates with the same origin $O$ but units vectors $\bm{T}$ and $\bm{N}$ on left panel of Fig.~\ref{fig:monbotriangle}. 
Then,
\begin{equation}
\rho(\bm{X})={\tilde \rho}^0 + \nabla \rho^1.\bm{T}\ \bm{X}.\bm{T} + \nabla \rho^1.\bm{N}\ \bm{X}.\bm{N}
={\tilde \rho}^0 +\nabla \rho^1.\bm{T}\ t +\nabla \rho^1.\bm{N}\ n, 
\end{equation}
where $(t,n)$ are the coordinates of point $X$ in the new coordinate system. Then
\begin{eqnarray}
{\cal I}_{2D} =\int_{\cal T} \rho(\bm{X})\, {\rm d}^2\bm{X}= {\tilde \rho}^0\ {\cal S}_0 + \nabla \rho^1.\bm{T}\ {\cal S}_T + \nabla \rho^1.\bm{N}\ {\cal S}_N, \\
{\cal S}_0 \equiv \int_{\cal T} {\rm d}t\, {\rm d}n, \quad
{\cal S}_T \equiv \int_{\cal T}t\, {\rm d}t\, {\rm d}n, \quad
{\cal S}_N \equiv \int_{\cal T}n\, {\rm d}t\, {\rm d}n.
\end{eqnarray}
Setting $\alpha=n/n_A=n/n_B$ and applying Thales theorem we can rewrite
\begin{eqnarray}
{\cal S}_0&=&- n_A \int_0^1 {\rm d}\alpha \int_{\alpha t_A}^{\alpha t_C} {\rm d}t=\frac{1}{2} n_A(t_A-t_C)=
\frac{1}{2} (\bm{A}.\bm{N}_A\ \bm{A}.\bm{T}_A + \bm{C}.\bm{N}_C\  \bm{C}.\bm{T}_C),\\
{\cal S}_T &=&-n_A \int_0^1 {\rm d}\alpha  \int_{\alpha t_A}^{\alpha t_C} t\, {\rm d}t=\frac{1}{6} n_A(t_A-t_C) (t_A+t_C)=
\frac{1}{6} [\bm{A}.\bm{N}_A\ (\bm{A}.\bm{T}_A)^2 - \bm{C}.\bm{N}_C \ (\bm{C}.\bm{T}_C)^2],\\
{\cal S}_N&=&-n_A \int_0^1 \alpha\, n_A\, {\rm d}\alpha \int_{\alpha t_A}^{\alpha t_C} {\rm d}t=\frac{1}{3} n_A^2(t_A-t_C)=
\frac{1}{3} [(\bm{A}.\bm{N}_A)^2\ \bm{A}.\bm{T}_A + (\bm{C}.\bm{N}_C)^2\  \bm{C}.\bm{T}_C],
\end{eqnarray}
where we exploited the fact that 
\begin{eqnarray}
\bm{T}=\bm{T}_A =-\bm{T}_C,\\
\bm{N}=\bm{N}_A=\bm{N}_C.
\end{eqnarray}
In these equations we assumed an explicit orientation or, equivalently, direction of circulation along the polygon, namely that point $O$ is at the left of segment $AC$ while travelling from $A$ to $C$. Performing the circulation likewise over the whole polygon, we obtain easily equations (\ref{eq_firstOrder}), (\ref{eq_contrib}), (\ref{eq:E02D}) and (\ref{eq:E12D}). These equations do not depend on the fact that $O$ was chosen to be inside the polygon nor that the polygon ${\cal P}$ is convex, two assumptions which can now be dropped {\em a posteriori} by continuity.   

Extension of the calculation to the three-dimensional case is straightforward. Consider a convex polygon triangle ${\cal P}=ACD\ldots$ and assume that the origin $O$ of the system of coordinates is inside it. Let us focus on tetrahedron ${\cal T}=ACDO$ for a start. We define a new system of coordinates with the same origin $O$ but units vectors $\bm{T}$, $\bm{N}$, $\bm{B}$ on right panel of Fig.~\ref{fig:monbotriangle}. Then, 
\begin{eqnarray}
{\cal I}_{3D} =\int_{\cal T} \rho(\bm{X})\, {\rm d}^3\bm{X}= {\tilde \rho}^0\ {\cal V}_0 + \nabla \rho^1.\bm{T}\ {\cal V}_T + \nabla \rho^1.\bm{N}\ {\cal V}_N + \nabla \rho^1.\bm{B}\ {\cal V}_B, \\
{\cal V}_0 \equiv \int_{\cal T} {\rm d}t\, {\rm d}n\, {\rm d}b, \quad {\cal V}_T \equiv \int_{\cal T}t\, {\rm d}t \,{\rm d}n\, {\rm d}b, \quad
{\cal V}_N \equiv \int_{\cal T}n\, {\rm d}t\, {\rm d}n\, {\rm d}b, \quad
{\cal V}_B \equiv \int_{\cal T}b\, {\rm d}t\, {\rm d}n\, {\rm d}b, 
\end{eqnarray}
where $(t,n,b)$ are the coordinates of point $X$ in the new coordinate system.
Setting $\beta=b/b_A=b/b_C=b/b_D$ and defining the triangle ${\cal T}(\beta) \equiv A(\beta)\, C(\beta)\, D(\beta)$, with $\bm{A}(\beta)=\beta \bm{A}$ and likewise for $\bm{C}(\beta)$ and $\bm{D}(\beta)$, we can rewrite
\begin{eqnarray}
{\cal V}_0&=&- b_A \int_0^1 {\rm d}\beta \int_{{\cal T}(\beta)} {\rm d}t\, {\rm d}n = -b_A \int_0^1 \beta^2\, {\rm d}\beta \int_{{\cal T}(1)}{\rm d}t\, {\rm d}n= -\frac{1}{3} b_A\ \int_{{\cal T}(1)}{\rm d}t\, {\rm d}n, \\
{\cal V}_T &=&-b_A \int_0^1 {\rm d}\beta  \int_{{\cal T}(\beta)} t\, {\rm d}t\, {\rm d}n=-b_A \int_0^1 \beta^3\, {\rm d}\beta \int_{{\cal T}(1)}t\, {\rm d}t\, {\rm d}n=
-\frac{1}{4} b_A\ \int_{{\cal T}(1)}t\, {\rm d}t\, {\rm d}n, \\
{\cal V}_N&=&-b_A \int_0^1 {\rm d}\beta \int_{{\cal T}(\beta)} n\, {\rm d}t\, {\rm d}n=-b_A \int_0^1 \beta^3\, {\rm d}\beta \int_{{\cal T}(1)}n\, {\rm d}t\, {\rm d}n=-\frac{1}{4} b_A\ \int_{{\cal T}(1)}n\, {\rm d}t\, {\rm d}n, \\
{\cal V}_B&=&-b_A \int_0^1 \beta\, b_A\, {\rm d}\beta \int_{{\cal T}(\beta)} {\rm d}t\, {\rm d}n=-b_A^2 \int_0^1 \beta^3\, {\rm d}\beta \int_{{\cal T}(1)} {\rm d}t\, {\rm d}n=-\frac{1}{4} b_A^2 \int_{{\cal T}(1)} {\rm d}t\, {\rm d}n.
\end{eqnarray}
Exploiting the 2D results derived just above, we thus find
\begin{equation}
{\cal I}_{3D} =-\frac{1}{6} \sum _{{\cal T}(1)} \bm{P}.\bm{T} \ \bm{P}.\bm{N}\ \bm{P}.\bm{B}\  \left[{\tilde \rho}^0 +\frac{1}{4} \bm{P}.\bm{T}\ \nabla \rho^1.\bm{T} +\frac{1}{2} \bm{P}.\bm{N}\ \nabla \rho^1.\bm{N} +\frac{3}{4} \bm{P}.\bm{B} \ \nabla \rho^1.\bm{B} \right],
\end{equation}
which summed up over all the facets of the polygonal volume gives
equations (\ref{eq_firstOrder}), (\ref{eq_contrib}), (\ref{eq:E03D}) and (\ref{eq:E13D}). Here, we assumed that the vector $\bm{B}$ is always pointing inside the polygon, which is the case if $O$ is inside the polygon and if the polygon is convex.  Note that the restrictions that $O$ must be inside the polygon and that the polygon must be convex can again be dropped {\em a posteriori} by continuity.   

\section{Exact projection algorithm}
\label{app_exactProjImpl}
In this appendix, we give a more detailed description of the implementation of the exact projection algorithm in section \ref{sec_exactProj} (see algorithm \ref{algo_projectShort}). The reader can refer to this section for details on the notations.

We start by noting that for corner types B, C and D (see section \ref{sec_exactProjImpl}), all associated quadruplets have symmetric counterparts generated by corners symmetric with respect to the simplex facets. The anti-symmetry of equation (\ref{eq_contrib}) with respect to $\T$, $\N$ and $\B$ can be exploited to significantly accelerate the computation of the corresponding contributions. Indeed, let $f$ be a facet of $M$ shared by simplices $M_i$ and $M_j$. Then any quadruplet $\left(\VP,\T,\N,\B\right)$ associated to $\left(v,e,f\right)$ within a given corner has a symmetric counterpart $\left(\VP,\T^\prime,\N^\prime,\B^\prime\right)$ associated to the same  $\left(v,e,f\right)$ but generated by the corner opposed to it with respect to $f$. The contributions from these two quadruplets can therefore be advantageously fused as they are identical except for exactly one of the $\T^\prime$, $\N^\prime$ or $\B^\prime$ vectors which has the opposite sign of its counterpart, and we have:
\begin{equation}
  \label{eq_contribFused}
  C_{M_i}\left(v,e,f\right) + C_{M_j}\left(v,e,f\right) = E_0 \left(\delta_f\rho^0 + {\mathbf E_1}.{\mathbf \delta_f\nabla \rho^1}\right)
\end{equation}
with
\begin{align}
\label{eq_diffRho}
  \delta_f\rho^0 &= \rho^0(M_i)-\rho^0(M_j),\\
  \delta_f\rho^1 &= {\mathbf \nabla \rho^1(M_i)}-{\mathbf \nabla \rho^1(M_j)}
\end{align}
and $E_0$ and ${\mathbf E_1}$ are associated to the corner generated by $M_i$.

\begin{myalgo}{Exact projection}{A simplicial mesh $M$, a function $\rho_i=\rho\left(\vertex{i}\right)$ defined for each vertex $v_i\in M$ and an (adaptively refined) uniform grid $G$.}{The exact projection of the linear interpolation of $\rho_i$ over $M$ onto the voxels of $G$.}
\label{algo_project}
\item For each simplex $M_i$ in $M$, check if it falls entirely inside a voxel $G_j$ of $G$. If it does, add its entire contribution to $G_i$, set $\rho\left(M_i\right)=0$ and tag the simplex as already projected. 
\item For each vertex $\vertex{i}$ in $M$:
\begin{enumerate}
  \item Create a list $L_e\left(\vertex{i}\right)$ of edges $\edge{i}{j}$ incident to $\vertex{i}$ and such that $j<i$. This can be easily achieved using a precomputed simplex incidence array (see the end of section \ref{sec_exactProjImpl}).
  \item For each edge $\edge{i}{j}$ in $L_e\left(\vertex{i}\right)$:
    \begin{enumerate}  
    \item Store $\VP$ and $\T$ associated to $\edge{i}{j}$.
    \item retrieve an ordered list $L_f(\edge{i}{j})$ of facets $\facet{i}{j}{k}$ incident to $\edge{i}{j}$ and such that two consecutive items in the list are facets of a common simplex. This list can be obtained by starting from any simplex incident to $\edge{i}{j}$, finding one of its facets incident to $\edge{i}{j}$ that is not yet in $L_f\left(e\right)$, adding it to $L_f\left(e\right)$ and repeating the operation from the neighbouring simplex sharing that facet until one loops back to the first simplex again. The set of pink simplex facets on figure \ref{fig_cubeTetInter_c} illustrates such a list. Exclude from this list any facet for which the two incident simplices are tagged as projected.
    \item \label{item_wei}For each facet $\facet{i}{j}{k}$ in $L_f(\edge{i}{j})$, compute the $\N$ and $\B$ vectors such that $\B$ is oriented toward the next facet in the list. Also store the differential weights $\delta_f\rho^0$ and $\delta_f\rho^1$ introduced in equation (\ref{eq_diffRho}).
    \end{enumerate}
  \item Retrieve which voxel $G(\vertex{i})$ vertex $\vertex{i}$ falls into and add the contributions from all previously computed quadruplets [i.e. from all facets in $L_f(\edge{i}{j})$ incident to all edges $\edge{i}{j}$ in $L_e(\vertex{i})$; these are type B contributions, see figure \ref{fig_cubeTetInter_b}]. 
  \item For each edge $\edge{i}{j}$ in $L_e(\vertex{i})$:
    \begin{enumerate}
    \item Locate which voxel $G(\vertex{j})$ vertex $v_j$ falls into and add the contribution from all facets in $L_f(\edge{i}{j})$, taking advantage of fused contribution equation (\ref{eq_contribFused}) (these are type B contributions, see figure \ref{fig_cubeTetInter_b}).
    \item Raytrace segment $\edge{i}{j}$ in $G$ from voxel $G(\vertex{i})$ to voxel $G(\vertex{j})$. Taking advantage of the symmetries and for each intersection of the ray with a voxel facet, compute all the contributions from the quadruplets associated to each facet $L_f(\edge{i}{j})$ as well as from their symmetric counterparts with respect to the voxel facet (these are the type C contributions, see figure \ref{fig_cubeTetInter_c}).
    \end{enumerate}
  \item For each simplex $M_n=\simplex{i}{j}{k}{l}$ incident to $\vertex{i}$ such that $i>j$, $i>k$ and $i>l$ and $M_n$ is not tagged as projected:
    \begin{enumerate}
      \item Retrieve the set of voxels $L_G(M_n)$ that overlap with $M_n$.
      \item For each voxel corner vertex in $L_G(M_n)$, test if it falls inside $M_n$ (see section \ref{sec_robustness} and \ref{app_predicates}) and if it does, compute and add the contribution of type A (figure \ref{fig_cubeTetInter_a}) to the $8$ voxels surrounding it.
      \item Build a list $L_f(M_n)$ of facets of $M_n$ such that the index $n^{\prime}$ of the simplex $M_{n^{\prime}}$ symmetric to $M_n$ with respect to $f$ is such that $n^{\prime} < n$.
      \item for each facet $f_M$ in $L_f(M_n)$ and each edge $e_G$ belonging to a voxel of $L_G(M_n)$, test if they intersect (see section \ref{sec_robustness} and \ref{app_predicates}) and if they do, compute the type D contributions (figure \ref{fig_cubeTetInter_d}) to the $4$ voxels surrounding $e_G$ and from both sides of $f_M$. A total of $48$ quadruplets are contributing, but most of the computations can be factorised taking symmetries into account.
    \end{enumerate}    
\end{enumerate}
\end{myalgo}

\section{Predicates implementation}
\label{app_predicates}
\subsection{Point in tetrahedron}
Point in tetrahedron (and in general, point in simplex) predicates are implemented by testing whether the point $\bm{p}=(p_x,p_y,p_z)$ lies on the same side of all the (properly oriented) $4$ facets of a tetrahedron. Let  $(\bm{q},\bm{r},\bm{s})$ be one such oriented facet of tetrahedron $(\bm{q},\bm{r},\bm{s},\bm{t})$, with $\bm{q}=(q_x,q_y,q_z)$, $\bm{r}=(r_x,r_y,r_z)$ and $\bm{s}=(s_x,s_y,s_z)$. Then the side on which $\bm{p}$ lies is defined by the orientation of $(\bm{p},\bm{q},\bm{r},\bm{s})$ which is in turn given by the sign of the determinant ${\cal D}$ of the following matrix:
\begin{equation}
\label{eq_orientation}
{\rm orientation}(\bm{p},\bm{q},\bm{r},\bm{s})={\rm sign}({\cal D}) ={\rm sign}
\left| \begin{array}{cccc}
1 & p_x & p_y & p_z\\
1 & q_x & q_y & q_z\\
1 & r_x & r_y & r_z\\
1 & s_x & s_y & s_z\\
\end{array} \right| = {\rm sign}\left| \begin{array}{ccc}
 q_x-p_x & q_y-p_y & q_z-p_z\\
 r_x-p_x & r_y-p_y & r_z-p_z\\
 s_x-p_x & s_y-p_y & s_z-p_z\\
\end{array} \right|.
\end{equation}
This determinant can be easily developed into a sum over the appropriate triplets $\left(i,j,k\right)$:
\begin{equation}
{\rm orientation}(\bm{p},\bm{q},\bm{r},\bm{s})={\rm sign}\left[ \sum_{\left(i\neq j \neq k\right)}\,\pm (q_i-p_i) (r_j-p_j) (s_k-p_k) \right].
\end{equation}
Following the implementation of the {\tt CGAL} \cite{CGAL_LG} geometric library, an upper bound on the error on the value of the determinant is therefore given as 
\begin{align}
\label{eq_orientation_err}
E &\leq \alpha.{\rm max}_{i\neq j \neq k}\,\left(\left|\left(q_i-p_i\right) \left(r_j-p_j\right) \left(s_k-p_k\right)\right|\right)\nonumber\\
&\leq \alpha \prod_{i=1}^3{\rm max}\left(\left|q_i-p_i\right|,\left|r_i-p_i\right|,\left|s_i-p_i\right|\right),
\end{align}
where $\alpha$ is the expected precision for a given finite precision floating point type and set of operations \citep[see e.g.][for a short review on floating point arithmetic]{Burnikel98} and the second expression is used to simplify the filter calculation. We use a safe value of $\alpha=5.10^{-14}$ for double precision computations in our particular implementation and switch to exact calculation of the orientation predicate if the absolute value of the determinant evaluated in equation (\ref{eq_orientation}) is lower than the error computed from equation (\ref{eq_orientation_err}).

Degenerate cases of the point in tetrahedron predicate happen when the point lies exactly on the boundary of a simplex. Adopting the simulation of simplicity convention of equation (\ref{eq_sos}), we perturb $\bm{p}$ as:
\begin{equation}
  \bm{p}=(p_x,p_y,p_z) \mapsto  \bm{p}(\epsilon) = (p_x-\epsilon^2,p_y-\epsilon^3,p_z-\epsilon^4).
\end{equation}
The determinant ${\cal D}$ from equation (\ref{eq_orientation}) becomes:\footnote{see \cite{Edelsbrunner90} for the general case.}
\begin{align}
\label{eq_orientation_sos}
{\cal D}(\epsilon)&=\left|\begin{array}{ccc}
 q_x-p_x+\epsilon^2 & q_y-p_y+\epsilon^3 & q_z-p_z+\epsilon^4\\
 r_x-p_x+\epsilon^2 & r_y-p_y+\epsilon^3 & r_z-p_z+\epsilon^4\\
 s_x-p_x+\epsilon^2 & s_y-p_y+\epsilon^3 & s_z-p_z+\epsilon^4\\
\end{array} \right|\nonumber\\
&={\cal D}(\epsilon=0)+
\epsilon^2 \left| \begin{array}{ccc}
 1 & q_y-p_y & q_z-p_z\\
 1 & r_y-p_y & r_z-p_z\\
 1 & s_y-p_y & s_z-p_z\\
\end{array} \right|\nonumber\\
&+\epsilon^3 \left| \begin{array}{ccc}
 q_x-p_x & 1 & q_z-p_z\\
 r_x-p_x & 1 & r_z-p_z\\
 s_x-p_x & 1 & s_z-p_z\\
\end{array} \right|+
\epsilon^4 \left| \begin{array}{ccc}
 q_x-p_x & q_y-p_y & 1\\
 r_x-p_x & r_y-p_y & 1\\
 s_x-p_x & s_y-p_y & 1\\
\end{array} \right|\nonumber\\
&+ ...
\end{align}
and the first non null term in this expansion can be taken to consistently determine the sign of determinant (\ref{eq_orientation}) even though the matrix is degenerate.

\begin{myalgo}{Point in tetrahedron}{A point $\bm{p}$ and a tetrahedron $T$.}{{\em True} if the point lies inside the tetrahedron, {\em false} otherwise}
\label{algo_pointTetrahedron}
\begin{enumerate}
  \item Count the number of facets of $T$ for which the orientation of the tetrahedron formed by $\bm{p}$ and that (consistently oriented) facet is positive using the filtered exact predicate discussed above.
  \item If this number is $4$ or $0$ return {\em true}, else return {\em false}.   
\end{enumerate}
\end{myalgo}

\subsection{Triangle and segment intersection}
Our implementation of the triangle and segment intersection test in 3D is very similar to that of the point in simplex test: for a segment to intersect a triangle in 3D, its extremities should lie on different sides of the plane defined by this triangle. This can be robustly tested using the orientation predicate discussed in previous section. But this condition is not sufficient as the segment may intersect the plane of the triangle outside of it. In the case of our exact projection algorithm (see algorithm \ref{algo_projectShort} in section \ref{sec_exactProjImpl} and algorithm \ref{algo_project} in \ref{app_exactProjImpl}), segments are edges of a regular grid and are therefore always axis aligned. We take advantage of this fact by simply projecting the triangle and edge along the axis of the segment as doing so, the problem reduces to a 2D point in triangle test.

\begin{myalgo}{Axis aligned segment and triangle intersection}{An axis aligned segment and a triangle}{{\em True} if the segment and triangle intersect, {\em false} otherwise}
\label{algo_segmentTriangle}
\begin{enumerate}
  \item Project the triangle and segment along the axis of the segment. The segment reduces to a point $\bm{p}$.
  \item Test if $\bm{p}$ is inside the projected triangle. If it is not, return {\em false}.
  \item Test on which side of the 3D triangle each extremity of the segment lie using the filtered exact predicate discussed in \ref{app_predicates}. If it is the same, return {\em false}, else return {\em true}.
\end{enumerate}
\end{myalgo}

\subsection{Exact segment raytracing}
What we mean here by ``exact'' raytracing is that for a given ray and in the general case (i.e. excluding exact degeneracies), all the predicates on whether it crosses a given facet of a given voxel $G_i$ of $G$ are exact, as well as the order in which it does so. Note however that this does not necessarily imply that the coordinates of these crossings are also computed exactly.

Consider a segment $[\bm{p},\bm{q}]$ with extremities $\bm{p}=(p_x,p_y,p_z)$ and $\bm{q}=(q_x,q_y,q_z)$. Let us propagate the ray from $\bm{p}$ toward $\bm{q}$. Then we start from the voxel $G_i\in G$ which contains $\bm{p}$ and need to reliably decide which facet of $G_i$ segment $[\bm{p},\bm{q}]$ will intersect next. Because $G$ is regular, facets of voxels in $G$ are orthogonal to the axis of the orthonormal base vectors and we can denote $f^\pm_k(G_i)$ the facet of $G_i$ orthogonal to axis $k$ and with highest/lowest coordinate along this axis respectively. Then we know that $[\bm{p},\bm{q}]$ may only cross $f^{{\rm sign}(q_k-p_k)}_k(G_i)$ if $p_k\neq q_k$, and neither of $f^\pm_k(G_i)$ otherwise. Let $\bm{c}=(c_x,c_y,c_z)$ be the corner of $G_i$ at the intersection of facets $f^{{\rm sign}(q_k-p_k)}_k(G_i)$ (we formally set $c_k=\infty$ if $p_k=q_k$). The first facet the ray will cross is the one orthogonal to the direction that minimize $t_k$ in the following equation:
\begin{equation}
p_k + (q_k-p_k) t_k = c_k \implies t_k = \frac{c_k-p_k}{q_k-p_k}.
\end{equation}
From a practical point of view, as by definition $t_k \geq 0$ for all $k$, 
\begin{equation}
\label{eq_raytraceMin}
t_i<t_j \Leftrightarrow \frac{c_i-p_i}{q_i-p_i} < \frac{c_j-p_j}{q_j-p_j} \Leftrightarrow \left|(c_i-p_i)(q_j-p_j)\right|<\left|(c_j-p_j)(q_i-p_i)\right|,
\end{equation}
where the right inequality can be evaluated correctly using finite precision floating point arithmetic within an error margin of:
\begin{equation}
\label{eq_raytraceErr}
E \leq \beta. \left(|c_i-p_i| |q_j-p_j|+|c_j-p_j||q_i-p_i|\right),
\end{equation}
with $\beta$ the expected precision for a given finite precision floating point type and set of operations \citep[e.g.][]{Burnikel98}.
We therefore find the value of $k$ that minimize $t_k$, check that it is minimal within the confidence level of equation (\ref{eq_raytraceErr}) and reevaluate it using multi-precision arithmetic if necessary. The process can then be repeated until we reach voxel $G_j\in G$ that contains $\bm{q}$.

Finally, we deal with exact degeneracies that happen whenever the ray exactly intersects a voxel edge or a voxel corner by adopting the simulation of simplicity convention of equation (\ref{eq_sos}):
\begin{equation}
  \bm{c}=(c_x,c_y,c_z) \mapsto  \bm{c}(\epsilon) = (c_x-\epsilon^2,c_y-\epsilon^3,c_z-\epsilon^4).
\end{equation}
Following this convention, whenever the ray intersects more than one facet of a voxel at a given point, priority is given to :
\begin{enumerate}
\item the facet that is orthogonal to axis $i$ along which $q_i-p_i$ is positive and for which $i$ is the lowest;
\item if there is no such facet, the facet orthogonal to axis $i$ for which $i$ is the highest.
\end{enumerate}
\section{Cosmology}
\label{app:cosmology}
In this appendix, we recall how the equations (\ref{eq:vlaeq}) and (\ref{eq:poieq}) are modified in the standard cosmological framework, then we provide details about initial condition settings, time step calculation, energy conservation and comment on Poincar\'e invariant.
\subsection{Supercomoving coordinates}
\label{app:supcom}
Our choice of internal code units in the cosmological case is the same as in \cite{Teyssier2002}. It makes use of ``supercomoving'' coordinates \cite{Martel1998}, first introduced by \cite{Doroshkevich1973}:
\begin{eqnarray}
{\rm d}\tau &\equiv& H_0  \frac{{\rm d}t}{a^2}, \label{eq:superco1}\\
\tilde{\bm{x}} &\equiv & \frac{\bm{r}}{a L},\\
\tilde{\bm{u}} &\equiv & a \frac{\bm{u}-H \bm{r}}{H_0 L},\\
{\tilde \rho} & \equiv & a^3 \frac{\rho}{\Omega_0 \rho_{\rm c}}. \label{eq:superco4}
\end{eqnarray}
In these equations, $a$ is the expansion factor of the Universe supposed to be equal to unity at present time and the Hubble parameter $H \equiv (1/a)\ {\rm d}a/{\rm d}t$ drives the expansion rate of the Universe,
\begin{equation}
\left(\frac{H}{H_0}\right)^2=\Omega_0 a^{-3}+(1-\Omega_0-\Omega_{\rm L})
a^{-2}+\Omega_{\rm L}, \label{eq:hsurh0}
\end{equation}
with $\Omega_0$ the total matter density parameter of the Universe, $\Omega_{\rm L}$ the
cosmological constant and $H_0$ the Hubble constant, 
\begin{equation}
H_0  \simeq  100\,h\,{\rm km/s/Mpc},\quad h \simeq 0.7.
\end{equation}
In addition, $L$ is the comoving size of the simulation box, $\tilde{\bm{x}}$ the ``comoving coordinate'', $\tilde{\bm{u}}$ the ``comoving'' peculiar velocity, with $\bm{u}_{\rm pec} \equiv \bm{u}-H \bm{r}$ being the actual peculiar velocity, i.e. the velocity of the element of fluid subtracted from the velocity of the expansion of the Universe.  Finally, ${\tilde \rho}$ is the comoving density, $\rho_{\rm c}$ the critical density of the Universe at present time: $\rho_{\rm c} \equiv  3 H_0^2/(8 \pi G)$, where $G$ is the gravitational constant.  Note therefore that conveniently, 
\begin{equation}
\langle {\tilde \rho} \rangle=1.
\end{equation}

With the new coordinates and the new time, Vlasov-Poisson equations (\ref{eq:vlaeq}) and (\ref{eq:poieq}) remain nearly unchanged, except that one has to replace $\phi$ with a new gravitational potential $\psi$ obeying the following modified Poisson equation 
\begin{equation}
\Delta_{\tilde{\bm{x}}} \psi=\frac{3}{2} a\,\Omega_0\,[{\tilde
  \rho}(\tilde{\bm{x}})-1].
\label{eq:poissonfinal}
\end{equation}
This equation is valid both in 3 and 2 dimensions.  In the later case, we
resolve the gravitational interaction between infinite parallel lines in the expanding universe. 

\subsection{Initial conditions}
\label{app:inic}
While {\tt ColDICE} can read a standard {\tt GADGET} particle file \cite{Springel2001} to set up the initial positions and velocities of the vertices and tracers of the simplicial mesh, it can also compute them by using fastest growing mode of first or second order Lagrangian perturbation theory. In this framework, the positions and velocities of the nodes are related to Lagrangian coordinates $\bm{q} \equiv \tilde{\bm{x}}(a=0)$ through the following expressions \citep[see, e.g.][]{Bouchet1995}:
\begin{eqnarray}
\tilde{\bm{x}}({\bm q},\tau) &=& \bm{q}+D_1^+(\tau) \bm{P}_1(\bm{q})+D_2^+(\tau)\bm{P}_2(\bm{q}),\\
\tilde{\bm{u}}({\bm q},\tau) &=& \frac{{\rm d}D_1^+}{{\rm d}\tau} \bm{P}_1(\bm{q})+\frac{{\rm d}D_2^+}{{\rm d}\tau}\bm{P}_2(\bm{q}),
\end{eqnarray}
with 
\begin{equation}
D_1^+(\tau_{\rm i}) \nabla_{\bm{q}}.\bm{P}_1=-\delta_{\rm i}(\bm{q}),\quad
\nabla_{\bm{q}} \times \bm{P}_1=0, \quad
\nabla_{\bm{q}}.\bm{P}_2=\frac{1}{2}\sum_{i \neq j} \left[\frac{\partial P_{1,i}}{\partial q_i}\frac{\partial P_{1,j}}{\partial q_j} 
- \frac{\partial P_{1,j}}{\partial q_i}\frac{\partial P_{1,i}}{\partial q_j}\right], \quad
\nabla_{\bm{q}} \times \bm{P}_2 =0, \label{eq:P2a}
\end{equation}
where $\delta_{\rm i}$ is the initial density contrast at the beginning of the simulation, 
$D_1^+$ is the linear growing mode and $D_2^+$ is the second order growing mode. To compute $D_1^+$ we use the following approximation by \cite{Lahav1991}
\begin{eqnarray}
D_1^+ \simeq \frac{5}{2} \Omega a \left[ \Omega^{4/7}-\Lambda+\left(1+\frac{1}{2}\Omega\right)
\left( 1+\frac{1}{70} \Lambda \right) \right]^{-1}.
\end{eqnarray}
where $\Omega$ and $\Lambda$ are the values of the cosmological density and cosmological constant at the time considered
\begin{equation}
\Omega(a) =\frac{\Omega_0}{a+\Omega_0(1-a)+\Omega_{\rm L}(a^3-a)}, \quad
\Lambda(a) = \frac{a^3 \Omega_{\rm L}}{a+\Omega_0 (1-a)+\Omega_{\rm L}(a^3-a)}.
\end{equation}
To estimate $D_2^+$ and time derivatives of $D_1^+$ and $D_2^+$, we use the following approximations proposed in \citep{Bouchet1995} in the case $\Omega_0+\Omega_{\rm L}=1$ 
\begin{equation}
f_1(\Omega) \equiv \frac{a}{D_1^+} \frac{{\rm d}D_1^+}{{\rm d}a} \simeq \Omega^{5/9}, \quad
f_2(\Omega) \equiv \frac{a}{D_2^+} \frac{{\rm d}D_2^+}{{\rm d}a} \simeq \Omega^{6/11}, \quad
D_2^+ \equiv -\frac{3}{7} \Omega^{-1/143} (D_1^+)^2.
\end{equation}
\subsection{Time step}
\label{sec:timestepappcosm}
As displayed by equation (\ref{eq:poissonfinal}), expansion of the Universe induces a screening effect on the gravitational force and introduces an $a$ dependance on the source term of Poisson equation. As a result our dynamical time step constraint, equation (\ref{eq:DYN}), is modified as follows:
\begin{equation}
\Delta \tau \leq \frac{C_{\rm dyn}}{\sqrt{\frac{3}{2} \Omega_0\,a\,{\tilde \rho}_{\rm max}}},
\end{equation}
while CFL condition remains unchanged if applied to the new time and new velocity coordinates. In addition, because of the $1/a^2$ factor in equation (\ref{eq:superco1}), the time step $\Delta \tau$ can correspond to exaggeratedly large variations of the expansion factor. Since this latter coincides at early times with the linear growing mode, $D_1^+ \simeq a$, this might be troublesome at early stages of the simulation. To avoid this, we introduce an additional constraint on the time step bounding relative variation of the expansion factor 
\begin{equation}
\Delta \tau \leq C_{a}\,a\, \left[ \frac{{\rm d}a}{{\rm d}\tau} \right]^{-1}, \quad {\rm with}\ C_a=0.1.
\end{equation}
\subsection{Energy conservation}
\label{app:cosener}
Energy conservation involves an additional term in the cosmological case compared to the standard physical one, due to the expansion of the Universe, through the Layer-Irvine-Dmitriev-Zel'dovich equation \citep[see, e.g.][]{Peebles1980}. In our system of coordinates, total energy can be written,
\begin{equation}
E_{\rm tot}=E_{\rm k}(a)+E_{\rm p}(a)+E_{\rm exp}(a),
\end{equation}
with
\begin{eqnarray}
E_{\rm k}(a) &= &\frac{1}{2}\int \rho(\tilde{\bm{x}}) [{\tilde u}(\tilde{\bm{x}})]^2 {\rm d}\tilde{\bm{x}},\\
E_{\rm p}(a) &=& \frac{1}{2} \int \rho(\tilde{\bm{x}}) \psi(\tilde{\bm{x}}) {\rm d}\tilde{\bm{x}}, \\
E_{\rm exp}(a) &=& -a \int_0^a \frac{1}{a'} E_{\rm p}(a') {\rm d}a',
\end{eqnarray}
where $E_{\rm k}$, $E_{\rm p}$ and $E_{\rm exp}$ are respectively the total kinetic energy, total potential energy and the term due to the expansion of the Universe. 
\subsection{Note on the calculation of Poincar\'e invariant constraint}
In the new system of coordinates, the system remains Hamiltonian so the local Poincar\'e invariant constraint does not change fundamentally and equation (\ref{eq:refcritt}) is still valid, but applied to coordinates $\tilde{\bm{x}}$ and $\tilde{\bm{u}}$. Furthermore, a natural choice for the scales of the system is simply $L_{\tilde{x}}=L_{\tilde{v}}=1$ stemming from assuming  $L_{\tilde{x}}$ to be equal to the simulation box size and $L_{\tilde{v}}$ to correspond to the Hubble flow across the simulation box at present time. 
\section{Influence of refinement on inter-vertex distance and simplices count}
\subsection{Refinement and inter-vertex distance}
\label{app:refanddis}
Here, we try to estimate the consequence on inter-vertex distance of applying the refinement criterion proposed in \S~\ref{sec_refinementCriterion}.  To this end, it is interesting to estimate the modification on the contour integral (\ref{eq:poinc}) induced by adding a virtual refinement point $P$ along a segment $[AB]$ of a triangle $ABC$. Given our local second order description of the phase-space sheet, this refinement point, which corresponds to a tracer, is not exactly on $[AB]$ but still very close to it and is nearly equidistant to $A$ and $B$, $d(PA) \simeq d(PB)=d/2$, where $d=d(AB)$ is the distance between $A$ and $B$. Furthermore, since this point is very close to $[AB]$ this means that $d \ll R$, where $R$ is the local curvature along direction defined by $[AB]$. 

Defining $I_{\cal P}=\oint_{\cal P} \bm{u} {\rm d}\bm{x}(s)$, where ${\cal P}$ is the piecewise linear curve coinciding with the closed polygon ${\cal P}$, we have the following equalities
\begin{equation}
I_{APBC} =I_{ABC} + I_{PBA}=I_{APC}+I_{BCP}.
\end{equation}
The symplectic area of triangle $PBA$ is given by
\begin{equation}
I_{PBA} = \sum_{i=x,y (z)} s_i S_i,
\label{eq:invpa}
\end{equation}
where $s_i$ is a sign depending on the direction of circulation and $S_i$ the surface of the projection
of triangle $APB$ onto subspace $i=(x,v_x)$, $(y,v_y)$, $(z,v_z)$ [or $i=(x,v_x)$ and $(y,v_y)$ in 4 dimensional phase-space]. Hence
\begin{equation}
I_{PBA}= \gamma S,
\end{equation}
where $S$ is the surface of the triangle in phase-space and $\gamma$ is a form factor depending on the orientation in phase-space of the 2D plane containing the triangle. 
In the limit $d \ll R$, we have
\begin{equation}
S \simeq \frac{1}{8} \frac{d^3}{R}.
\end{equation}
Therefore, $I_{PBA}$ scales like
\begin{equation}
I_{PBA} \sim \frac{1}{8} \gamma\frac{ d^3}{R}.
\label{eq:su}
\end{equation}

Proceeding again by adding one virtual refinement point $U$ between $A$ and $P$ and one virtual refinement point $V$ between $P$ and $B$ while keeping them in the same plane as $APB$, we obtain the equality
\begin{equation}
I_{AUPVBC}=I_{ABC}+I_{PBA}+I_{VBP}+I_{UPA}.
\end{equation}
Using the fact that equation (\ref{eq:su}) applies as well to $I_{VBP}$ and $I_{UPA}$ but with a reduction of $d$ by a factor 2, we find that if
\begin{equation}
1/4 |I_{PBA}| \simeq |I_{VBP}+I_{UPA}|=|I_{AUPBVC}-I_{APBC}| \leq |I_{AUPVBC}|+|I_{APBC}|  \lesssim 4 \epsilon_I,
\label{eq:ccc}
\end{equation}
refinement is not necessary.  This assumes that refinement reduces violation from Poincar\'e invariance, $|I_{AUPVBC}|  \lesssim |I_{APBC}=I_{APC}+I_{BCP}| \leq |I_{APC}|+|I_{BCP}| \lesssim 2 \epsilon_I$, the latter inequality stemming directly from equation (\ref{eq_invariantRefined}). So as long as
\begin{equation}
d^3 \lesssim \beta \epsilon_I R,
\label{eq:refde}
\end{equation}
where $\beta$ is some number, refinement is not needed. 
This means that the local vertex separation should scale like  $(R \epsilon_I)^{1/3}$ to maintain $J < \epsilon_I$ in equation (\ref{eq_invariantRefined}), where $R$ is the local curvature radius along the direction examined. This result extends to 4 and 6 dimensional phase-space the earlier findings of \cite{Colombi2014} in 2 dimensional phase-space. 

\subsection{Refinement and simplicess count in the strongly anisotropic case}
\label{sec:refstrans}
\begin{figure}
\centering
\includegraphics[width=0.6\linewidth]{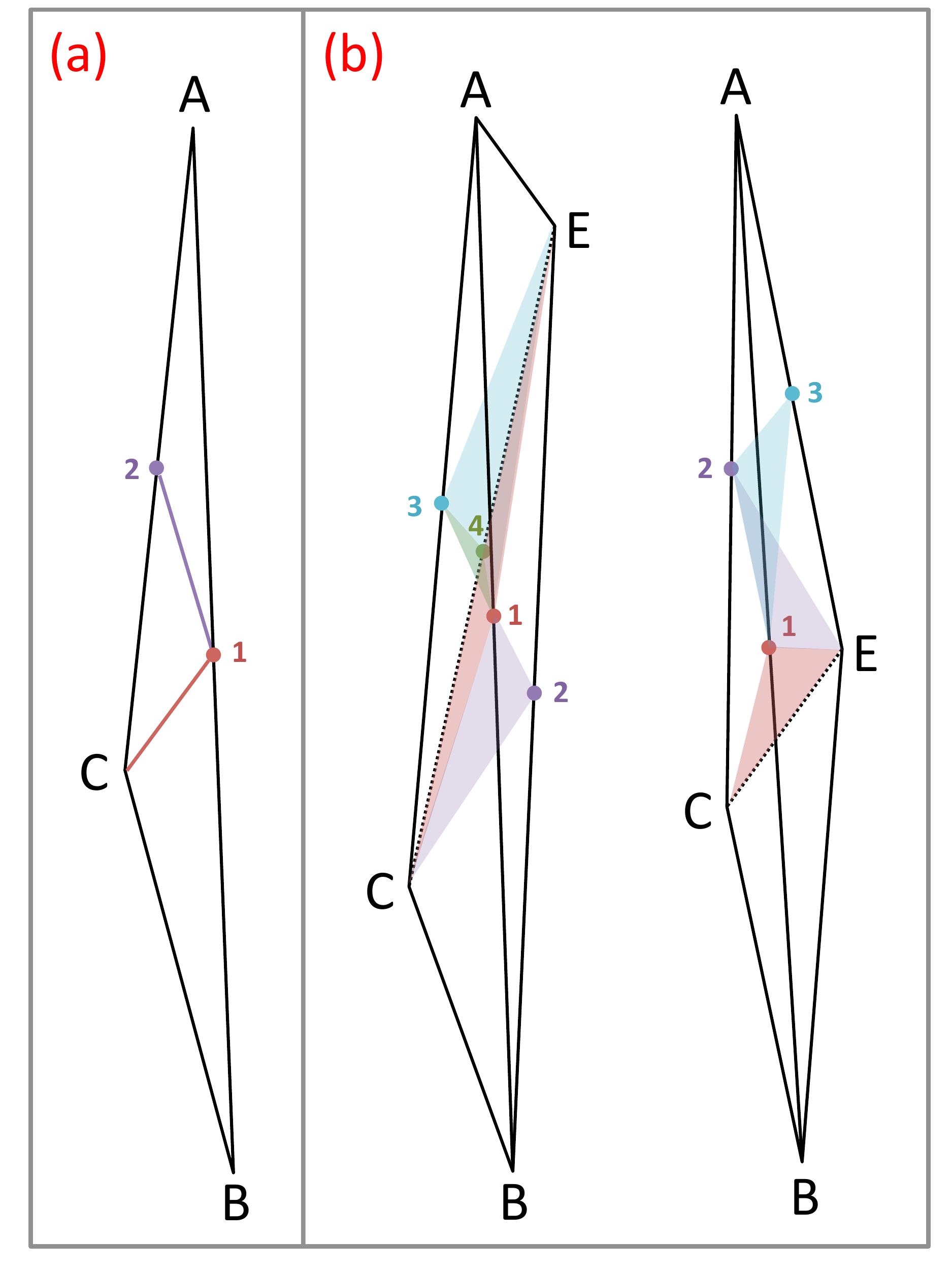}
\caption{Illustration of refinement in the case the simplex is very elongated. The left panel (a) corresponds to the two-dimensional case, while the right panel (b) shows the two typical configurations expected in the three-dimensional case. On this figure, refinement is triggered if the length of a segment exceeds a threshold $d_I$. The numbers indicate in which order bisection is performed. }
\label{fig:longsim}
\end{figure}

In this section, we try to estimate how the simplices count depends on the value of refinement criterion $\epsilon_I$ in the strongly anisotropic regime. Two cases which lead roughly to the same result are considered: extreme stretching of the phase-space sheet or/and large anisotropy of local curvature. With the appropriate choice of a new local system of coordinates aligned along the direction for which $\beta R$ in equation (\ref{eq:refde}) is minimal, followed by the appropriate scaling along each axis, one can reduce the refinement procedure to the condition 
\begin{equation}
d \lesssim d_I, \quad d_I^3 =\gamma \epsilon_I,
\end{equation}
where $\gamma$ is now a constant. In the new system of coordinates, the vertex is very elongated along one direction, e.g. $x$,  as illustrated by Fig.~\ref{fig:longsim}. If extreme stretching is the dominant phenomenon, we suppose that the initial tessellation is sufficiently dense so that refinement is not needed until significant elongation of the phase-space sheet has taken place (but the approximate results derived below are probably roughly correct even without this assumption). Note that in the presence of stretching, the direction $x$ is not necessarily aligned with the new system of coordinates. 

We consider a simplex $i$ that was already obtained from the refinement of a parent. With our refinement criterion, the longest side of $i$ is thus larger than $d_I$ and shorter than $2d_I$. We now examine what happens if refinement threshold $\epsilon_I$ is divided by a factor 8 i.e. if $d_I$ is divided by a factor two, $d'_I=d_I/2$. In two dimensions, simplex $i$ will be refined once or twice as illustrated by left panel of Fig.~\ref{fig:longsim}. It cannot be refined more than twice, because it is easy to convince oneself that the sides of its children are  shorter than $d'_I$. Analogously, in three dimensions (right panel of Fig.~\ref{fig:longsim}), simplex $i$ will be refined once, twice, three times or four times. Assuming extreme stretching, the simplex can be roughly reduced to a line parallel to $x$. Let $x_A$ be the coordinate of leftmost point $A$ of the triangle on this line and $x_B$ the coordinate of the rightmost point $B$. We have, from the arguments above
\begin{equation}
x_A+d < x_B \leq 2d.
\end{equation}
The last point(s), $C$ and $E$ verify
\begin{eqnarray}
x_A \leq x_C \leq x_B,\\
x_A \leq x_E \leq x_B.
\end{eqnarray}
We now estimate the respective probabilities $p_i$ that the simplex is refined once and twice in the 2D case, once, twice, three times and four times in the 3D case. To be able to compute these probabilities, we need to have access to probability distributions of relative positions $b\equiv (x_B-x_A)/d$, $c \equiv (x_C-x_A)/d$ and $e \equiv (x_E-x_A)/d$, a rather non trivial exercise. To simplify the calculations, we make the very strong but natural assumption that maximum length of simplices varies uniformly between $d$ and $2d$, so the density probability of $b$ is simply
\begin{eqnarray}
p(b) =\begin{cases} 1 \quad {\rm  if}\ 1 < b \le 2,\\
      0 \quad {\rm otherwise}.
\end{cases}
\end{eqnarray}
Similarly, we assume that the points $C$ and $E$ are uniformly distributed along segment $[A,B]$, so
\begin{eqnarray}
p(e) =p(c) =\begin{cases} \frac{1}{b} \quad {\rm if}\ 0 \leq c,e \le b,\\
                0 \quad  {\rm otherwise}.
               \end{cases}
\end{eqnarray}
Hence, 
\begin{eqnarray}
p_2^{2D} &=& \int_1^2 p(b) {\rm d} b\left[ {\int_1^b p(c) {\rm d}c +\int_0^{b-1} p(c) {\rm d}c} \right]=2 -\ln 4\simeq 0.61, \\
p_1^{2D} &=& 1-p_2=\ln 4 -1 \simeq  0.39.
\end{eqnarray}
Therefore, reducing $d_I$ by a factor two implies an increase of the simplex count by
\begin{equation}
\alpha_{2D} \simeq 2 p_1^{2D} +3 p_2^{2D}=4 -\ln 4 \simeq 2.6,
\end{equation}
implying a simplices count $n$ finally scaling like
\begin{equation}
n \propto \epsilon_I^{-\ln \alpha_{2D}/\ln 8} \sim \epsilon_I^{-0.46} .
\label{eq:nfound2d}
\end{equation}
Of course this result relies on strong assumptions so it should be only a very rough approximation of reality.  Note that a naive 
calculation simply assuming that refining once is as likely as refining twice, 
$p_1^{2D}=p_2^{2D}=0.5$ just gives $\alpha_{2D} = 5/2$ and $N \propto \epsilon_I^{-0.44}$, a result very
close to equation (\ref{eq:nfound2d}).  In the worse possible case, $p_2^{2D} \simeq 1$, we still have
\begin{equation}
n < n_{\rm worse}^{2D} \propto \epsilon_I^{\ln 3/\ln 8} \simeq \epsilon_I^{-0.53},
\end{equation}
a scaling still better than what would be given by isotropic refinement $N \propto \epsilon_I^{-2/3}$ (see \S~\ref{sec:2dchaos}). 

Likewise, the calculations can be performed in the 3D case, they are just slightly more cumbersome:
the probability of having four refinement points is equal to that of having a refinement point on segment $[CE]$, hence
\begin{equation}
p_4^{3D}=2\int_{1}^{2} p(b) {\rm d}b \int_0^{b-1} p(c) {\rm d}c \int_{c+1}^b p(e) {\rm d}e=\frac{3}{2} - 2 \ln 2 \simeq 0.11.
\end{equation}
The probability of having three refinement points exactly is,
\begin{equation}
p_{3}^{3D}=\int_{1}^{2} p(b)  {\rm d}b \left[ \int_0^{b-1} p(c) {\rm d}c +\int_{1}^b p(c) {\rm d}c \right]^2-p_4^{3D}=\frac{9}{2}-6 \ln 2 \simeq 0.34.
\end{equation}
Similarly,
\begin{equation}
p_2^{3D}=2 \int_{1}^{2} p(b) {\rm d} b  \left[ \int_0^{b-1} p(c) {\rm d}c +\int_{1}^b p(c) {\rm d}c \right] \int_{b-1}^1 p(c) {\rm d}c=12 \ln 2 - 8 \simeq 0.32,
\end{equation}
and finally,
\begin{equation}
p_1^{3D}=1-p_2^{3D}-p_3^{3D}-p_4^{3D}=3-4 \ln 2 \simeq 0.23.
\end{equation}
So at the end, decreasing the refinement threshold by a factor 8 roughly increases the number of simplices by a factor
\begin{equation}
\alpha_{3D} \simeq \sum_{j=1}^{4} p_j^{3D} (j+1)=\frac{15}{2}-6 \ln 2 \simeq 3.34,
\end{equation}
to be compared to the naive average $\alpha_{3D}=(2+3+4+5)/4=3.5$. As a result, in the three-dimensional case, we
can expect the following typical scaling for the number of simplices in the strongly anisotropic case:
\begin{equation}
n \propto \epsilon_I^{-\ln \alpha_{3D}/\ln 8} \sim \epsilon_I^{-0.58}.
\end{equation}
While this equation is expected to be a reasonable approximation of the simplices count, its calculation relies on strong assumptions.
Examining the worse possible case ($p_4^{3D} \simeq 1$)  we can derive a more firm bound on $n$,
\begin{equation}
n < n_{\rm worse}^{3D} \propto \epsilon_I^{-\ln 5/\ln 8} \simeq \epsilon_I^{-0.77},
\end{equation}
in the strongly anisotropic limit, a scaling still better than what is obtained with isotropic refinement $n < \epsilon_I^{-1}$
(see \S~\ref{sec:2dchaos}). 


\section*{References}
\bibliographystyle{elsarticle-harv} 
\bibliography{mybib}

\end{document}